\documentclass[aps,pre,reprint,twocolumn,superscriptaddress,showpacs]{revtex4-1}
\usepackage{amssymb,amsmath}
\usepackage[table]{xcolor}
\usepackage{graphicx}
\usepackage{mathtools}
\usepackage[export]{adjustbox}
\usepackage{overpic}
\usepackage[colorlinks=true,
 linkcolor=black,
 urlcolor=blue,
 citecolor=blue]{hyperref}
\usepackage{nicefrac}
\usepackage{multirow}
\usepackage{lipsum} 
\usepackage[normalem]{ulem}
\usepackage{pbox}
\newcolumntype{C}[1]{>{\centering\let\newline\\\arraybackslash\hspace{0pt}}m{#1}}

\usepackage{enumitem}

\usepackage{framed,color}
\definecolor{shadecolor}{rgb}{0.85,0.80,0.80}

\definecolor{myorange}{RGB}{253, 184, 99}
\definecolor{mypurple}{RGB}{178, 171, 210}

\newcommand{\comments}[1]{}
\usepackage{multirow}

\usepackage{float}

\newcommand{\avg}[1]{\left\langle{#1}\right\rangle}

\newcommand{\beq}{\begin{equation}}
\newcommand{\eeq}{\end{equation}}
\newcommand{\bal}{\begin{aligned}}
\newcommand{\eal}{\end{aligned}}

\newcommand{\be}{\begin{equation}}
\newcommand{\ee}{\end{equation}}
\newcommand{\bd}{\begin{displaymath}}
\newcommand{\ed}{\end{displaymath}}
\newcommand{\BE}{\begin{eqnarray}}
\newcommand{\EE}{\end{eqnarray}}

\allowdisplaybreaks
\begin{document}
\title{Stochastic fluctuations and quasi-pattern formation in reaction-diffusion systems with anomalous transport}
\author{Joseph W. Baron}
\email{joseph.baron@postgrad.manchester.ac.uk}
\affiliation{Theoretical Physics, School of Physics and Astronomy,
The University of Manchester, Manchester M13 9PL, United Kingdom}

\author{Tobias Galla}
\email{tobias.galla@manchester.ac.uk}
\affiliation{Theoretical Physics, School of Physics and Astronomy,
The University of Manchester, Manchester M13 9PL, United Kingdom}
 
\begin{abstract}
Many approaches to modelling reaction-diffusion systems with anomalous transport rely on deterministic equations and ignore fluctuations arising due to finite particle numbers. Starting from an individual-based model we use a generating-functional approach to derive a Gaussian approximation for this intrinsic noise in subdiffusive systems. This results in corrections to the deterministic fractional reaction-diffusion equations. Using this analytical approach, we study the onset of noise-driven quasi-patterns in reaction-subdiffusion systems. We find that subdiffusion can be conducive to the formation of both deterministic and stochastic patterns. Our analysis shows that the combination of subdiffusion and intrinsic stochasticity can reduce the threshold ratio of the effective diffusion coefficients required for pattern formation to a greater degree than either effect on its own.
\end{abstract}


\maketitle

\section{Introduction}
Reaction-diffusion schemes are a well-established tool for modelling non-linear pattern-forming phenomena in a wide variety of systems, ranging from developmental biology \cite{gierer,murray1,murray2} to chemical reactions \cite{bzreaction}, from the self-organisation of slime mold \cite{slimemold} and the distribution of mussels \cite{vandekoppel} to Rayleigh-B\' enard convection in fluids \cite{rayleigh}. Most notably, Turing showed in his seminal work \cite{turingchemical}, that systems which have a stable fixed point when diffusion is not present could exhibit an instability in the presence of diffusion. He showed that diffusion, normally considered a stabilising, equilibrating influence, combined with activator-inhibitor reaction dynamics, could provide a simple mechanism for pattern formation.

A difficulty that arises in the Turing interpretation of pattern formation is that the diffusion constants of the activator and inhibitor species are often required to be quite disparate in order for patterns to be found \cite{murraybook}. These large ratios of diffusion constants are unphysical.

One potential solution to this is that a system may not have to undergo a deterministic  Turing instability in order to exhibit pattern formation. Stochastic quasi-patterns, which arise due to the intrinsic noise of individual-based systems, may be sufficient to explain the spatial ordering observed in some natural systems \cite{butlergoldenfeld1,butlergoldenfeld2}. The emergence of such quasi-patterns requires a lower threshold ratio of diffusion coefficients than the emergence of deterministic patterns \cite{biancalani2,biancalani}. Noise-induced quasi-patterns have been observed recently in experimental biological systems \cite{stochpattexp,dipatti}.

Another proposed solution to this problem is modelling one or more of the components as subdiffusing \cite{randomwalkguide, metzlerklafter2}; that is, diffusing with a mean squared displacement which is sub-linear in time. This has also been shown to reduce the ratio of the effective diffusion coefficients required for pattern formation \cite{yadavmilu,barongalla}. Furthermore, subdiffusion has been observed in many experimental systems with low particle numbers, such as proteins moving in the cell membrane \cite{ghoshwebb, schwille} and in other transport phenomena involving obstacles or binding sites \cite{saxton1994,saxton1996}. It has been argued that morphogens could be subject to similar binding and trapping effects and move subdiffusively as a result \cite{hornung, yustabadlindenberg, fedotovfalconer2}. 

However, as of yet, the treatment of reaction-subdiffusion systems has been restricted to deterministic equations which ignore the stochastic effects due to the finite numbers of particles involved. Examples of the deterministic treatment include \cite{fedotovfalconer2, fedotovfalconer3, yustabadlindenberg}. Stochastic effects due to demographic noise are important particularly in biological systems, where subdiffusion has been observed, and where the particle numbers involved can be sufficiently small for noise to be non-negligible. From a more theoretical perspective, subdiffusion is a non-Markovian phenomenon. Because of this, reaction-diffusion systems with subdiffusing components provide a good opportunity to study the combination of memory effects and intrinsic noise.

In this work, we use an individual-based approach, which explicitly takes into account intrinsic stochasticity, to study systems with reactions and anomalous diffusion. We carry out a path-integral (or generating-functional) analysis of the stochastic dynamics, combining transport and reactions, and perform an expansion in the strength of the intrinsic noise. To lowest order, effectively neglecting fluctuations, we recover the familiar deterministic reaction-subdiffusion equations. Taking into account sub-leading orders of fluctuations, we find additional coloured Gaussian noise terms in these equations, encapsulating the intrinsic stochasticity of the individual-based system. We show that the analytical expressions for this noise can be used to characterise the emergence of stochastic quasi-patterns in reaction systems with subdiffusing components. Specifically, we use both the Brusselator \cite{brusselator} and Lengyel-Epstein \cite{lengyelepstein1,lengyelepstein2} systems as examples. We conclude that the combination of the two effects, noise-driven quasi-patterns and subdiffusion, can lower the critical ratio of diffusion coefficients required for pattern formation significantly, and to a greater degree than either effect on its own.

The remainder of this paper is set out as follows: In Section \ref{model}, we describe the individual-based model in detail and briefly present some background material related to subdiffusion and to the Brusselator and Lengyel-Epstein models. Section \ref{genfuncts} contains an outline of our analysis of the individual-based system and the generating-functional approach to approximating the noise; further details of this calculation can be found in the Supplemental Material. In Section \ref{reactdiffeqs}, we show that this approach can be used to derive the familiar reaction-diffusion equations with fractional diffusion, now with additional noise terms. In Section \ref{fluctuations}, we describe how one can then compute the fluctuations about the deterministic solution, using the linear-noise approximation. We verify our approach by comparing our theory predictions to stochastic simulations. In Section \ref{stochpatt}, we go on to use our theoretical approach to find the parameter regions where stochastic patterns are present and where deterministic patterns are present. Finally, in Section \ref{conc} we discuss our results and conclude.


\section{Model construction and background}\label{model}
In this section we introduce our model and the general notation, and we briefly summarise some background material. We also define the Brusselator and Lengyel-Epstein models, which we later use to illustrate the formation of quasi-patterns in stochastic subdiffusive systems.  

\subsection{Individual-based model}\label{modeldef}
We consider a general class of individual-based reaction-diffusion models. We refer to the individuals as `particles' from here on in, but these particles may represent biological entities or molecules of chemical reactants. 

Several species of particle react with one another and hop around on a discrete lattice. We use $\alpha$ to index the different species. Each lattice site can be occupied by multiple particles simultaneously, and we denote the number of particles of type $\alpha$ at site $i$ at time $t$ by $n^\alpha_{i,t}$. Reactions between particles occur locally within a given site of the lattice, and result in the annihilation and creation of particles, or the conversion of one type of particle into another. For the purposes of simplicity, we will confine ourselves to a discrete one-dimensional lattice with periodic boundary conditions. Most of our analysis could be generalised to other spatial arrangements or to higher dimensions, and indeed the continuum limit may be taken.

Particles are assigned a random waiting-time at birth (creation), drawn from a, as yet unspecified, waiting-time probability density function $\psi^\alpha\left(\tau\right)$. These may be different for the different species of particle, as indicated by the superscript. The particle then hops to a new location once it has waited for the assigned time, assuming that it has not been eliminated in a reaction. A new waiting-time is drawn from $\psi^\alpha\left(\tau\right)$ once the particle has hopped and the process begins again. This means that a particle which has been at its current site for an amount of time $\tau$ hops with a rate $h^\alpha_\tau = \frac{\psi^\alpha\left(\tau\right)}{\Psi^\alpha\left(\tau\right)}$, where $\Psi^\alpha\left(\tau\right) = 1-\int^\tau_0\psi^\alpha\left(\tau'\right)d\tau'$ is the survival probability. The quantity $h^\alpha_\tau$ is often referred to as the hazard rate \cite{randomwalkguide}; we use subscript notation for $\tau$ for later convenience. When the particle does hop, its new location is drawn from the hopping kernel $\phi_{i,i'}$. Here $\phi_{i,i'}$ is the probability that the particle hops from location $i'$ to $i$, given that a hopping event occurs. For our purposes, $\phi_{i,i'}$ will be a function of $\lvert i'-i \rvert$ only, so as to ensure the translational invariance of the problem. In principle the hopping kernel could also be different for the different species. In order to keep the model simple we use the same hopping kernel for all species; it is straightforward to extend the model to the more general case.

During the particles' sojourn periods at a given site, they may undergo reactions. We index the various reaction types with $r$. The rate $\lambda_{i,r,t}$ at which reactions of type $r$ occur at site $i$ is in general dependent on the number of particles of the various types currently at site $i$.  We denote the number of particles of type $\alpha$ that are produced or annihilated in a reaction of type $r$ by $\nu^\alpha_r$, which can be positive or negative. The constants $\nu^\alpha_r$ are the so-called stoichiometric coefficients \cite{chemicalterminology}. If a particle is annihilated in a reaction then the hop which was scheduled to occur for that particle no longer occurs. As a result of this and of the fact that the reaction rates depend on local concentrations, the hopping process and the reactions are interdependent and cannot easily be separated as in conventional reaction-diffusion equations (such as those used by Turing \cite{turingchemical}).

\subsection{Subdiffusion}\label{section:anomalousdiffusion}
We first discuss the phenomenon of subdiffusion in the case where particles undergo hopping events but no reactions. Particles are said to undergo subdiffusion if the mean-squared displacement for a single particle behaves as follows in the long term: $\langle x(t)^2 \rangle \sim t^\gamma$, where $0< \gamma < 1$. The behaviour approaches normal diffusion as $\gamma \to 1$. Subdiffusive transport may be produced by choosing a long-tailed waiting-time distribution $\psi\left(t\right)$ for particles which hop around as described in the previous section (for the time being we suppress the dependence on the particle species $\alpha$). The distribution we will use in this paper to model subdiffusion is that of Mittag-Leffler, which produces the desired behaviour and is particularly convenient for the theoretical analysis. That is, we choose $\psi\left(t\right) = -\frac{d}{dt} E_\gamma\left[ - \left( \frac{t}{t_0}\right)^\gamma\right]$ where $E_\gamma\left[ x \right]$ is the Mittag-Leffler function \cite{gorenflo}. The parameter $t_0$ sets the overall scale of the hopping process. This distribution has the convenient property that its Laplace transform is given by $\hat\psi\left(u\right) = \frac{1}{1+\left(t_0u\right)^{\gamma}}$. Notably, for $\gamma = 1$ the Mittag-Leffler function reduces to an exponential such that $\psi\left(t\right) = \frac{1}{t_0} e^{-\frac{t}{t_0}}$. 

For a single species of diffusing particle, which undergoes no reactions, one may use the Mittag-Leffler waiting-time distribution, along with the Montroll-Weiss formula \cite{montrollweiss,montrollweissoriginal},  to recover the fractional diffusion equation \cite{anomaloustransport, randomwalkguide, mendezfedotov}
\begin{gather}
\frac{\partial P_{i,t}}{\partial t} = \sum_{i'} \left(\phi_{i,i' } - \delta_{i,i'}\right) t_0^{-\gamma} {}_0D_t^{1-\gamma} P_{i,t}  , \label{fractionaldiff}
\end{gather}
where $P_{i,t}$ is the probability of finding a particular such particle at position $i$ at time $t$. The Riemann-Liouville fractional derivative ${}_0D_t^{1-\gamma}$ is defined as 
\begin{equation}
{}_0D_t^{1-\gamma} f\left(t\right)= \frac{1}{\Gamma\left(\gamma\right)} \frac{\partial}{\partial t} \int_0^t \frac{f\left(t'\right)}{\left(t-t'\right)^{1-\gamma}}dt' ,
\end{equation}
where $\Gamma(\cdot)$ denotes the gamma function. The fractional derivative has the property
\begin{equation}
\mathcal{L}_t\left\{ {}_0D_t^{1-\gamma} f\left(t\right)\right\}\left(u\right)= u^{1-\gamma} \hat f \left(u\right) ,
\end{equation}
where we use $\mathcal{L}_t$ to denote the Laplace transform with respect to $t$. One can show from Eq. (\ref{fractionaldiff}) that $\langle i^2 \rangle = \frac{2}{3 \Gamma\left(1+\gamma\right)} \left(\frac{t}{t_0}\right)^{\gamma}$, if one uses the symmetrical hopping kernel
\begin{equation}
\phi_{i,i'} =
\begin{cases}
	\frac{1}{3},				& \mathrm{if} \,\,\, i = i'+1\\
	\frac{1}{3},                & \mathrm{if} \,\,\, i = i' \\
	\frac{1}{3}, 			    & \mathrm{if} \,\,\, i = i'-1	
\end{cases} \;\;\; .
\end{equation}
In our system, different species of particles may hop with different typical waiting times $t_0^\alpha$ and different anomalous exponents $\gamma^\alpha$. Precisely, the waiting-time distributions for each species are given by
\begin{align}
\psi^{\alpha}\left(t\right) &= -\frac{d}{dt} \left\{ E_{\gamma^\alpha}\left[ - \left( \frac{t}{t_0^{\alpha}}\right)^{\gamma^\alpha}\right] \right\}. \label{waitingtimes}
\end{align}
\subsection{Reaction schemes}\label{reactscheme}
As example systems, we will use the Brusselator \cite{brusselator} and the Lengyel-Epstein \cite{lengyelepstein1, lengyelepstein2} models. The Brusselator is an activator-inhibitor model, first conceived of to describe oscillatory chemical reactions such as the Belousov-Zhabotinsky reaction \cite{tysonbelousov}. The reactions between the two species of particle involved, $A$ and $B$, are given by
\begin{align}
\emptyset &\stackrel{aN}{\longrightarrow} X_A , \nonumber \\
2X_A + X_B &\stackrel{\frac{1}{N^2}}{\longrightarrow} 3 X_A , \nonumber \\
X_A &\stackrel{b}{\longrightarrow} X_B ,\nonumber \\
X_A &\stackrel{1}{\longrightarrow} \emptyset. \label{stoich}
\end{align}

Without hopping the system has a homogeneous stable deterministic fixed point at $\bar n^{\left(A\right)} = aN, \bar n^{\left(B\right)} = \frac{b}{a}N$ so long as $b<1+ a^2$, where $\bar n^{\left(\alpha\right)}$ is the number of particles of type $\alpha$ at the fixed point.

The Lengyel-Epstein model was introduced primarily as a way of modelling the $\textrm{ClO}_2^-$-$\textrm{I}^-$-$\textrm{MA}$ reaction, which exhibits Turing patterns experimentally \cite{castets}. We use a simplified two-species version of the full model, which effectively assumes that the concentrations of the remaining components are constant. The reactions are given by
\begin{align}
\emptyset &\stackrel{aN}{\longrightarrow} X_A , \nonumber \\
X_A &\stackrel{b}{\longrightarrow} X_B ,\nonumber \\
4X_A + X_B &\stackrel{R}{\longrightarrow} \emptyset, \label{stoichle}
\end{align}
where the rate $R= \frac{1}{\left(n^{\left(A\right)}\right)^3}\frac{cN}{dN^2 + \left(n^{\left(A\right)}\right)^2}$ is dependent on the concentration of type-A particles. The Lengyel-Epstein system has a homogeneous deterministic fixed point at $\bar n^{(A)} = \frac{aN}{5b}, \bar n^{(B)} = \frac{bdN}{c}\left[1+\left(\frac{\bar n^{(A)}}{\sqrt{d}N}\right)^2 \right]$ so long as $ca>\frac{3}{5}a^2 - 25b^2d$.

In both models, species $A$ is the activator and species $B$ is the inhibitor. The parameter $N$ in both systems characterises the typical number of particles per site and will become useful to us later when we perform a system-size expansion in order to analyse noise in the these systems. The reaction rates (the number of reactions which occur per unit time) are calculated according to the usual mass action kinetics \cite{erdi}. For example, the rate at which the last reaction in Eq.~(\ref{stoichle}) occurs is $R\times\left(n^{\left(A\right)}\right)^4 \times n^{\left(B\right)} = \frac{cN n^{\left(A\right)} n^{\left(B\right)}}{dN^2 + \left(n^{\left(A\right)}\right)^2}$.


\section{Approximation of the fluctuations in particle number using Gaussian noise}\label{genfuncts}
\subsection{Formulation of the problem}
In order to capture the noise-driven effects in the models described in section \ref{modeldef}, we carry out an expansion in the inverse system-size. The idea is similar to the principles underlying the Kramers--Moyal expansion or the system-size expansion by van Kampen for Markovian systems \cite{vankampen}. However, there are also conceptual differences due to the non-Markovian nature of subdiffusion. 

To carry out the expansion, we use the reciprocal of the parameter $N$ in Eqs.~(\ref{stoich}) and (\ref{stoichle}); $N$ characterises the typical particle number per site. More specifically, ensemble-averaged particle numbers are of the order $N$ at each lattice site, whereas the fluctuations in particle numbers are of the order $\sqrt{N}$. Thus, for large $N$, the noise is small in relative terms. The limit $N\to \infty$ reproduces the deterministic behaviour, and when $N$ is large but finite the expansion can be expected to accurately describe stochastic corrections. By performing the expansion, we obtain a set of stochastic differential equations (SDEs) \cite{gardiner} with Gaussian noise terms. These SDEs encapsulate not only the deterministic trajectory of the system but also the next-order stochastic noise corrections.

A complicating feature of systems where the waiting-time distribution for the hops is non-exponential is that the hazard rate for hopping is non-constant, meaning that what happens at a given point in time is dependent on the history of the system. A sensible and common way to deconvolute the problem is to introduce the age coordinate $\tau$, which denotes the length of time that a particle has resided at a particular location since its last hop \cite{cox}. That is, we denote the number of particles of type $\alpha$ which have resided at site $i$ for a time between $\tau_1$ and $\tau_2$ by $\int_{t_1}^{t_2}n^{\alpha}_{i,\tau,t}d\tau$. This effectively recasts the problem as Markovian. The quantity $n^\alpha_{i,\tau,t}$ is a density of particles per time $\tau$.

In order to develop the formalism we discretise time into steps of size $\Delta$. We assume that all reaction rates and hopping rates remain constant during each step, similar to the $\tau$-leaping approach to Gillespie simulations in discrete time \cite{tauleap}. Eventually, we will take the limit $\Delta \to 0$ to restore continuous time.

For our system, the quantity $n^{\alpha}_{i,\tau,t}$ may change due to two effects in each time step: reactions and hopping. We represent the changes due to these effects by the following quantities respectively: $k^{\left(R\right) \alpha}_{i,r, \tau, t}$ and $k^{\left(H\right) \alpha}_{i,i', \tau, t}$. That is, $k^{\left(R\right) \alpha}_{i,r, \tau, t}$ is the number of particles of type $\alpha$ and age $\tau$ at position $i$ which are annihilated in the time step from $t$ to $t+\Delta$ due to reactions of type $r$. We note that newly created particles have age zero; this will be dealt with separately below. Similarly, $k^{\left(H\right) \alpha}_{i,i', \tau, t}$ is the number of particles of age $\tau$ hopping away from position $i'$ to $i$ at time $t$. The sets of integer variables $\{k^{\left(R\right) \alpha}_{i,r, \tau, t}\}$ and $\{k^{\left(H\right) \alpha}_{i,i', \tau, t}\}$ are stochastic, that is they take different values for every realisation of the system. 

We then have 
\begin{align}
n^{\alpha}_{i,\tau+\Delta,t+\Delta} - n^{\alpha}_{i,\tau,t} =& -\sum_{r} \frac{ k^{\left(R\right) \alpha}_{i,r, \tau, t} \theta\left(- \nu^{\alpha}_{r}\right)}{\Delta}\nonumber \\
 &- \sum_{i'} \frac{k^{\left(H\right) \alpha}_{i',i, \tau, t} }{\Delta} , \label{discrete1}
\end{align}

where $\theta\left(\cdot \right)$ denotes the Heaviside function. We note that, in the discrete-time setup, the total number of particles at site $i$ at time $t$ is given by $n^\alpha_{i,t}=\Delta\sum_{m} n^\alpha_{i,\tau=m\Delta,t}$; the quantity $n^\alpha_{i,\tau,t}$ therefore has dimensions of inverse time, in-line with the expression on the right-hand side of Eq. (\ref{discrete1}). 

Eq.~(\ref{discrete1}) only includes reactions which annihilate particles, i.e. it describes the outflux of particles from position $i$. When particles are produced, or when they newly arrive at a location after hopping, they have age $\tau = 0$. Therefore, the influx of particles to position $i$ is given by
\begin{gather}
n^{\alpha}_{i,0,t+\Delta} = \sum_{r} \frac{\ell_{i,r,t}^{\left(R\right) }\nu^{\alpha}_{r} \theta \left(\nu^{\alpha}_{r}\right) }{\Delta } + \sum_{i'} \frac{k^{\left(H\right) \alpha}_{i,i',\tau,t}}{\Delta }, \label{discrete2}
\end{gather}
where $\ell_{i,r,t}^{\left(R\right)}$ is the number of reactions of type $r$ firing in the time window $t$ to $t+\Delta$ at position $i$. We note also that $\theta\left(-\nu_r^\alpha\right) \ell^{\left(R\right)}_{i,r,t} \lvert \nu_r^\alpha\rvert= \sum_{\tau}k^{\left(R\right) \alpha}_{i,r, \tau, t}$.  Furthermore, particles of a given species which are annihilated due to a reaction at a given site are selected at random from all particles of this species at the site, irrespective of age. 

\subsection{Generating functional approach to the system-size expansion}
To approximate the stochastic fluctuations in particle number with Gaussian noise, one would normally write down a master equation and expand in powers of $N^{-1}$ in order to obtain a Fokker-Planck equation, see for example \cite{risken,gardiner, mckanenewman,biancalani}. From this Fokker-Planck equation one could read off the corresponding SDE. Since the hopping of the particles in our system is history-dependent, there is no straightforward way to write down a master equation, even after the introduction of the additional age coordinate $\tau$. The most natural way to analyse a non-Markovian stochastic system is with the Martin-Siggia-Rose-Janssen-De Dominicis (MSRJD) path integral, which takes into account all possible histories of the system \cite{MSR,J,D}. Such an approach allows one to perform the expansion in inverse powers of the system-size without writing down a master equation explicitly \cite{brettgalla, brettgalla2}.\\
In order to reflect the system-size dependence in our calculation we introduce $x^{\alpha}_{i,\tau,t} = \frac{n^{\alpha}_{i,\tau,t}}{N}$. These quantities are of order $N^0$. The MSRJD generating functional then takes the form of a path integral over all possible trajectories of the variables $\{x^{\alpha}_{i, \tau, t}\}$ \cite{altlandsimons}. It can be written as
\be
Z\Big[ \{ \Xi_{i,\tau,t}^\alpha \} \Big] = \avg{ \exp\left(i\sum_{i,\alpha,\tau,t} \Xi^{\alpha}_{i,\tau, t} x^{\alpha}_{i,\tau, t}\right)}_{\{x^{\alpha}_{i, \tau, t}\}},\label{generatingfunctional}
\ee
where $\avg{\dots}_{\{x^{\alpha}_{i, \tau, t}\}}$ denotes an average over all trajectories of the system. The $\{\Xi^{\alpha}_{i, \tau, t}\}$ are source variables. The procedure for performing the expansion in $N^{-1}$ is similar to that used in \cite{brettgalla,brettgalla2}: We find the joint probability distribution for the sets of variables $\{k^{\left(R\right) \alpha}_{i,r, \tau, t}\}$ and $\{k^{\left(H\right) \alpha}_{i,i', \tau, t}\}$ and rewrite the generating functional in term of these quantities. We then average these random numbers against their joint distribution, to obtain the generating functional in terms of only the coordinates $\{x^{\alpha}_{i, \tau, t}\}$  and the model parameters. We then carry out an expansion in $N^{-1}$ up to and including sub-leading order. This approximate generating functional is recognised as that of an effective SDE. The leading-order terms in this SDE correspond to the deterministic dynamics; after further re-arrangement it reproduces the well-known deterministic reaction-subdiffusion equation, as shown in Sec. \ref{reactdiffeqs}. This deterministic approximation is accurate in the limit $N\to \infty$. The next-order terms correspond to Gaussian noise corrections, with a standard deviation of order $N^{-1/2}$.

The details of this procedure are provided in the Supplementary Material (Sections S1 and S2); here we only quote the final result, in which we have restored continuous time. It is given by the following stochastic equations
\begin{align}
\frac{\partial n^{\alpha}_{i,\tau , t} }{\partial t} + \frac{\partial n^{\alpha}_{i,\tau, t} }{\partial \tau} &= -  h^\alpha_{\tau} n_{i,\tau,t}^{\alpha} - p_{i,t}^{\alpha} n_{i,\tau,t}^\alpha + \xi^{\alpha}_{{i,\tau,t}} , \nonumber \\
n^{\alpha}_{i, 0 , t} &= \sum_{i'} \phi_{i,i'} \int_0^t  h^\alpha_{\tau} n_{i',\tau,t}^{\alpha} d\tau + \gamma_{i,t}^\alpha  + \xi^{\alpha}_{{i,0,t}} .\label{stochasticeqs}
\end{align}
In these expressions, $p_{i,t}^\alpha$ is the per capita removal rate for particles of species $\alpha$ at position $i$ and time $t$; $\gamma_{i,t}^\alpha$ is the total production rate for the particles of type $\alpha$ at $i$ and $t$, that is
\begin{align}
p_{i,t}^{\alpha} n_{i,\tau,t}^\alpha &= \sum_{r} \lambda_{i,r,t} \lvert \nu_{r}^{\alpha} \rvert \frac{n_{i,\tau,t}^\alpha}{n_{i,t}^\alpha} \theta\left(-\nu_{r}^{\alpha}\right) , \nonumber \\
\gamma_{i,t}^\alpha  &= \sum_{r} \lambda_{i,r,t} \lvert \nu_{r}^{\alpha} \rvert \theta\left(\nu_{r}^{\alpha}\right).\label{bdrates}
\end{align}
The quantities $\lambda_{i,r,t}$, $\nu_r^\alpha$, $\phi_{i,i'}$ and $h_\tau^\alpha$ are defined in Section \ref{modeldef}. The fraction $n_{i,\tau,t}^\alpha/n_{i,t}^\alpha$ in the first of the relations in Eq.~(\ref{bdrates}) reflects the fact that the particles which are to be removed are selected irrespective of their age $\tau$. The quantities $\{ \xi^{\alpha}_{i,\tau,t}\}$ in Eq. (\ref{stochasticeqs}) represent white Gaussian noise of zero mean. Eqs.~(\ref{stochasticeqs}) are approximations of the full random process given by Eqs.~(\ref{discrete1}) and (\ref{discrete2}), in which the randomness is discrete. While the noise $\{ \xi^{\alpha}_{i,\tau,t}\}$ is white, we note that the components are correlated across species and lattice sites. The expressions for these correlations are somewhat lengthy; we derive and give them in Section S2 of the Supplemental Material. It is also important to note that the noise is multiplicative, i.e. the elements in the covariance matrix of the $\{ \xi^{\alpha}_{i,\tau,t}\}$ depend on the variables $\{n^\alpha_{i,\tau,t}\}$. This is similar to the outcome of a Kramers--Moyal expansion of the master equation for conventional Markovian systems \cite{gardiner,risken}.


\section{Fractional reaction-diffusion equation with noise}\label{reactdiffeqs}
Using the waiting-time distributions discussed in Section \ref{section:anomalousdiffusion} in conjunction with Eqs.~(\ref{stochasticeqs}), we are able to obtain the fractional reaction-diffusion equation reported in the literature (see e.g. \cite{yadavmilu, vladross, yustabadlindenberg, fedotovfalconer2}) if we neglect the noise terms. Including these terms, we are able to capture effects driven by the stochasticity of the original individual-based dynamics.

One proceeds from Eqs.~(\ref{stochasticeqs}) by integrating out the age variables $\tau$, with the aim of finding a time-evolution equation in terms of only $n_{i,t}^\alpha$ and the noise. Using the Mittag-Leffler function from Eq.~(\ref{waitingtimes}) as the waiting-time distribution, one arrives at the following fractional reaction-diffusion equation for the particle number $n_{i,t}^\alpha$
\begin{widetext}
\begin{align}
\frac{\partial n_{i,t}^\alpha }{\partial t}  = \sum_{i'} \left\{\left( \phi_{i,i'} - \delta_{i,i'}\right)\left(t_0^\alpha\right)^{-\gamma^\alpha} e^{ -\int_0^t p_{i',T'}^{\alpha}dT' } {}_0D^{1-\gamma^\alpha}_t\left[ n_{i',t}^\alpha e^{ \int_0^t p_{i',T'}^{\alpha}dT'} \right]\right\} +f^\alpha_{i,t} +\eta^\alpha_{i,t}, \label{reactsubdiff}
\end{align}
\end{widetext}
where we have defined the total reaction rate $f^\alpha_{i,t} = \gamma^\alpha_{i,t} - p^\alpha_{i,t} n^\alpha_{i,t}$.

In the limit of continuous space, the sum involving the hopping kernel becomes a Laplacian operator (i.e. $\sum_{i'}\left( \phi_{i,i'} - \delta_{i,i'}\right) \to \sigma^2 \nabla^2 $, with a suitable constant $\sigma^2$, related to the variance of hopping distances). Thus, the fractional reaction-diffusion equation Eq.~(\ref{reactsubdiff}) corresponds to Eq.~(3) in \cite{yadavmilu}, but with the addition of a noise term. We recover the deterministic reaction-subdiffusion equation if we take the infinite system-size limit, $N \to \infty$, whereupon the noise becomes negligible.  

One notes that the reaction and diffusion terms in Eq.~(\ref{reactsubdiff}) are coupled. This is due to the non-Markovian nature of the hopping and the fact that particles may be annihilated in reactions and are thus unable to perform scheduled hops. The exponential factors in Eqs.~(\ref{reactsubdiff}) correspond to the probability of particles surviving without being annihilated in a reaction.

The noise also involves such exponential factors and is given by
\begin{widetext}
\begin{align}
\eta^\alpha_{i,t} &= \sum_{i'} \Bigg\{ \left( \phi_{i,i'} - \delta_{i,i'}\right) \Bigg[ \int_0^t \psi^\alpha\left(\tau\right) e^{ -\int^{t}_{t-\tau} p_{i',T'}^{\alpha}dT' }\int^\tau_0 \frac{\xi^\alpha_{i',T,T+t-\tau}}{\Psi^\alpha\left(T\right) e^{-\int_0^{T} p_{i',T'+t-\tau}^{\alpha}dT' }} dTd\tau \nonumber \\
&- e^{ \int^{t}_0 p_{i',T'}^{\alpha}dT' } \left(t_0^\alpha\right)^{-\gamma^\alpha} {}_0 D_t^{1-\gamma^\alpha}\Bigg(\int_0^t \Psi^\alpha\left(\tau\right) e^{ \int^{t-\tau}_0 p_{i',T'}^{\alpha}dT' }\int^\tau_0 \frac{\xi^\alpha_{i',T,T+t-\tau}}{\Psi^\alpha\left(T\right) e^{-\int_0^{T} p_{i',T'+t-\tau}^{\alpha}dT' }} dTd\tau\Bigg)\Bigg] \Bigg\} \nonumber \\
&+\int_0^t \xi^\alpha_{i,\tau,t} d\tau + \xi^\alpha_{i,0,t} ~~ . \label{noisetermsmt}
\end{align}
\end{widetext}
The derivation of Eqs.~(\ref{reactsubdiff}) and (\ref{noisetermsmt}) from Eqs.~(\ref{stochasticeqs}) broadly follows the method in \cite{vladross} or \cite{fedotovfalconer1} but also handles the additional noise terms. It is given in the Supplemental Material (Section S3). The noise variables $\{\eta^\alpha_{i,t}\}$ involve integrals of the $\{ \xi^{\alpha}_{i,\tau,t}\}$, and as a consequence they are correlated in time, in addition to their correlations across components and lattice sites. Given that the statistics of $\{ \xi^{\alpha}_{i,\tau,t}\}$ depend on the variables $\{n^\alpha_{i,\tau,t}\}$, this noise is multiplicative as well.\\
Although the expression for the noise $\eta^\alpha_{i,t}$ in Eq.~(\ref{noisetermsmt}) appears cumbersome at first glance, a great deal of simplification can be achieved in the regime where one considers small deviations about the homogeneous fixed point of the deterministic system. We discuss this in the next section.\\

One notes that for $\gamma \to 1$ we recover the normal reaction-diffusion equation where the reactions and diffusion are uncoupled:
\begin{align}
\frac{\partial n_{i,t}^\alpha }{\partial t}  = \sum_{i'} \left[\left( \phi_{i,i'} - \delta_{i,i'}\right)\left(t_0^\alpha\right)^{-1}   n_{i',t}^\alpha \right] +f^\alpha_{i,t} +\eta^\alpha_{i,t}.
\end{align}
In this case, the expression for the noise also simplifies greatly and contains no exponential factors:
\begin{align}
\eta^\alpha_{i,t} = \int_0^t \xi^\alpha_{i,\tau,t} d\tau + \xi^\alpha_{i,0,t} .
\end{align}
This can be seen from Eq.~(\ref{noisetermsmt}) by using the fact that the Mittag-Leffler function reduces to an exponential in the limit $\gamma \to 1$ and that the fractional derivative becomes an identity operator. In this special case, $\langle \xi^\alpha_{i,\tau,t} \xi^{\alpha}_{i',\tau',t'} \rangle \propto \delta\left(t-t'\right)$, and we then also have $\langle \eta^\alpha_{i,\tau,t} \eta^{\alpha}_{i',\tau',t'} \rangle \propto \delta\left(t-t'\right)$. Therefore, in the Markovian limit, we recover white noise. This is explored further in the Supplemental material (Section S2).


\section{Linear-noise approximation and comparison with simulations}\label{fluctuations}

\subsection{Linear-noise approximation}
We now derive explicit expressions for the deviations of the stochastic system (i.e., with a finite number of particles per lattice site) from the solution of the deterministic reaction-subdiffusion equation.

We write $\bar n^{\alpha}_{i,t}$ for the solution of Eq.~(\ref{reactsubdiff}) with the noise term removed, and define the deviation from this  deterministic solution through
\begin{align} 
n^{\alpha}_{i,t}=  \bar n^\alpha_{i,t} + \delta^\alpha_{i,t}. \label{fluct}
\end{align}
We focus on fluctuations about the steady state, i.e. we assume that $\bar n^{\alpha}_{i,t}\equiv \bar n^{\alpha}$ is the homogeneous fixed point of the deterministic system. From the results stated so far, we are able to compute the correlation matrix of the fluctuations $\langle \delta^\alpha_{i,t} \delta^{\alpha'}_{i',t'} \rangle$, which we then compare with results from stochastic simulations of the individual-based system.

The expansion in powers of $N^{-1}$ in Section~\ref{genfuncts} was based on the assumption that the fluctuations about the deterministic trajectory were small in comparison to the total numbers of particles. A more precise formulation of this assumption allows us to make further simplifications. Since the noise term $\eta^\alpha_{i,t}$ in Eq.~(\ref{reactsubdiff}) is of order $N^{1/2}$, it is also reasonable that the fluctuations would be such that $\delta^\alpha_{i,t} \sim \mathcal{O}(N^{1/2})$. This is similar to the assumption used by van Kampen \cite{vankampen} to perform the system-size expansion now named after him. Using this, one can expand Eq.~(\ref{reactsubdiff}) about the deterministic fixed point $\bar n^{\alpha}$.

We note that the $\delta^\alpha_{i,t}$ are small deviations about the homogeneous deterministic fixed point. The correlators $\langle \delta^\alpha_{i,t} \delta^{\alpha'}_{i',t'} \rangle$ therefore do not describe any any deterministic pattern-forming features of the system. Instead, they characterise stochastic phenomena induced by fluctuations due to finite system sizes $N$.

To leading order (i.e., sending $N\to\infty$) we obtain the deterministic dynamics already mentioned, and in which all noise terms are removed. The terms to sub-leading order result in a linear expression for the fluctuations $\delta^\alpha_{i,t}$ in terms of $\eta^\alpha_{i,t}$. Within this order of approximation the quantities dependent upon $ n^\alpha_{i,t}$ in the noise correlators $\langle \eta^\alpha_{i,t}\eta^\alpha_{i',t'} \rangle$ are evaluated at the fixed point $\bar n^{\alpha}$. The noise in the dynamics for the $\{\delta^\alpha_{i,t}\}$ is therefore now additive rather than multiplicative. The procedure is discussed further in Section S4 of the Supplemental Material.

\begin{figure*}[t!!]
\includegraphics[scale = 0.2]{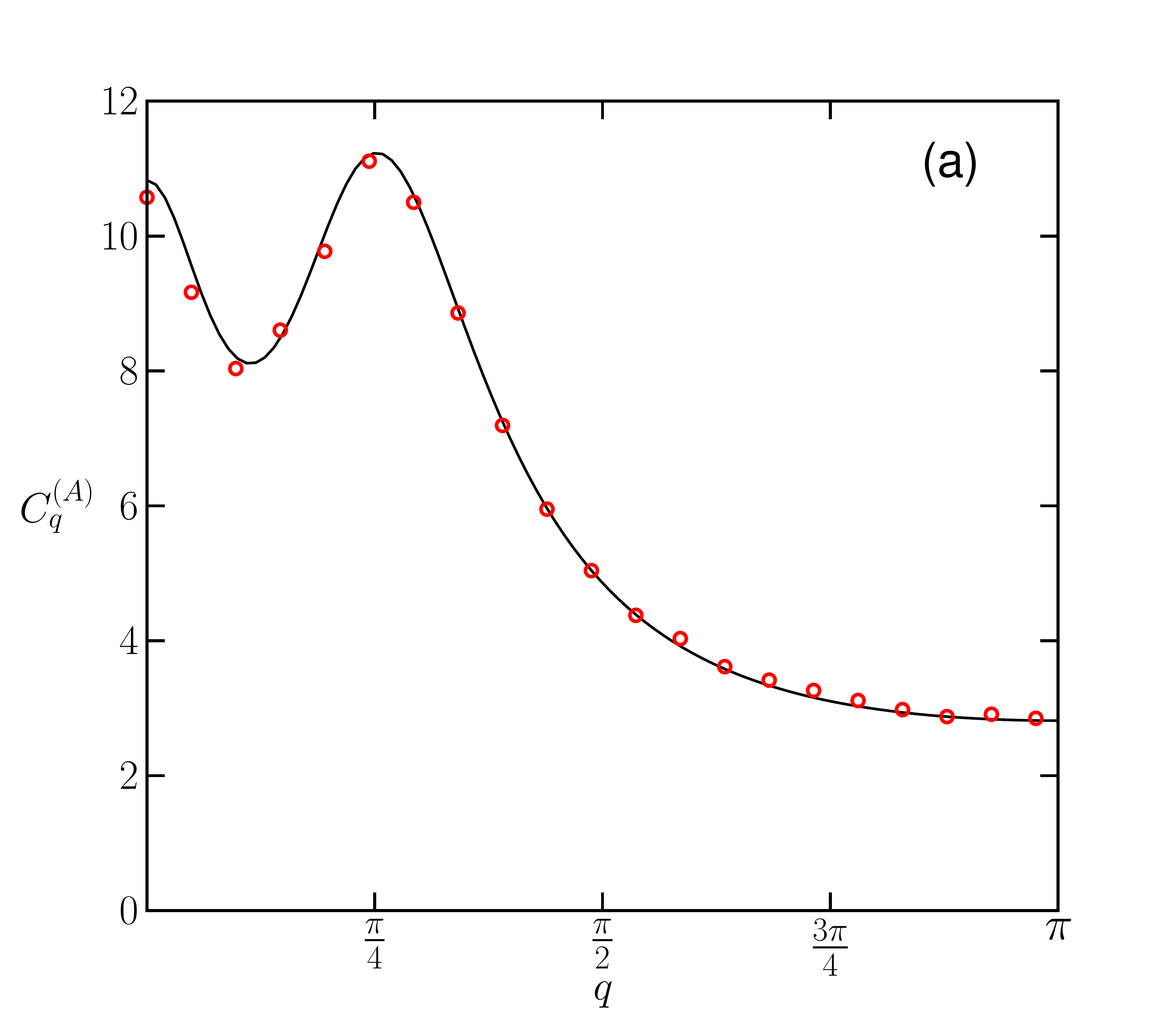}\hspace{2cm}
\includegraphics[scale = 0.2]{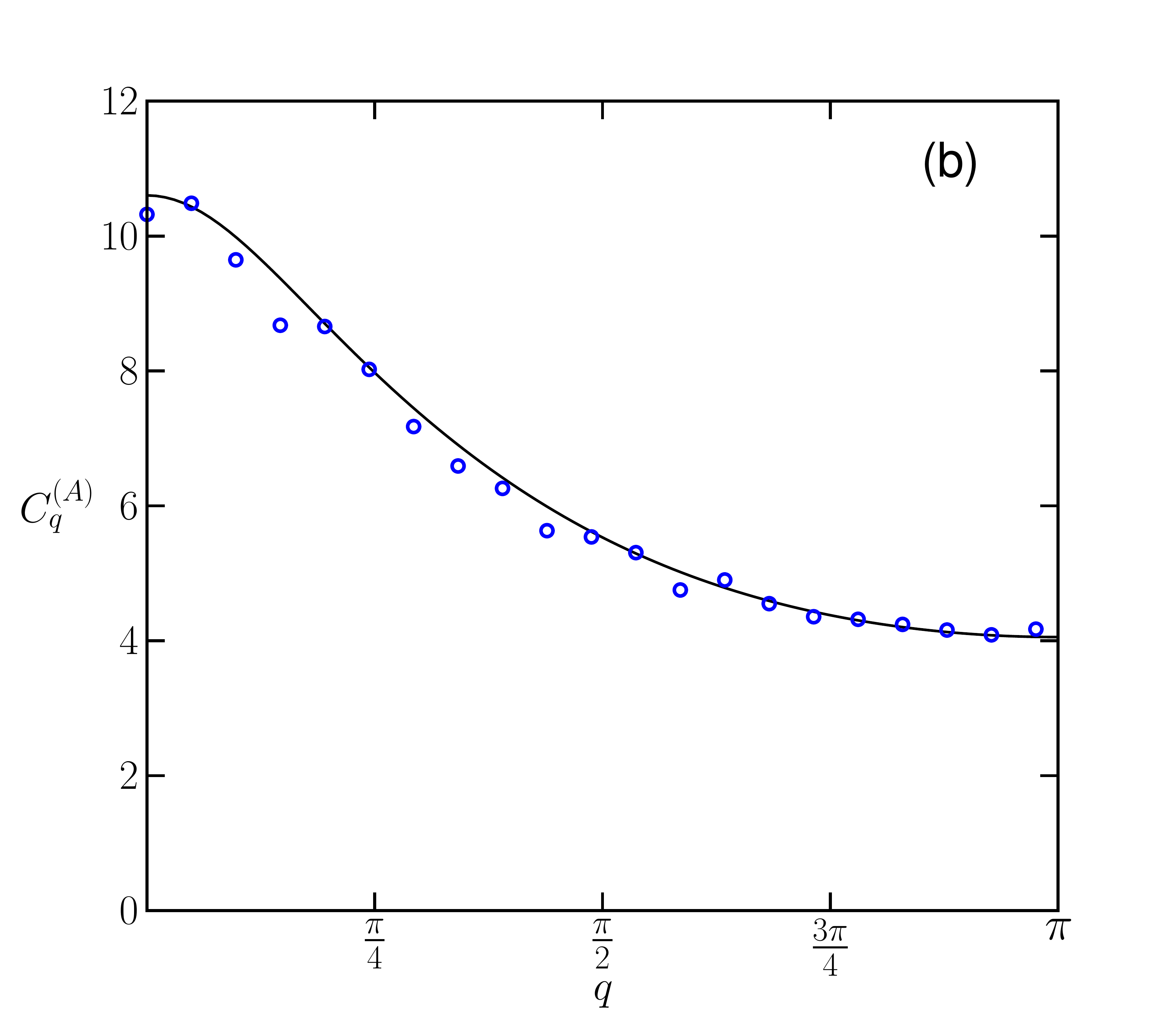}
 \caption{The correlator $C_q^{\left(A\right)}=\langle \lvert \tilde\delta^{\left(A\right)}_{q,t} \rvert^2\rangle/N$ for the Brusselator model with a subdiffusing activator (a) and a subdiffusing inhibitor (b). The markers represent simulation results and the solid lines depict results derived from the theory in the text. The model parameters are (a) $t_0^{\left(A\right)} = 0.6$ ,  $t_0^{\left(B\right)} = 0.1$, and (b) $t_0^{\left(A\right)} = 1.33$ ,  $t_0^{\left(B\right)} = 0.1$, giving an effective diffusion coefficient ratio of $\theta_\gamma = 4.63$ for both plots [see Eqs.~(\ref{effdiffcoeff1}) and (\ref{effdiffcoeff2})]. The remaining model parameters are $N = 4000$, $b= 1.8$ and $a = 1.1$. We have $\gamma = 0.5$ for the subdiffusing component in both figures. One observes that a range of non-zero Fourier modes $q$ are excited to a greater extent than the $q=0$ mode when the activator is subdiffusing (a) for this parameter set. This is not the case when inhibitor subdiffuses (b). The simulations were averaged over $1000$ trials with $41$ discrete positions on the lattice. Data was taken at $t = 20$ to ensure a stationary state had been reached.}
	\label{fig:equaltime}
\end{figure*}

\begin{figure*}[t!]
   
       \includegraphics[scale = 0.18]{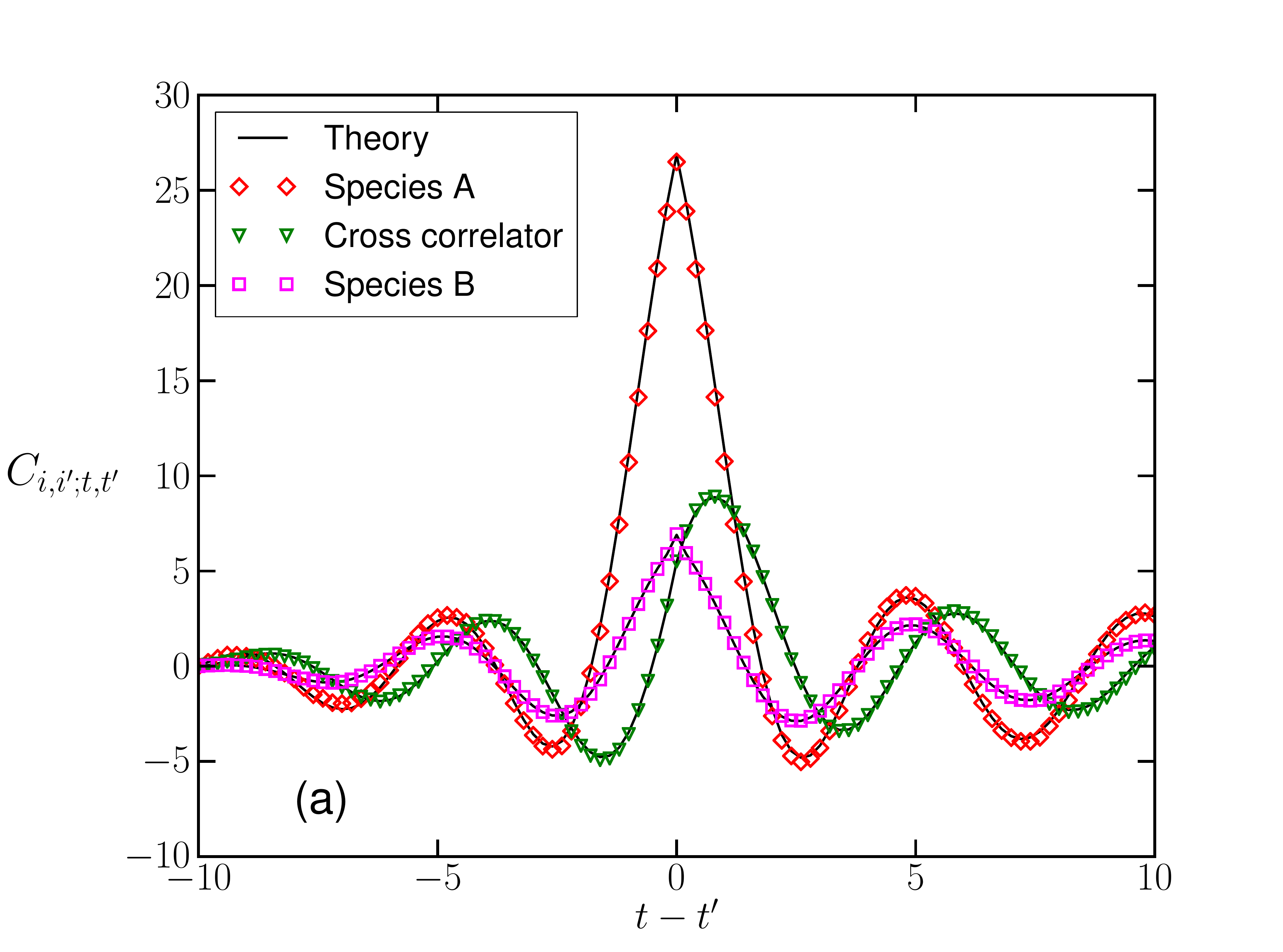}\hspace{2cm}
         \includegraphics[scale = 0.18]{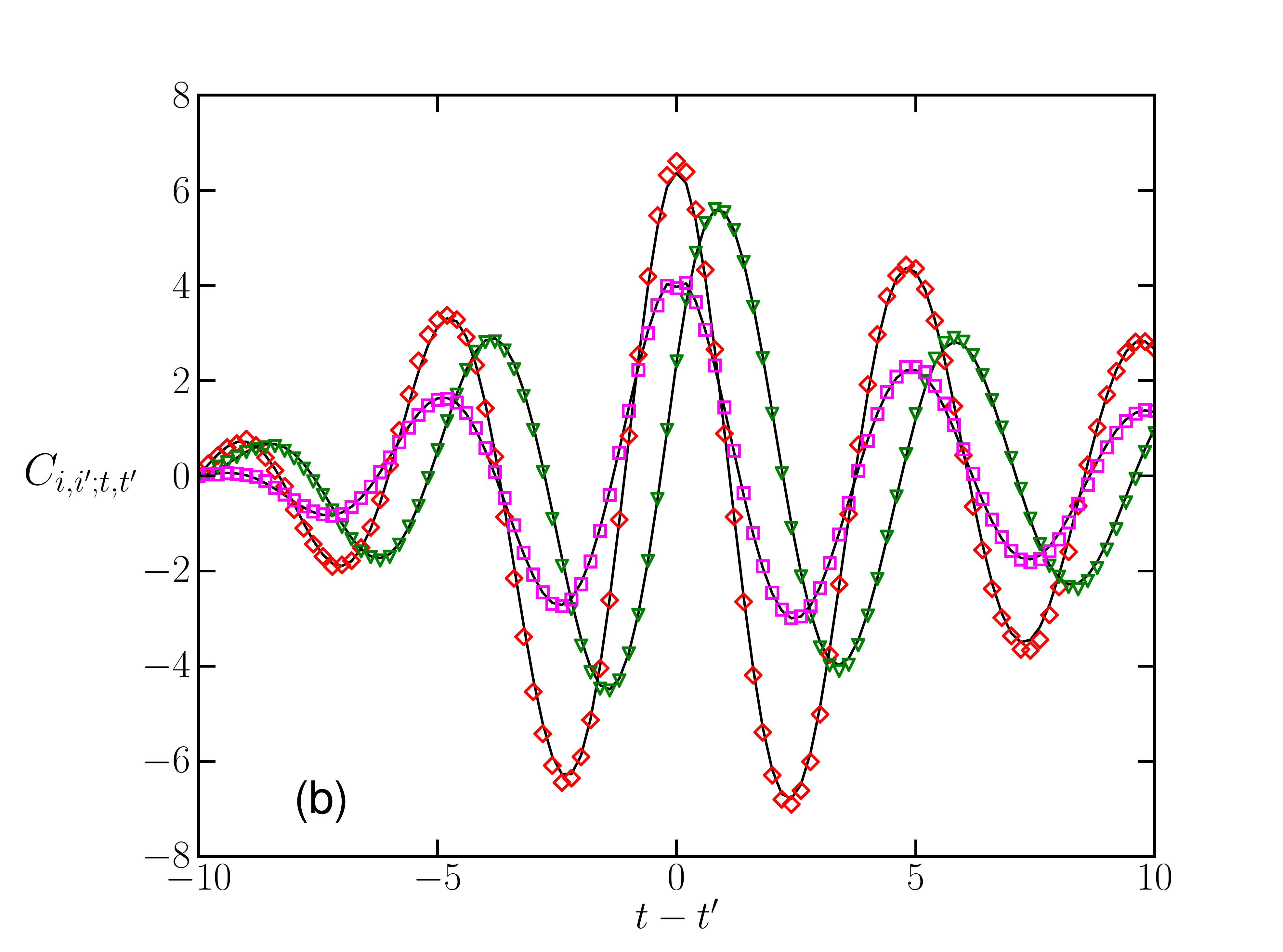}
         \\
        \includegraphics[scale = 0.18]{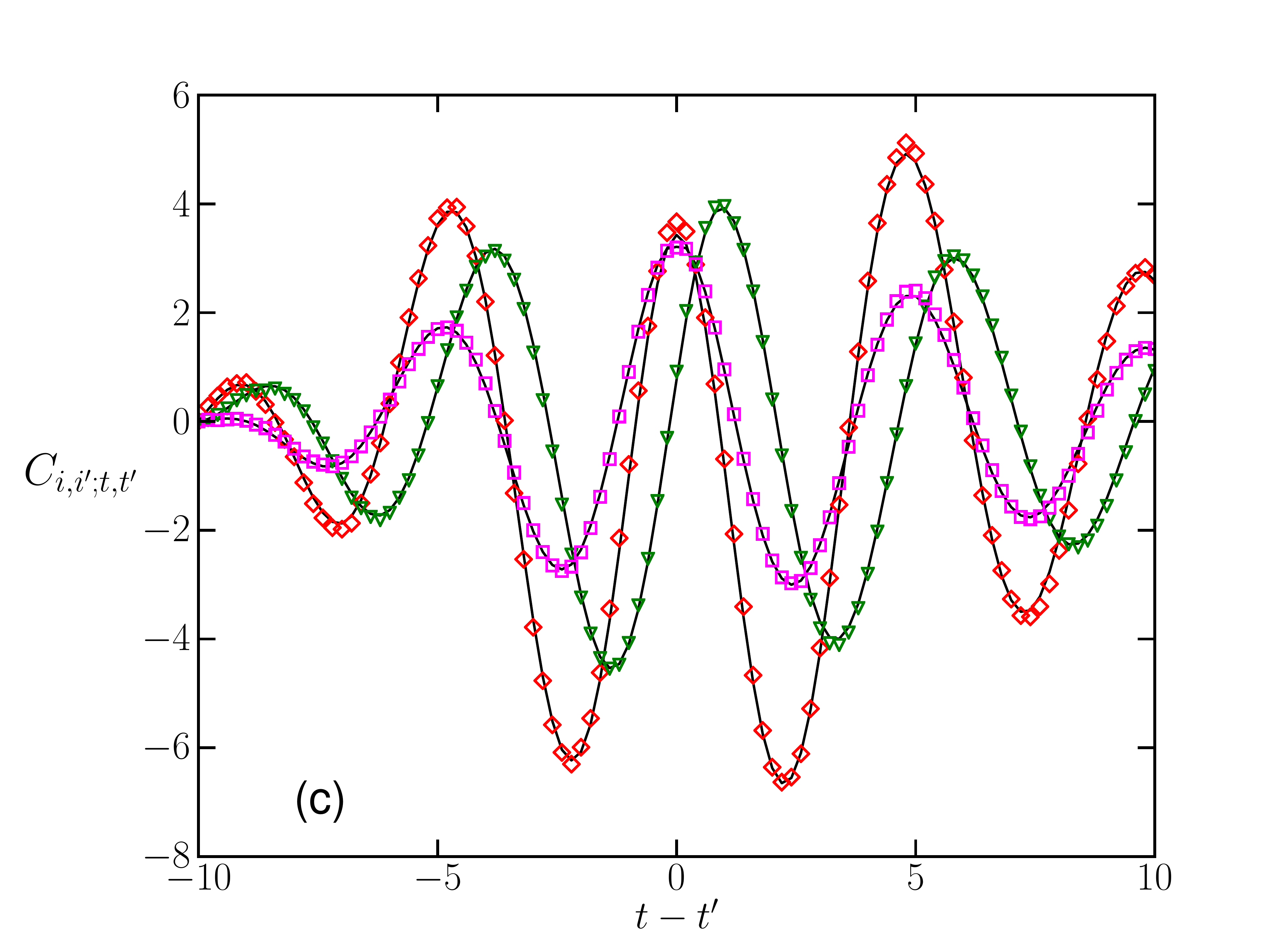}\hspace{2cm}
                \includegraphics[scale = 0.18]{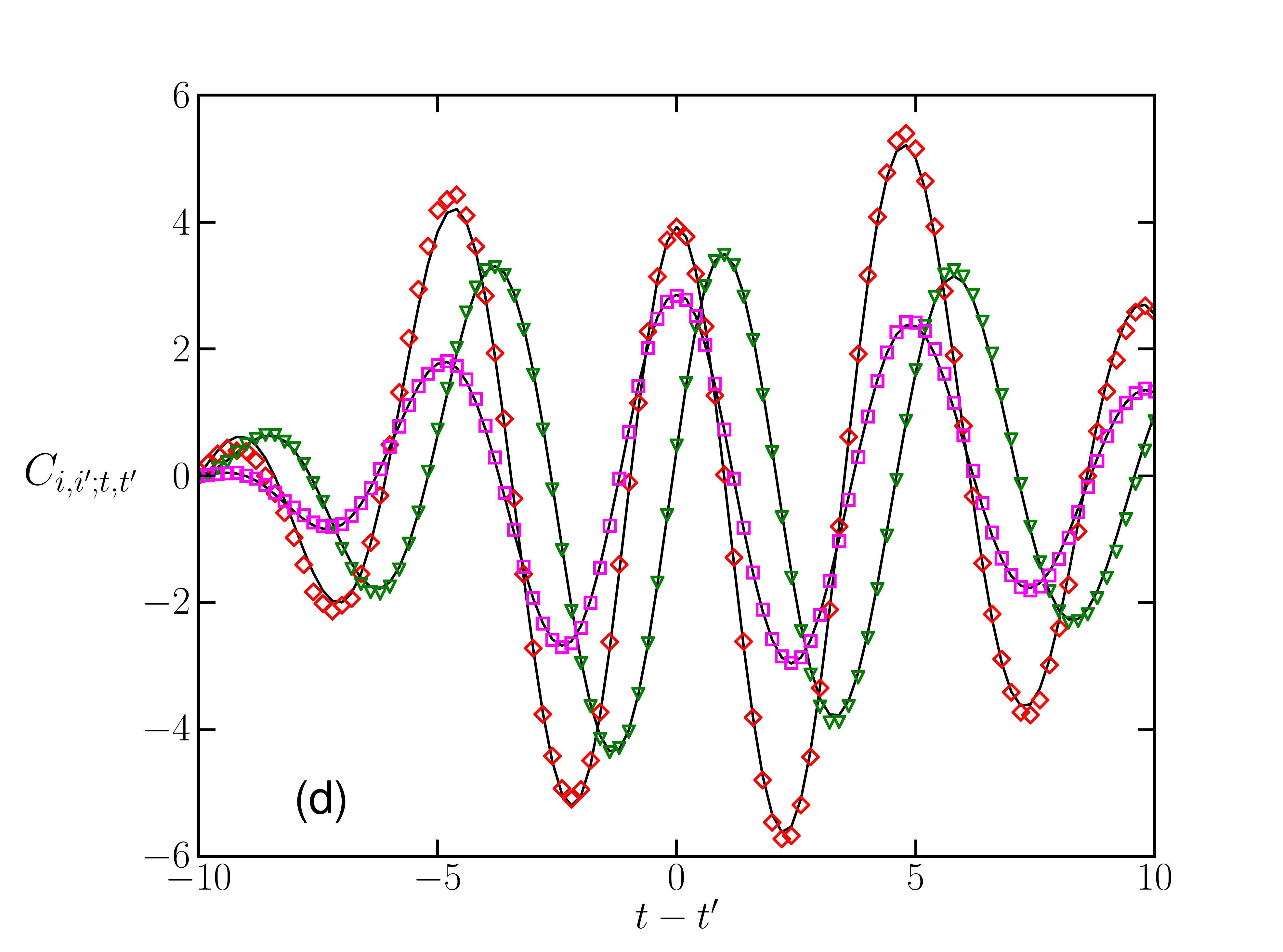}
       
\caption{The correlators $C_{i,i';t,t'}= \frac{\langle \delta^{\alpha}_{i,t} \delta^{\alpha'}_{i',t'} \rangle}{N}$ for various spatial and temporal separations for the Lengyel-Epstein model with both species subdiffusing. Monte-carlo simulation results are shown as hollow markers and the lines from the theory are solid black. The species A auto-correlator results are red diamonds, the species B auto-correlator results are magenta squares and the cross-correlator results are green triangles. The model parameters are $a = 5$, $b = 0.7$, $c = 1$, $d=1$, $t_0^{\left(1\right)} = 2.0$, $t_0^{\left(2\right)} = 0.05$, $\gamma^{\left(1\right)} = 0.5$ and $\gamma^{\left(2\right)} = 0.75$. The system-size is $N = 10000$ and the number of trials over which the simulations were averaged is $8000$. The simulations were performed on a discrete lattice with 7 sites and periodic boundary conditions. The spatial separations are (a): $i-i' = 0$; (b): $i-i' = 1$; (c): $i-i' = 2$; (d): $i-i' = 3$. }
	\label{fig:allle}
\end{figure*}

Once the linearised equations are found, the problem is most naturally handled by taking Fourier and Laplace transforms with respect to the spatial and temporal coordinates respectively. We use $\mathcal{F}_n \left\{ f_n\right\}\left(q\right) = \sum_n e^{inq} f_n =\tilde f_q$ to denote the discrete Fourier transform and $\mathcal{L}_t \left\{ g_t\right\}\left(u\right) =\int_0^t e^{-ut} g\left(t\right) dt = \hat g_u$ to denote the Laplace transform. We then obtain equations of the form
\begin{align}
\underline{\hat{\tilde{\delta}}}_{u,q} = \underline{\underline{\hat{\tilde{m}}}}_{u,q}  \underline{\hat{\tilde{\epsilon}}}_{u,q}, \label{matrixeq}
\end{align}
with suitable Gaussian noise variables $\{\hat{\tilde\epsilon}^\alpha_{u,q}\}$. In the time domain this noise is coloured. We use underscores to indicate vectors in the space of species. The square matrix $\underline{\underline{\hat{\tilde{m}}}}_{u,q}=(\hat{\tilde m}_{u,q})^{\alpha\beta}$ has dimension equal to the number of particle species. Its precise form is given in the Supplemental Material (Section S4), as are full expressions for the correlators $\langle \tilde\epsilon^\alpha_{q,t} \tilde\epsilon^{\alpha'}_{q',t'} \rangle$.

It then remains to invert the Fourier and Laplace transforms. Expressions for the resulting correlators $\langle \delta^{\alpha}_{i,t} \delta^{\alpha'}_{i',t'} \rangle$ are given in the Supplemental Material (Section S5). In practice, the evaluation of the correlators $\langle \delta^{\alpha}_{i,t} \delta^{\alpha'}_{i',t'} \rangle$ is performed using numerical inverse Laplace transform methods (such as that of Zakian \cite{halstedbrown}) and fast Fourier transform routines. This is because analytical expressions for the inverse transforms of the elements of $\underline{\underline{\hat{\tilde{m}}}}_{u,q}$ are in general difficult to find. That being said, the equal-time equal-wavenumber correlator, $\langle \tilde \delta^{(\alpha)}_{q,t} \tilde\delta^{(\alpha')\star}_{q,t} \rangle$, can indeed be found without the use of a numerical inverse Laplace transform technique (see Section S5 of the Supplemental Material).

\subsection{Comparison of theory to simulation results}

The theory hitherto presented can be expected to be accurate when fluctuations about the deterministic homogeneous fixed point are small enough so that the linear approximation is valid. The examples in Figs.~\ref{fig:equaltime} and \ref{fig:allle} show that one obtains good agreement with individual-based simulations of the full system with typical particle numbers per lattice site as low as $N \approx 4000$. 

Fig. \ref{fig:equaltime} demonstrates the accuracy of the theory for the steady-state power-spectrum of fluctuations $\langle \lvert \tilde\delta^{\left(A\right)}_{q,t} \rvert^2 \rangle$ for the case of the Brusselator model. We compare the results of simulations to our theory for the cases where the activator subdiffuses and the inhibitor diffuses normally, and vice versa. Fig.~\ref{fig:allle} again compares the theory predictions to Monte Carlo simulations but this time in the case of the Lengyel-Epstein model with a subdiffusing activator and a subdiffusing inhibitor. The auto-correlators $\langle \delta^{\left(A\right)}_{i,t} \delta^{\left(A\right)}_{i',t'} \rangle$ and $\langle \delta^{\left(B\right)}_{i,t} \delta^{\left(B\right)}_{i',t'} \rangle$ and the cross-correlator $\langle \delta^{\left(A\right)}_{i,t} \delta^{\left(B\right)}_{i',t'} \rangle$ are shown for different temporal separations $t-t'$ and spatial separations $i-i'$. The agreement in both Figs. \ref{fig:equaltime} and \ref{fig:allle} is good apart from some minor deviations. These arise from a combination of the stochastic nature of the simulations and the limits to the accuracy of the numerical inverse Laplace transform. The discrepancy reduces as the number of trials, over which the average is taken, is increased. 

To carry out the simulations, it was necessary to modify the Gillespie algorithm \cite{gillespie} in a similar way to \cite{anderson}. This was due to the non-Markovian nature of the hopping processes. Broadly, the simulations involve keeping a list of the scheduled hopping times of particles, carrying these out at appropriate times and in the right sequence, while performing reactions in the intermediate times. The full procedure is given in Appendix \ref{simulations}. To the best of our knowledge, individual-based simulations of reaction-diffusion systems with subdiffusive transport have not been performed in this way previously.

The full expressions for the correlators $\langle \delta^{\alpha}_{i,t} \delta^{\alpha'}_{i',t'} \rangle$, from which the theory lines in Figs.~\ref{fig:equaltime} and \ref{fig:allle} are derived, can be found in Section S5 of the Supplemental Material. In the Supplement we also explain how the equal-time correlators in Fig.~\ref{fig:equaltime} can be found without the use of a numerical inverse Laplace transform.

\section{Onset of stochastic and deterministic Turing patterns}\label{stochpatt}
\subsection{Description of the phenomena}

In a deterministic system which exhibits a Turing instability, a finite range of Fourier modes with non-zero wavenumbers is unstable. That is to say, if the system is perturbed from its uniform fixed point, the amplitudes of these modes will grow with time. A criterion for the Turing instability in systems with one subdiffusing component and one normally diffusing component is given in \cite{yadavmilu,barongalla}. In practice, particle numbers do not deviate infinitely far from the fixed point. Instead, the growth of the unstable modes is curtailed by the non-linearity of the reaction equations. If one simulates systems with such an instability and looks at the Fourier transform of deviations from the uniform state one finds the dominant peak at a non-zero wavenumber. It is this wavenumber which characterises the periodicity of the observed Turing patterns. It is important to note that noise is not required to sustain these patterns once the initial perturbation about the uniform state has been applied. In particular, the amplitude of these deterministic patterns is set by the nonlinearities of the reaction-(sub)diffusion system, and not by any source of stochasticity.
\begin{figure}[t!]
	\centering
	\includegraphics[scale = 0.2]{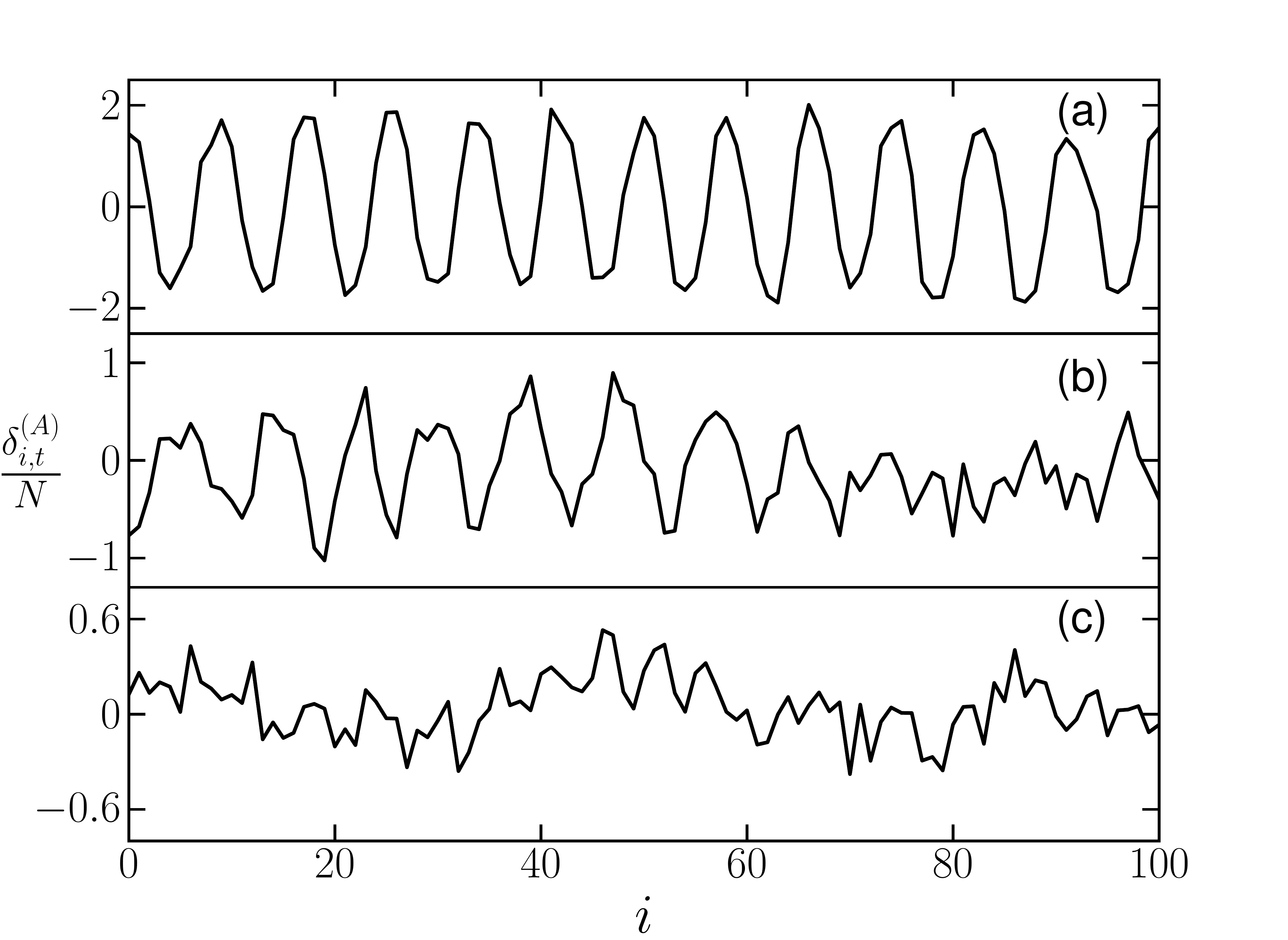}
	\caption{Comparison of the spatial patterns produced in the Lengyel-Epstein model. (a): Parameters are in the phase where deterministic pattern formation occurs ($t_0^{\left(A\right)} = 0.2$, $t_0^{\left(B\right)} = 0.1$). (b): Noise-induced patterns but no deterministic patterns ($t_0^{\left(A\right)} = 0.11$, $t_0^{\left(B\right)} = 0.1$). (c): No patterns are found at all ($t_0^{\left(A\right)} = 0.1$, $t_0^{\left(B\right)} = 0.2$). In these simulations, $N=1000$, the activator subdiffuses with $\gamma = 0.5$, the inhibitor diffuses normally. Reaction rates are given in the caption of Fig.~\ref{fig:phasele}. The parameter sets used in the three panels correspond to those marked by the three blue dots in Fig. \ref{fig:phasele}.}
	\label{fig:examples}
\end{figure}
\begin{figure*}[t!!]
	\centering
	\includegraphics[scale = 0.2]{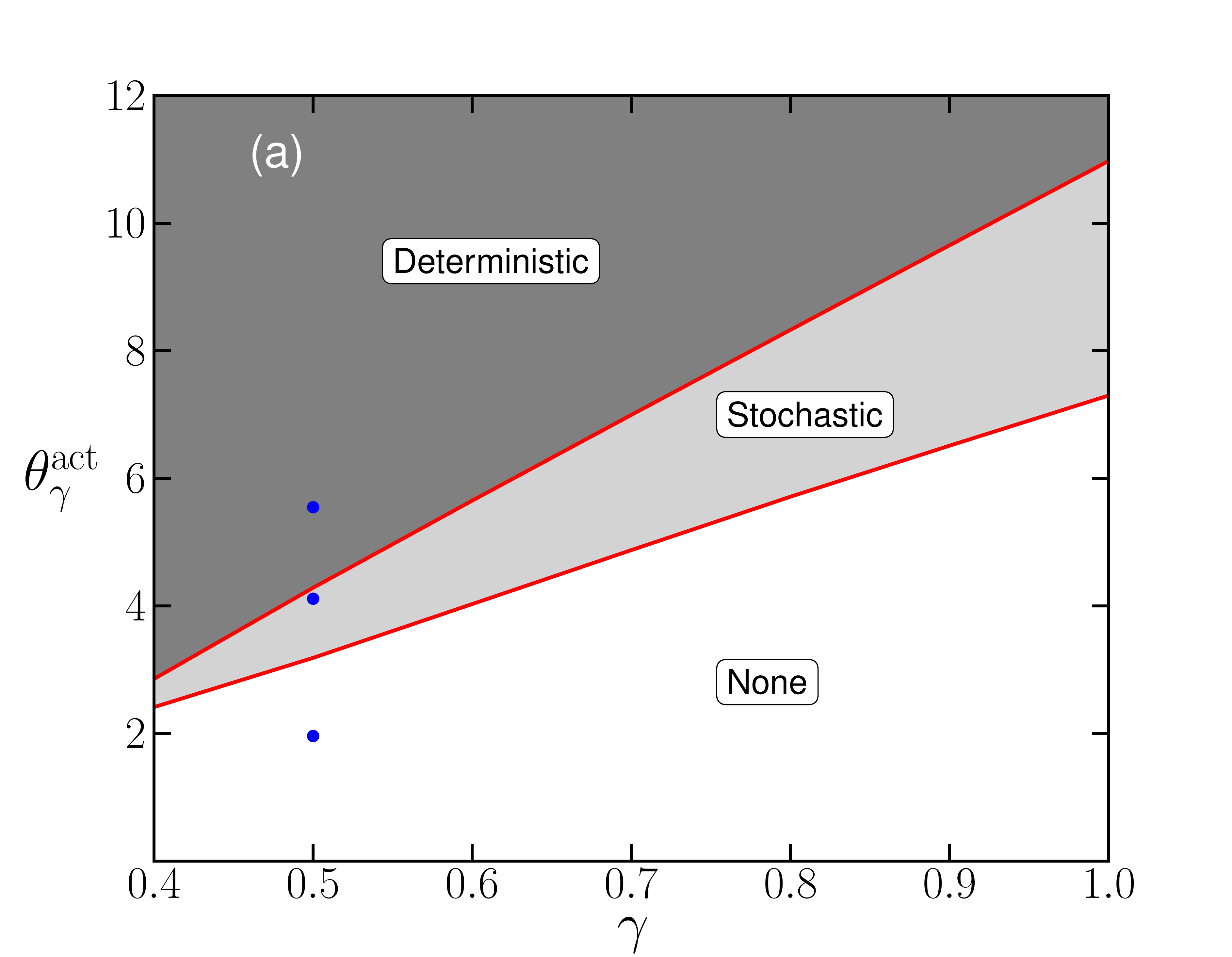}\hspace{2cm}
	\includegraphics[scale = 0.2]{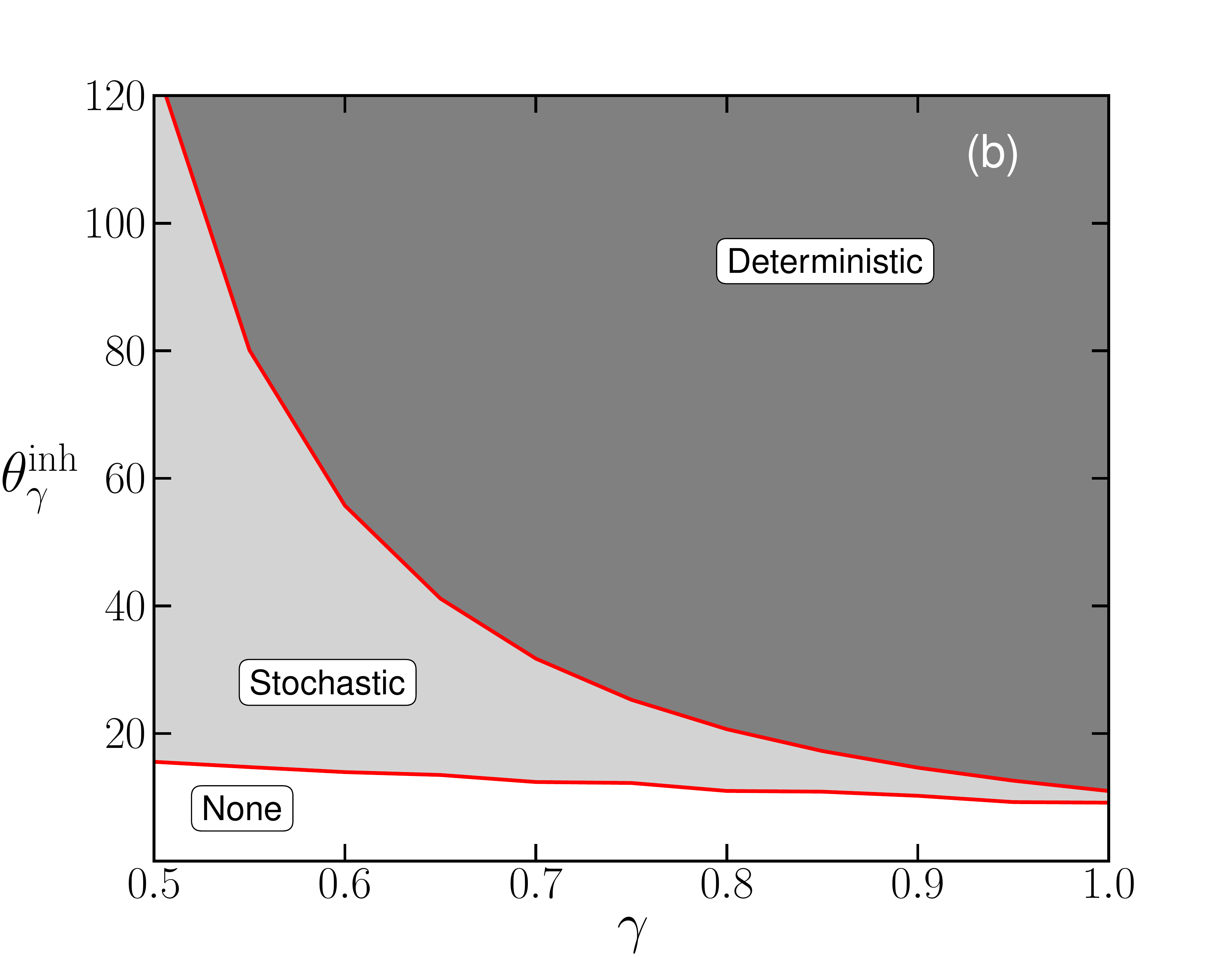}
	\caption{Phase diagrams showing the parameter regions where deterministic Turing patterns (dark grey shading) and stochastic (noise-driven) patterns (light grey) occur for the Lengyel-Epstein model. Figures (a) and (b) show the behaviour when the activator and the inhibitor subdiffuse respectively. The other component diffuses normally. As $\gamma$ decreases, and the diffusion becomes more anomalous, the critical values of $\theta_\gamma$ at which deterministic and stochastic patterns emerge decrease for the subdiffusing activator (a) and increase for the subdiffusing inhibitor (b). When $\gamma= 1$, the diffusion is normal and the critical values of $\theta_\gamma$ in both plots are the same. The remaining model parameters are $a = 2$, $b = 0.13$, $c =d = 1$. }
	\label{fig:phasele}
\end{figure*}
\begin{figure*}[t!!]
	\centering
	\includegraphics[scale = 0.2]{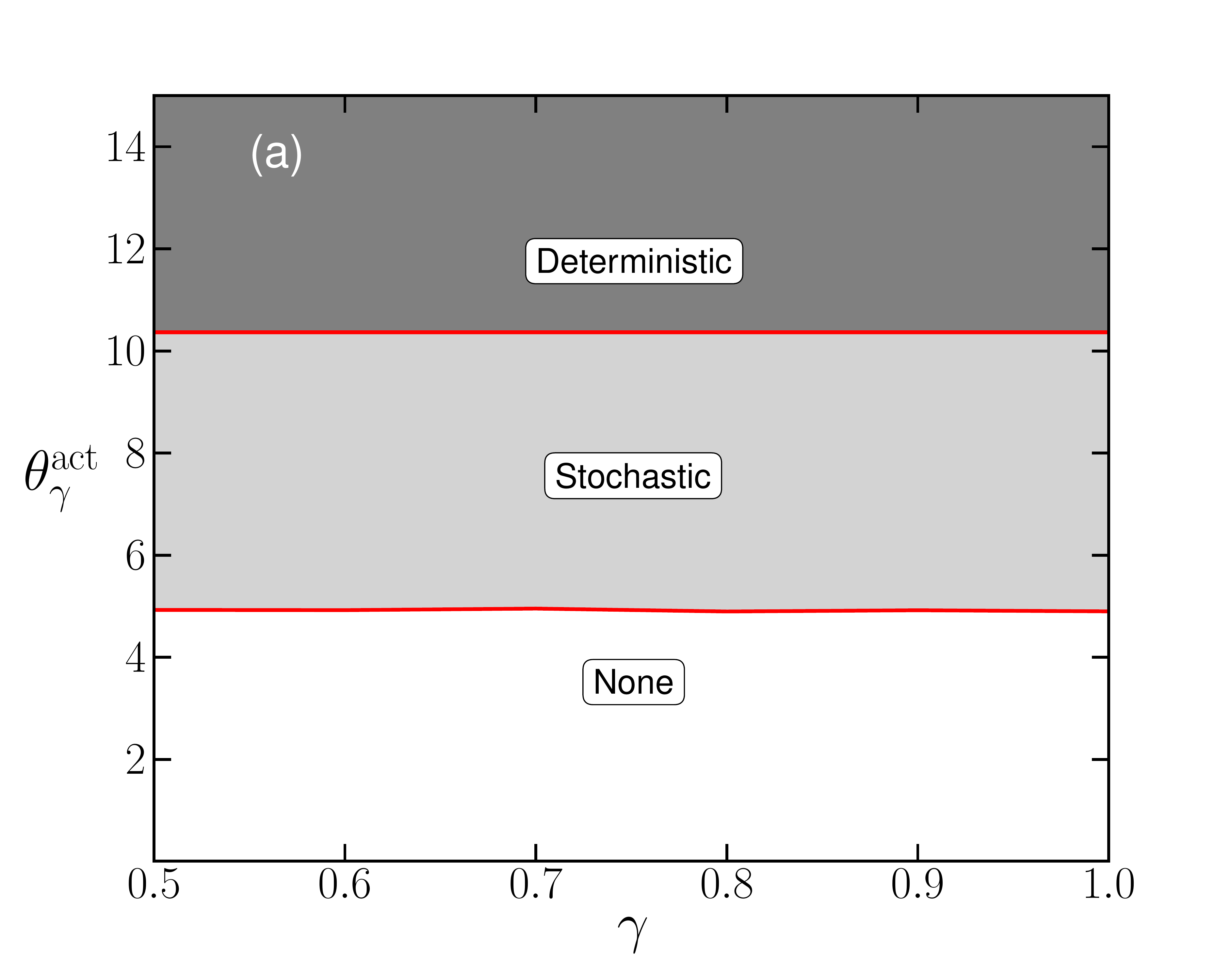}\hspace{2cm}
	\includegraphics[scale = 0.2]{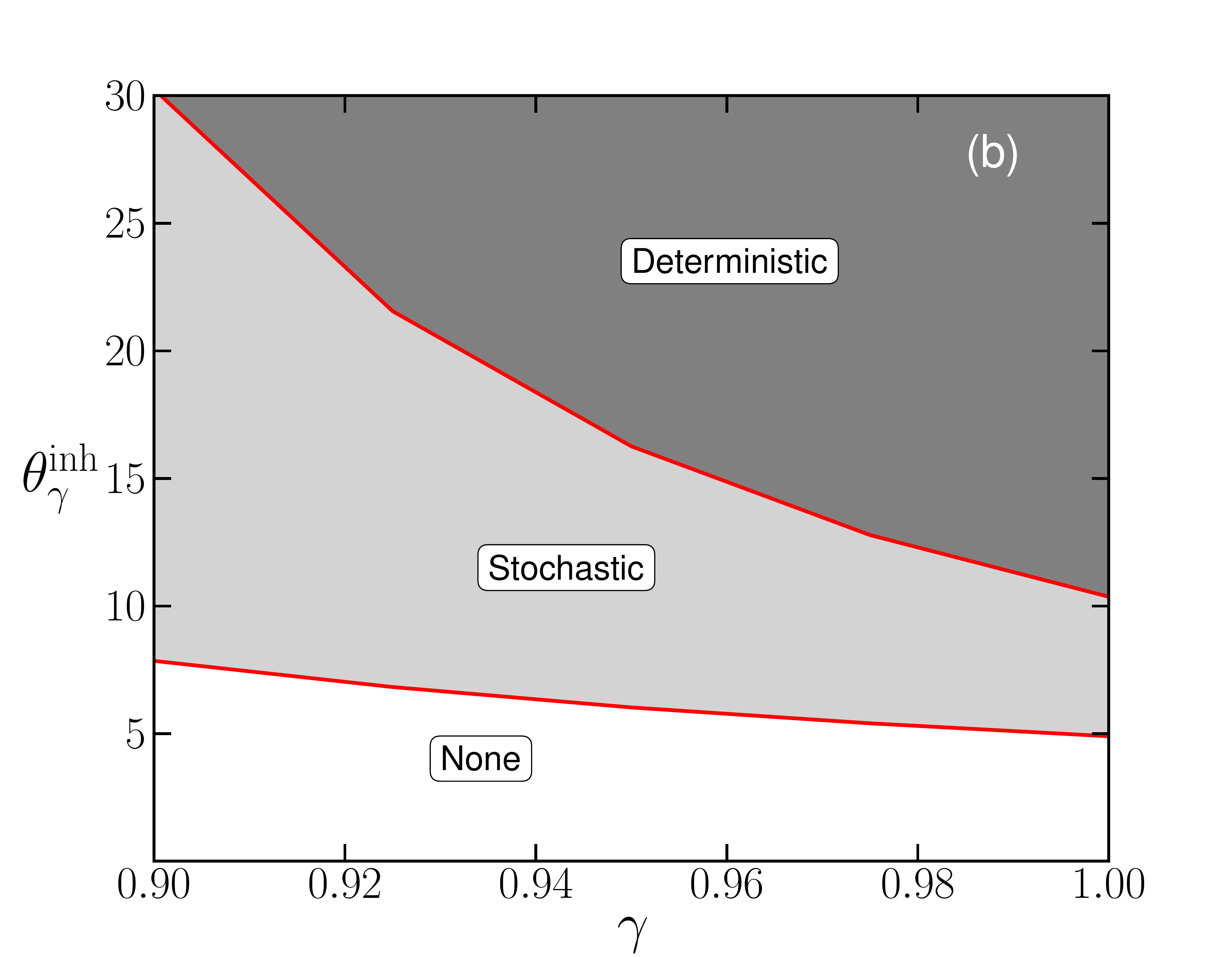}
	\caption{Phase diagrams showing the parameter regions where deterministic Turing patterns (dark grey shading) and stochastic (noise-driven) patterns (light grey) occur for the Brusselator model. Figures (a) and (b) show the behaviour when the activator and the inhibitor subdiffuse respectively. The other component diffuses normally. Due to the constant death-rate of the activator, the critical value of $\theta_\gamma$ for the onset of both stochastic and deterministic patterns does not vary with $\gamma$ when the activator subdiffuses (a). When the inhibitor (with concentration-dependent death rate) subdiffuses (b), the behaviour is qualitatively similar to that seen in the Lengyel-Epstein model in Fig. \ref{fig:phasele}. When $\gamma= 1$, the diffusion is normal and the critical values of $\theta_\gamma$ in both plots are the same. The remaining model parameters are $a = 1.1$, $b = 1.8$. }
	\label{fig:phasebruss}
\end{figure*}

However, one may also observe patterns in systems with finite particle number (i.e., systems with intrinsic noise), even when parameters are such that the purely deterministic system would not show any patterns \cite{butlergoldenfeld1,butlergoldenfeld2}. These patterns are noise-driven, and their amplitude is set by the strength of the noise; in individual-based systems the variance of the intrinsic noise in turn is determined by the inverse typical particle number $N^{-1}$. We refer to patterns formed in this way as  quasi-patterns or stochastic Turing patterns. They can be detected in the Fourier spectra of fluctuations about the uniform deterministic fixed point; these spectra show a peak at a characteristic nonzero wavenumber. In contrast to deterministic Turing patterns, this peak is not observed in the absence of noise. Such patterns have been analysed in systems with regular diffusion, see e.g. \cite{butlergoldenfeld1, butlergoldenfeld2, biancalani,stochpattexp}.

In Fig.~\ref{fig:examples} we show examples of Turing patterns in the Lenyel-Epstein system with subdiffusion. The data in the upper panel is from individual-based simulations in the parameter regime in which the deterministic model shows Turing patterns. The pattern is seen in the stochastic simulation as well, but it is important to stress that this is not a noise-driven pattern. The stochasticity in the individual-based model modulates the deterministic pattern, but the amplitude of the structure is independent of the noise, i.e., increasing the number of particles, $N$, per site does not change the relative magnitude of the pattern. In the central panel we show an example of a quasi-pattern. This data is taken in the regime in which the deterministic Lengyel-Epstein system does not have any instability; the pattern is purely noise-driven and its amplitude decreases with increasing particle number per lattice site. One further main difference between stochastic and deterministic patterns is that the stochastic patterns are not stationary. Instead, regions of high concentration for one chemical species continually assemble, shift and dissipate in such a manner that a typical level of periodicity is maintained. Deterministic patterns on the other hand are stationary in time. In the lower panel of Fig. \ref{fig:examples} finally, the parameters are such that neither deterministic nor stochastic Turing patterns are seen. The system fluctuates about the deterministic uniform fixed point, but no particular spatial structure emerges in these fluctuations. That being said, the fluctuations do appear to be quite large in magnitude, given the system size. Similarly, the magnitude of the noise-induced pattern in Fig. \ref{fig:examples}(b) is similar to that of the pattern in panel (a). This is not an unusual observation and is due to a phenomenon known as `stochastic amplification', which has been reported in the context of noise-driven cycles and patterns in normally diffusing systems (see e.g. \cite{mckanenewman, alonsomckane, butlergoldenfeld1, butlergoldenfeld2, biancalani}).

\subsection{Phase diagrams}
The theory we developed in the earlier sections allows us to find the equal-time correlator in Fourier space (i.e., the structure factor  $\langle \tilde \delta^{(\alpha)}_{q,t} \tilde\delta^{(\alpha')\star}_{q,t} \rangle$) of fluctuations without having to carry out time-consuming simulations. We can use these expressions to search parameter space and to identify the regimes in which stochastic patterns emerge, and the parameter ranges where they do not. This gives a more complete picture of the effect of subdiffusion on stochastic pattern formation. Our results are summarised in the phase diagrams in Figs. \ref{fig:phasele} and \ref{fig:phasebruss}, which we discuss in more detail below. While the expressions from the theory have to be evaluated numerically, we note that determining the phase behaviour from the analytical expressions is very efficient; accurately establishing the phase diagrams from individual-based simulations on the other hand would require unrealistic computing time.

It is well-known that the emergence of deterministic Turing patterns relies on the slow diffusion of the activator and the comparatively fast diffusion of the inhibitor. In previous work \cite{barongalla} we showed that it is possible to define effective diffusion coefficients for systems in which one reactant undergoes subdiffusion and the other normal diffusion; see also \cite{yadavmilu}. To characterise the degree to which the rates of (sub)diffusion of the two particle species differ, we define the ratio of these effective diffusion constants, $\theta_\gamma$, see again \cite{barongalla,yadavmilu} for details. The detailed definition depends on which one of the components (activator or inhibitor) is subdiffusing. We have 
\be
\theta_\gamma^{\mathrm{act}} = \frac{\left(t_0^{\left(\mathrm{act}\right)}\right)^\gamma}{t_0^{\left(\mathrm{inh}\right)}\left(\bar p^{\left(\mathrm{act}\right)}\right)^{1-\gamma}},\label{effdiffcoeff1}
\ee
when the activator subdiffuses, and 
\be
\theta_\gamma^{\mathrm{inh}} = \frac{t_0^{\left(\mathrm{act}\right)}\left(\bar p^{\left(\mathrm{inh}\right)}\right)^{1-\gamma}}{\left(t_0^{\left(\mathrm{inh}\right)}\right)^\gamma } , \label{effdiffcoeff2}
\ee
when the inhibitor subdiffuses. The quantity $\bar p$ in these expressions denotes the per capita removal rate of the relevant substance at the uniform deterministic fixed point.

Figs. \ref{fig:phasele} and \ref{fig:phasebruss} show the critical ratios of the effective diffusion constants of the activator and the inhibitor for the onset of stochastic and deterministic Turing patterns in the Lengyel-Epstein and Brusselator models respectively. For any fixed value of the anomalous exponent $\gamma$ the following behaviour is observed in Figs. \ref{fig:phasele} and \ref{fig:phasebruss}. For low values of $\theta_\gamma$ neither deterministic nor stochastic patterns emerge. As the ratio of effective diffusion coefficients is increased and crosses a first critical threshold, $\theta_\gamma^s$, a phase is entered in which stochastic patterns are found, but where the deterministic system shows no patterns. As $\theta_\gamma$ is increased further the system undergoes a full deterministic Turing instability at $\theta_\gamma^d$. 

Additionally, we find that in general, if the subdiffusion of a particular reactant is conducive to the formation of deterministic patterns, then it is also conducive to the formation of stochastic patterns. Similarly if subdiffusion of a reactant makes deterministic pattern formation more difficult then it also reduces the parameter range in which noise-driven patterns are found. In other words, if the critical value $\theta_\gamma^d$ for the onset deterministic patterns increases (decreases) with $\gamma$, then $\theta_\gamma^s$ also increases (decreases) with $\gamma$.

One notes that due to the concentration-independent decay rate of the activator in the Brusselator model, both $\theta_\gamma^d$ and $\theta_\gamma^s$ do not vary with the anomalous exponent $\gamma$ when the activator is subdiffusing (left-hand panel in Fig.~\ref{fig:phasebruss}). However, if the death rate of the subdiffusing component is not constant, one finds that, in our examples, as $\gamma$ decreases, so do both $\theta_\gamma^d$ and $\theta_\gamma^s$ if the activator is subdiffusing (left-hand panel of Fig.~\ref{fig:phasele}). Conversely, if the inhibitor is subdiffusing and the inhibitor death rate is concentration-dependent, a reduced value of $\gamma$ makes it more difficult it is for Turing patterns to form (right-hand panels of Figs.~\ref{fig:phasele} and \ref{fig:phasebruss}).

One caveat to the approach we have presented is that analytical results are only valid for large system sizes $N$. That being said, the qualitative conclusions we have reached (that the combination of stochasticity and subdiffusion lowers the threshold for pattern formation) can be extended to the regime of moderate $N$, where the theory is not as accurate. We have verified this in simulations, and discuss this point briefly in Section 6 of the Supplemental Material.


\section{Summary and conclusions}\label{conc}
We have successfully extended the description of reaction-diffusion systems with anomalous transport to include the intrinsic noise which comes about  when the number of particles in the system is finite. To do this, we carried out a Gaussian approximation of the individual-based system, using a generating-functional approach. We also provided a prescription for finding the fluctuations about the stable homogeneous fixed point of reaction-diffusion systems with anomalous diffusion. This was done in the example cases of the Brusselator and Lengyel-Epstein systems, and our theoretical predictions were successfully tested against computer simulations. Finally, we used this theory to determine the parameter regimes where one could expect to observe stochastic Turing patterns. We found examples for which a subdiffusing activator encouraged both stochastic and deterministic pattern formation, whereas a subdiffusing inhibitor hindered both stochastic and deterministic pattern formation. This is exemplified in Figs.~\ref{fig:equaltime}, \ref{fig:phasele} and \ref{fig:phasebruss}.

The theory developed in this paper is entirely general and would be equally applicable to any reaction scheme with anomalously diffusing reactants; the two examples discussed here were mainly chosen for illustration. Systems such as the Lotka-Volterra \cite{lotka} dynamics, the Oregonator \cite{fieldnoyes, tyson, jahnkeskaggs} or the Schnakenberg model \cite{schnakenberg} with subdiffusion would have been equally valid candidates for the study of noise-driven patterns. We could also have studied systems with different waiting-time distributions or hopping kernels.

A well-known problem with using Turing's mechanism as an explanation for pattern formation in nature is the often unphysically large ratio of the diffusion constants required for the instability to occur. Both subdiffusion and stochastic pattern formation have been proposed separately as potential remedies for this problem. In this paper, we have shown that a combination of the two can lower the threshold for pattern formation to a greater degree than either mechanism on its own. This is demonstrated by the fact that the critical ratio of the diffusion constants is always lower in Figs.~\ref{fig:phasele} and \ref{fig:phasebruss} for the onset of stochastic patterns than it is for deterministic patterns, and by the fact that this critical ratio can reduce as the activator becomes more subdiffusive.

Recently, the existence of noise-induced Turing patterns in bacterial cultures has been examined \cite{stochpattexp}. However, these experiments were concerned with a mechanism for pattern formation for which the diffusion was presumed to be Markovian. A possibility for further experiment would be to analyse stochastic pattern formation in a system in which the diffusion of the reactants was known to be subdiffusive. Our approach makes predictions for the way in which noise-driven pattern formation is affected by subdiffusion. 

Another question for future study would be to ask how other stochastic phenomena in reaction-diffusion systems, such as waves \cite{gallawaves}, would be affected by anomalous diffusion. Conversely, one might also wonder about how such phenomena are affected by superdiffusion. The theoretical approach we have developed is sufficiently general, and can be applied to the study of such systems.

{\em Acknowledgements.} JWB thanks the Engineering and Physical Sciences Research Council (EPSRC) for funding
(PhD studentship, EP/N509565/1).

\begin{appendix}

\section{Individual-based simulations}\label{simulations}
The non-Markovian nature of subdiffusive hopping requires us to modify the traditional Gillespie algorithm \cite{gillespie}, in a similar way to \cite{anderson}. The statistics of the processes laid out in Section \ref{modeldef} are preserved by using the following algorithm:

\begin{itemize}
\item[1.] Set $t=0$ and initialise the system. (In our simulations, we start from $n^{\left(1\right)}_{i} = \bar n^{\left(1\right)}$ and $n^{\left(2\right)}_{i} = \bar n^{\left(2\right)}$.) For each particle draw a waiting time until its next hop from the appropriate distribution, and create a list of all scheduled hopping events.
\item[2.] Take note of the earliest hopping time and the position of the particle associated with this time; we label the time at which the next hop occurs $\tau^{\left(H\right)}$.\\
\item[3.] Based on the numbers of particles at each position, calculate the reaction rates for each type of reaction and each point on the lattice, $\lambda_{i,r,t}$. Calculate the sum of the reaction rates $\lambda_t = \sum_{r,i} \lambda_{i,r,t} $ and draw the time at which the next reaction is due to occur: $\tau^{\left(R\right)} = t -\frac{1}{\lambda_t}\ln\left(s\right)$, where $s$ is an independent random number from the uniform distribution over $(0,1]$. This reaction can occur at any position on the lattice.
\item[4.] (a) If $\tau^{\left(R\right)} < \tau^{\left(H\right)}$, perform one single reaction. Choose this to be a reaction of type $r$ at site $i$ with probability $\lambda_{i,r,t}/\lambda_t$. If particles are removed during this reaction delete the scheduled hopping events for these particles from the list. If particles are created, draw hopping times from the appropriate distribution, and add these to the list of scheduled hops. Set $t = \tau^{\left(R\right)}$. Go to 2.\\

(b) If $\tau^{\left(R\right)} > \tau^{\left(H\right)}$, perform the hop associated with $\tau^{\left(H\right)}$ in step 2, and discard the reaction event from step 3. To do this draw a random hopping distance from the hopping kernel $\phi_{i,i'}$, and re-locate the particle. Draw a new waiting time until the next hop of this particle, and add this time to the list. Set $t = \tau^{\left(H\right)}$. Go to 2. 
\end{itemize}
To draw waiting times from the Mittag-Leffler distribution in Eq.~(\ref{waitingtimes}), one can use the following formula $t = -t_0^\alpha\ln\left(u\right)\left(\frac{\sin\left(\gamma^\alpha\pi\right)}{\tan\left(\gamma^\alpha \pi v \right)}- \cos\left(\gamma^\alpha\pi \right)\right)^{\frac{1}{\gamma^\alpha}}$ \cite{fulger, kozubowski}, where $u$ and $v$ are independent random numbers from a uniform distribution on $(0,1]$. \\

\end{appendix}

\pagebreak

\onecolumngrid

\begin{titlepage}
   \vspace*{\stretch{1.0}}
   \begin{center}
      \Large\textbf{Stochastic fluctuations and quasi-pattern formation in reaction-diffusion systems with anomalous transport\\~\\
Supplemental Material}\\
   \end{center}
   \vspace*{\stretch{2.0}}
\end{titlepage}

\onecolumngrid
 \pagebreak
One main result of our work is a Gaussian approximation of the fluctuations in a reaction-diffusion system with anomalous transport (i.e. generalised waiting-time distributions between hopping events). To derive this approximation, we find the generating functional for the reaction-diffusion system, and then expand this functional to sub-leading order in $N^{-1}$, where the parameter $N$ characterises the typical number of particles per site in the discrete lattice. This approximate path-integral is recognised as the generating functional of a stochastic different equation (SDE), with a special form, to be specified below. One can then compare the two generating functionals and read off the effective SDE for the reaction-diffusion system. This supplement sets out the different steps of the calculation, and subsequent analysis.

\setcounter{section}{0}
\setcounter{page}{1}
\setcounter{equation}{1}
 \renewcommand{\thesection}{S\arabic{section}} 		
\renewcommand{\thepage}{S\arabic{page}} 			
\renewcommand*{\theequation}{S\arabic{equation}}  	
\renewcommand{\thefigure}{S\arabic{figure}}  		
\renewcommand{\thetable}{S\arabic{table}} 
\setcounter{section}{0}
\setcounter{page}{1}
\setcounter{equation}{1}
 \renewcommand{\thesection}{S\arabic{section}} 		
\renewcommand{\thepage}{S\arabic{page}} 			
\renewcommand*{\theequation}{S\arabic{equation}}  	
\renewcommand{\thefigure}{S\arabic{figure}}  		
\renewcommand{\thetable}{S\arabic{table}} 
\section{Generating functional for an SDE with age variables}\label{genericsde}
We first present a method of calculating the generating functional for a stochastic process with Gaussian noise. The choice of the form of the equations for which we find the corresponding generating functional is important and is informed by the type of equations we expect for the reaction-subdiffusion systems in the main paper. 

To start with, we discretise time using step size $\Delta$. We will eventually take the continuous limit in time. The set of stochastic equations has the discretised form
\begin{align}
x^{\alpha}_{i,\tau+\Delta, t+\Delta} - x^{\alpha}_{i,\tau, t} - \frac{\Delta}{N} f^{\alpha}_{i,\tau,t}\left( \{x^{\alpha}_{i,\tau, t}\} \right) - \frac{\sqrt{\Delta}}{N} \chi^{\alpha}_{i, \tau, t} &= 0 , \label{discretisedapp1} \\
x^{\alpha}_{i,0, t+\Delta} - \frac{1}{N}f^{\alpha}_{i,0,t}\left( \{x^{\alpha}_{i,\tau, t}\} \right) - \frac{\chi^{\alpha}_{i, 0, t}}{N\sqrt{\Delta}} &= 0 . \label{discretisedapp2}
\end{align}
where $f^{\alpha}_{i,\tau,t}\left( \{x^{\alpha}_{i,\tau, t}\}\right)$ is an arbitrary function of the coordinates $\{x^{\alpha}_{i, \tau, t}\}$ that informs the deterministic trajectory of the system. The $\{\chi^{\alpha}_{i, \tau, t}\}$ are Gaussian random variables, which encode the fluctuations about the deterministic trajectory. We write their correlations as\begin{equation} 
 C^{\alpha, \alpha'}_{i,i',\tau,\tau', t, t'} \equiv \big<\chi^{\alpha}_{i, \tau, t} \chi^{\alpha'}_{i', \tau', t'} \big>. \label{cdef}
\end{equation}
The matrix $C^{\alpha, \alpha'}_{i,i',\tau,\tau', t, t'} $ is symmetric with respect to the pairs of indices $(\alpha, \alpha'$), $(i, i')$, $(t, t')$ and $(\tau, \tau')$.

In continuous time ($\Delta \to 0$) the expressions in Eqs.~(\ref{discretisedapp1},\ref{discretisedapp2}) correspond to
\BE
\frac{\partial n^\alpha_{i,\tau,t}}{\partial \tau} + \frac{\partial n^\alpha_{i,\tau,t}}{\partial t} &=& f^{\alpha}_{i,\tau,t}\left( \{x^{\alpha}_{i,\tau, t}\} \right) + \xi^{\alpha}_{i, \tau, t},\label{eq:h1}\\
n^{\alpha}_{i,0, t} &=& f^{\alpha}_{i,0,t}\left( \{x^{\alpha}_{i,\tau, t}\} \right) +\xi^{\alpha}_{i, 0, t}, \label{eq:h2}
\EE
where $\xi^{\alpha}_{i, \tau, t} = \lim_{\Delta \to 0 } \frac{\chi^{\alpha}_{i, \tau, t}}{\sqrt{\Delta}}$. We note that the first of these is a stochastic partial differential equation. The second relation is technically not a differential equation (it contains no derivatives with respect to time). We will nevertheless refer to the combination of Eqs.~(\ref{eq:h1}) and (\ref{eq:h2}) as a stochastic differential equation (SDE).

We begin by writing down the generating functional 
\be
Z\Big[ \{ \Xi_{i,\tau,t}^\alpha \} \Big] = \int \prod_{i,\alpha,\tau,t} \Big\{ dx^{\alpha}_{i,\tau, t} \Big\} \left[ \exp\left(i\sum_{i,\alpha,\tau,t} \Xi^{\alpha}_{i,\tau, t} x^{\alpha}_{i,\tau, t}\right) \mathcal{P}\left(\{x^{\alpha}_{i, \tau, t}\}\right)\right] \label{generatingfunctionalsuppl}
\ee
for the process in Eqs. (\ref{discretisedapp1},\ref{discretisedapp2}). The quantity $\mathcal{P}\left(\{x^{\alpha}_{i, \tau, t}\}\right)$ is the probability measure in the space of trajectories for the system. Assuming that the correspondence of the set of noise variables $\{\chi^{\alpha}_{i, \tau, t}\}$ to a particular trajectory $\{x^{\alpha}_{i, \tau, t}\}$ is one-to-one, we can write $\mathcal{P}\left(\{x^{\alpha}_{i, \tau, t}\}\right) = \langle \delta \left(\boldsymbol x - \boldsymbol x\left( \boldsymbol \chi \right) \right) \rangle_{\{\chi^{\alpha}_{i, \tau, t}\}}$, where bold face notation indicates entire trajectories.

Then letting $X^\alpha_{i,\tau+\Delta,t+\Delta} \equiv  x^{\alpha}_{i,\tau+\Delta, t+\Delta} - x^{\alpha}_{i,\tau, t} - \frac{\Delta}{N} f^{\alpha}_{i,\tau,t}\left( \{x^{\alpha}_{i,\tau, t}\} \right) - \frac{\sqrt{\Delta}}{N} \chi^{\alpha}_{i, \tau, t} $ and $X^\alpha_{i,0,t+\Delta} \equiv x^{\alpha}_{i,0, t+\Delta} -\frac{1}{N} f^{\alpha}_{i,0,t}\left( \{x^{\alpha}_{i,\tau, t}\} \right) - \frac{\chi^{\alpha}_{i, 0, t}}{ N\sqrt{\Delta}}$ we have 
\begin{align}
\delta \left(\boldsymbol x - \boldsymbol x\left( \boldsymbol \chi \right) \right) &= \lvert \frac{\delta \boldsymbol X}{\delta \boldsymbol x}\rvert \delta \left( \boldsymbol X\right) = \delta \left( \boldsymbol X\right) \nonumber \\
&=\prod_{i,\alpha,\tau, t} \delta\left(X^\alpha_{i,\tau,t} \right)\prod_{i,\alpha, t} \delta\left(X^\alpha_{i,0,t} \right) ,
\end{align}
where the Jacobian determinant $\lvert \frac{\delta \boldsymbol X}{\delta \boldsymbol x}\rvert = 1$ because the Jacobian itself is a triangular matrix with diagonal elements each equal to unity \cite{altlandsimons}. \\
The generating functional thus takes the following form 
\begin{align}
Z\Big[ \{ \Xi^{\alpha}_{i,\tau, t} \} \Big] =& \int \prod_{i,\alpha,\tau,t} \Big\{ dx^{\alpha}_{i,\tau, t} d\chi^{\alpha}_{i, \tau, t} \Big\} \Bigg[ \exp\left(i \Xi^{\alpha}_{i,\tau, t} x^{\alpha}_{i,\tau, t}\right) P\left(\{\chi^{\alpha}_{i, \tau, t}\}\right) \nonumber \\ 
&\times \prod_{i,\alpha,\tau,t} \Bigg\{ \delta\left( x^{\alpha}_{i,\tau+\Delta, t+\Delta} - x^{\alpha}_{i,\tau, t} - \frac{\Delta}{N} f^{\alpha}_{i,\tau,t}\left( \{x^{\alpha}_{i,\tau, t}\} \right) - \frac{\sqrt{\Delta}}{N} \chi^{\alpha}_{i, \tau, t} \right) \Bigg\}  \nonumber \\ 
&\times \prod_{i,\alpha,t} \Bigg\{ \delta\left(x^{\alpha}_{i,0, t+\Delta} -\frac{1}{N} f^{\alpha}_{i,0,t}\left( \{x^{\alpha}_{i,\tau, t}\} \right) - \frac{\chi^{\alpha}_{i, 0, t}}{ N\sqrt{\Delta}} \right) \Bigg\} \Bigg],
\end{align}
where the $\delta$-functions enforce the stochastic dynamics in Eqs.~(\ref{discretisedapp1},\ref{discretisedapp2}). For the sake of brevity, we have adopted the Einstein summation convention for sums over repeated indices in the exponential, and have omitted the summation symbols $\sum$. The quantity $P\left(\{\chi^{\alpha}_{i, \tau, t}\}\right)$ is the distribution of the noise variables $\{\chi^{\alpha}_{i, \tau, t}\}$; we keep in mind that these may depend on the variables $\{x^{\alpha}_{i,\tau, t}\}$.

Introducing conjugate variables $\{\hat{x}^{\alpha}_{i,\tau, t}\}$ to rewrite the $\delta$-functions in their exponential representation one obtains
\begin{align}
Z\Big[ \{ \Xi^{\alpha}_{i,\tau, t} \} \Big] =& \int \prod_{i,\alpha,\tau,t} \Big\{ \frac{dx^{\alpha}_{i,\tau, t} d \hat{x}^{\alpha}_{i,\tau, t} d \chi^{\alpha}_{i, \tau, t} }{2\pi}  \Big\}  \exp\left[i \Xi^{\alpha}_{i,\tau, t} x^{\alpha}_{i,\tau, t} \right]  \nonumber \\   
&\times \exp\left[i \hat{x}^{\alpha}_{i,\tau, t} \left( x^{\alpha}_{i,\tau+\Delta, t+\Delta} - x^{\alpha}_{i,\tau, t} - \frac{\Delta}{N} f^{\alpha}_{i,\tau,t}\left( \{x^{\alpha}_{i,\tau, t}\} \right) \right) \right]\nonumber \\
& \times \exp\left[i \hat{x}^{\alpha}_{i,0, t} \left( x^{\alpha}_{i,0, t+\Delta} -  \frac{1}{N}f^{\alpha}_{i,0,t}\left( \{x^{\alpha}_{i,\tau, t}\} \right) \right) \right] \nonumber \\  
&\times \exp\left[-i  \frac{\sqrt{\Delta}}{N}  \hat{x}^{\alpha}_{i,\tau, t}  \chi^{\alpha}_{i, \tau, t}   \right] \exp\left[-i \hat{x}^{\alpha}_{i,0, t}  \frac{\chi^{\alpha}_{i, 0, t}}{ N\sqrt{ \Delta}}   \right] P\left(\{\chi^{\alpha}_{i, \tau, t}\}\right).
\end{align}
We can now carry out the Gaussian integrals over the $\{\chi^{\alpha}_{i, \tau, t}\}$, using their joint distribution
\begin{gather}
P\left(\{\chi^{\alpha}_{i, \tau, t}\}\right) = \frac{1}{\Omega} \exp\left[-\frac{1}{2} \chi^{\alpha}_{i, \tau, t} \left( C^{-1}\right)^{\alpha, \alpha'}_{i,i',\tau,\tau', t, t'} \chi^{\alpha'}_{i', \tau', t'}\right] ,
\end{gather}
where $\Omega$ is a normalisation constant. We define the inverse of $C^{\alpha, \alpha'}_{i,i',\tau,\tau', t, t'}$ such that 
\be
C^{\alpha, \alpha'}_{i,i',\tau,\tau', t, t'} \left( C^{-1}\right)^{\alpha', \alpha''}_{i',i'',\tau',\tau'', t', t''}  = \delta_{\alpha, \alpha''}\delta_{i,i'}\delta_{\tau,\tau''}\delta_{t,t''}.
\ee

The integral over $\{\chi^{\alpha}_{i, \tau, t}\}$ can be written as follows:
\begin{align}
&\int \prod_{i,\alpha,\tau,t} \Big\{ d \chi^{\alpha}_{i, \tau, t}   \Big\} P\left(\{\chi^{\alpha}_{i, \tau, t}\}\right)  \exp\left[-i \frac{\sqrt{\Delta}}{N}  \hat{x}^{\alpha}_{i,\tau, t}  \chi^{\alpha}_{i, \tau, t}   \right] \nonumber \\  = &\int \prod_{i,\alpha,\tau,t} \Big\{ d \chi^{\alpha}_{i, \tau, t}   \Big\} \frac{1}{\Omega} \exp\left[-i \frac{ \sqrt{\Delta} }{N} \hat{x}^{\alpha}_{i,\tau, t}  \chi^{\alpha}_{i, \tau, t} - \frac{1}{2}  \chi^{\alpha}_{i, \tau, t} \left( C^{-1}\right)^{\alpha, \alpha'}_{i,i',\tau,\tau', t, t'} \chi^{\alpha'}_{i', \tau', t'}  \right].
\end{align}
We now complete the square in the exponent and carry out the integral over the noise variables. To do this let $\chi^{\alpha}_{i, \tau, t} = -i \frac{\sqrt{\Delta}}{N} C^{\alpha, \alpha'}_{i,i',\tau,\tau', t, t'} \hat{x}^{\alpha'}_{i',\tau', t'} + y^{\alpha}_{i, \tau, t}$. Then,
\begin{align}
&i \frac{\sqrt{\Delta}}{N}  \hat{x}^{\alpha}_{i,\tau, t}  \chi^{\alpha}_{i, \tau, t} + \frac{1}{2}  \chi^{\alpha}_{i, \tau, t} \left( C^{-1}\right)^{\alpha, \alpha'}_{i,i',\tau,\tau', t, t'} \chi^{\alpha'}_{i', \tau', t'} \nonumber \\
&= \frac{1}{2} y^{\alpha}_{i, \tau, t} \left( C^{-1}\right)^{\alpha, \alpha'}_{i,i',\tau,\tau', t, t'} y^{\alpha'}_{i', \tau', t'} +\frac{ \Delta}{ 2N^2}\hat{x}^{\alpha}_{i,\tau, t} C^{\alpha, \alpha'}_{i,i',\tau,\tau', t, t'} \hat{x}^{\alpha'}_{i',\tau', t'}.
\end{align}
Thus, if we change the set of integration variables to $\Big\{y^{\alpha}_{i, \tau, t}\Big\}$, the integral becomes
\begin{align}
&\int \prod_{i,\alpha,\tau,t} \Big\{ d \chi^{\alpha}_{i, \tau, t}   \Big\} p\left(\{\chi^{\alpha}_{i, \tau, t}\}\right)  \exp\left[-i  \sqrt{\Delta}  \hat{x}^{\alpha}_{i,\tau, t}  \chi^{\alpha}_{i, \tau, t}   \right] \nonumber \\
&= \exp\left[-\frac{ \Delta}{ 2N^2}\hat{x}^{\alpha}_{i,\tau, t} C^{\alpha, \alpha'}_{i,i',\tau,\tau', t, t'} \hat{x}^{\alpha'}_{i',\tau', t'} \right]
\int \prod_{i,\alpha,\tau,t} \Big\{ d y^{\alpha}_{i, \tau, t}   \Big\} \frac{1}{\Omega} \exp\left[-\frac{1}{2} y^{\alpha}_{i, \tau, t} \left( C^{-1}\right)^{\alpha, \alpha'}_{i,i',\tau,\tau', t, t'} y^{\alpha'}_{i', \tau', t'}\right] \nonumber \\ 
&= \exp\left[-\frac{ \Delta}{ 2N^2}\hat{x}^{\alpha}_{i,\tau, t} C^{\alpha, \alpha'}_{i,i',\tau,\tau', t, t'} \hat{x}^{\alpha'}_{i',\tau', t'} \right].
\end{align}
Therefore, we can rewrite the generating functional as
\begin{align} 
Z\Big[ \{ \Xi^{\alpha}_{i,\tau,t}\}\Big] =&  \int \prod_{i,\alpha,\tau,t} \Big\{ \frac{dx^{\alpha}_{i,\tau, t} d \hat{x}^{\alpha}_{i,\tau, t}  }{2\pi}  \Big\} \exp\left[i \Xi^{\alpha}_{i,\tau, t} x^{\alpha}_{i,\tau, t}\right] \nonumber \\   
&\times \exp\left[i \hat{x}^{\alpha}_{i,\tau, t} \left( x^{\alpha}_{i,\tau+\Delta, t+\Delta} - x^{\alpha}_{i,\tau, t} - \frac{ \Delta}{ N}f^{\alpha}_{i,\tau,t}\left( \{x^{\alpha}_{i,\tau, t}\} \right) \right) \right]\nonumber \\
&\times \exp\left[i \hat{x}^{\alpha}_{i,0, t} \left( x^{\alpha}_{i,0, t+\Delta} -  \frac{ 1}{ N}f^{\alpha}_{i,0,t}\left( \{x^{\alpha}_{i,\tau, t}\} \right) \right) \right]  \nonumber \\
&\times  \exp\left[- \frac{ \Delta}{ 2 N^2}\hat{x}^{\alpha}_{i,\tau, t} C^{\alpha, \alpha'}_{i,i',\tau,\tau', t, t'} \hat{x}^{\alpha'}_{i',\tau', t'}\right]\exp\left[-\frac{1}{2N^2}\hat{x}^{\alpha}_{i,0, t} C^{\alpha, \alpha'}_{i,i',0,\tau', t, t'} \hat{x}^{\alpha'}_{i',\tau', t'}\right]\nonumber \\
&\times\exp\left[-\frac{1}{2N^2}\hat{x}^{\alpha}_{i,\tau, t} C^{\alpha, \alpha'}_{i,i',\tau,0, t, t'} \hat{x}^{\alpha'}_{i',0, t'}\right]\exp\left[-\frac{1}{2 \Delta N^2}\hat{x}^{\alpha}_{i,0, t} C^{\alpha, \alpha'}_{i,i',0,0, t, t'} \hat{x}^{\alpha'}_{i',0, t'}\right]. 
\label{generatingfunctionalsde}
\end{align}
In Eq. (\ref{generatingfunctionalsde}), sums over repeated indices are implied. Note that the sum over the residence time variables $\tau$ are not to include $\tau = 0$; this special case is dealt with separately.

Thus, we have recast the generating functional in terms of only the coordinates and their conjugates, eliminating the noise variables in favour of their correlations. Now that we have the general form for the generating functional for a process of the form in Eq. (\ref{discretisedapp1},\ref{discretisedapp2}), we can compute the generating functional for the reaction-subdiffusion system and compare the results. This allows us to write down an effective SDE for the reaction-subdiffusion system.

It should be noted that terms in the exponent in Eq. (\ref{generatingfunctionalsde}) which are quadratic in the conjugate variables are associated with the noise, and the coefficients are the correlations between the noise variables. The terms which are linear in the conjugate variables arise due to the deterministic part of the dynamics.

\section{Generating functional for a reaction-diffusion system with non-constant hazard rate}\label{appendix:correlatorsrdsystem}
\subsection{Construction of the generating functional}
\setcounter{equation}{15}
In order to obtain effective SDEs for the type of (anomalous) reaction-diffusion systems in the main paper, we seek to write down the generating functional for the process and reduce it to a form similar to that in Eq.~(\ref{generatingfunctionalsde}). We begin by reminding ourselves of Eqs.~(8) and (9) from the main paper; recalling the notation $x_{i,\tau,t}^\alpha=n_{i,\tau,t}^\alpha/N$ these can be written as
\begin{align}
x^{\alpha}_{i,\tau+\Delta,t+\Delta} - x^{\alpha}_{i,\tau,t} = -\sum_{r} \frac{ k^{\left(R\right) \alpha}_{i,r, \tau, t} \theta\left(- \nu^{\alpha}_{r}\right)}{\Delta N}- \sum_{i'} \frac{k^{\left(H\right) \alpha}_{i',i, \tau, t} }{\Delta N} , \label{discrete1supp}
\end{align}
\begin{gather}
x^{\alpha}_{i,0,t+\Delta} = \sum_{r} \frac{\ell_{i,r,t}^{\left(R\right)}\nu^{\alpha}_{r} \theta \left(\nu^{\alpha}_{r}\right) }{\Delta N} + \sum_{i'} \frac{k^{\left(H\right) \alpha}_{i,i',\tau,t}}{\Delta N} , \label{discrete2supp}
\end{gather}
where  $k^{\left(R\right) \alpha}_{i,r, \tau, t}$ is number of particles of type $\alpha$ and age $\tau$ at position $i$ which are annihilated in the time step from $t$ to $t+\Delta$ due to reactions of type $r$. Similarly, $k^{\left(H\right) \alpha}_{i,i', \tau, t}$ is the number of particles of age $\tau$ hopping away from position $i'$ to $i$ at time $t$. Finally, $\ell_{i,r,t}^{\left(R\right)}$ is the number of reactions of type $r$ firing in the time window $t$ to $t+\Delta$ at position $i$. It should be noted that $k^{\left(R\right) \alpha}_{i,r, \tau, t}$, $k^{\left(H\right) \alpha}_{i,i', \tau, t}$ and $\ell_{i,r,t}^{\left(R\right)}$ are always non-negative integers and $\theta\left(-\nu_r^\alpha\right) \ell^{\left(R\right)}_{i,r,t} \lvert \nu_r^\alpha\rvert= \sum_{\tau}k^{\left(R\right) \alpha}_{i,r, \tau, t}$.

Following a procedure similar to that in Section \ref{genericsde}, we can write Eq.~(\ref{generatingfunctionalsuppl}) in the following form
\begin{align}
Z\Big[\{ \Xi^{\alpha}_{i,\tau,t} \} \Big] =& \sum_{\{ k \}} \int  \prod_{i,\alpha,\tau,t} \Big\{ dx^{\alpha}_{i,\tau, t} \Big\}
\exp\left[i \Xi^{\alpha}_{i,\tau, t} x^{\alpha}_{i,\tau, t}\right] \nonumber \\
&\times \prod_{i,\alpha,\tau,t} \delta \left( x^{\alpha}_{i,\tau+\Delta,t+\Delta} - x^{\alpha}_{i,\tau,t}  +\sum_{r} \frac{ k^{\left(R\right) \alpha}_{i,r, \tau, t} \theta\left(- \nu^{\alpha}_{r}\right)}{ \Delta N} + \sum_{i'} \frac{k^{\left(H\right) \alpha}_{i,i', \tau, t} }{ \Delta N} \right) \nonumber \\
&\times \prod_{i,\alpha,t}\delta \left(  x^{\alpha}_{i,0,t} - \sum_{r} \frac{\ell_{i,r,t}^{\left(R\right)}\nu^{\alpha}_{r} \theta \left(\nu^{\alpha}_{r}\right) }{\Delta N} - \sum_{i'} \frac{k^{\left(H\right) \alpha}_{i,i',\tau,t}}{ \Delta N} \right) \mathcal{P}\left( \{k\} \vert  \{x\}\right) .
\end{align}
Here, we use $\{k\}$ as a shorthand for the set of all the variables $\left\{\ell^{\left(R\right)}_{i,r,t}\right\}$ ,$\left\{k^{\left(R\right) \alpha}_{i,r, \tau, t}\right\}$ and $\left\{k^{\left(H\right) \alpha}_{i,i', \tau, t}\right\}$ and $ \{x\}$ as shorthand for the set $\{ x_{i,\tau,t}^\alpha\}$. The statistics of these variables is informed by the reaction rates, which in turn depend on the $\{x\}$. The joint probability distribution for the set $\{k\}$, given the set $\{x\}$, is denoted by $\mathcal{P}\left( \{k\} \vert  \{x\}\right)$. The precise form of this distribution is discussed below.

Similarly to the previous section, we write the delta function as the integral of a complex exponential, 
\begin{align}
Z\Big[\{ \Xi^{\alpha}_{i,\tau,t} \} \Big] =& \sum_{\{ k \}} \int  \prod_{i,\alpha,\tau,t} \Bigg\{ \frac{dx^{\alpha}_{i,\tau, t} d\hat{x}^{\alpha}_{i,\tau, t}}{2\pi} 
\exp\left[i \Xi^{\alpha}_{i,\tau, t} x^{\alpha}_{i,\tau, t}\right] \nonumber \\ 
&\times \exp \left[ i\hat{x}^{\alpha}_{i,\tau,t} \left(x^{\alpha}_{i,\tau+\Delta,t+\Delta} - x^{\alpha}_{i,\tau,t}  +\sum_{r} \frac{ k^{\left(R\right) \alpha}_{i,r, \tau, t} \theta\left(- \nu^{\alpha}_{r}\right)}{ \Delta N} + \sum_{i'} \frac{k^{\left(H\right) \alpha}_{i,i', \tau, t} }{\Delta N} \right)\right] \nonumber \\
&\times \exp \left[ i\hat{x}^{\alpha}_{i,0,t} \left(x^{\alpha}_{i,0,t+\Delta} - \sum_{r} \frac{\ell_{i,r,t}^{\left(R\right)}\nu^{\alpha}_{r} \theta \left(\nu^{\alpha}_{r}\right) }{\Delta N} - \sum_{i'} \frac{k^{\left(H\right) \alpha}_{i',i,\tau,t}}{\Delta N} \right)\right] \Bigg\}\mathcal{P}\left( \{k\} \vert  \{x\}\right) .
\label{generating1}
\end{align}
Our aim is to carry out the sum over $\{k\}$ in order to obtain an expression for the generating functional in terms of only the coordinates $\{x^{\alpha}_{i,\tau, t}\}$ and the model parameters. Using this, we can then perform an expansion in $N^{-1}$ to obtain an approximate expression which has the same form as Eq.~(\ref{generatingfunctionalsde}). This will allow us to read off the set of effective SDEs. It should be noted that the values $\{k\}$ are of order $N^0$, as are the $\{x^{\alpha}_{i,\tau, t}\}$. 

We must now identify the precise nature of $ \mathcal{P}\left( \{k\} \vert  \{x\}\right) $ in order to evaluate the summation over $\{k\}$. First, due to the way that we have deconvolved the problem by introducing the age-variable $\tau$, we observe that the values of $\{k\}$ at different times $t$ are independent. Further, the hopping events are independent of the reactions. However, the values of $k^{\left(R\right) \alpha}_{i,r, \tau, t}$ at a particular time $t$ are dependent upon the number of firings of the different reaction types $\ell^{\left(R\right)}_{i,r,t}$. Introducing the notation $\{ \cdot\}_t$ to mean the set of all variables for a fixed value of $t$, we write
\BE
\mathcal{P}\left(\{k\} \vert  \{x\}\right) &=& \prod_t P\left(\{k\}_t \vert \{x_{i,\tau,t}^\alpha\}_t  \right)\nonumber \\
& =& \prod_t P\left(\left\{k^{\left(H\right) \alpha}_{i,i', \tau, t}\right\}_t \vert \{x_{i,\tau,t}^\alpha\}_t \right) P\left(\left\{k^{\left(R\right) \alpha}_{i,r, \tau, t}\right\}_t \vert \left\{\ell^{\left(R\right)}_{i,r,t}\right\}_t , \{x_{i,\tau,t}^\alpha\}_t \right) P\left(\left\{\ell^{\left(R\right)}_{i,r,t}\right\}_t \vert \{x_{i,\tau,t}^\alpha\}_t \right).\nonumber \\ \label{sep1}
\EE
We can simplify matters further by noting that the hops of particles of different flavours $\alpha$, ages $\tau$ and starting/ending positions $i$ and $i'$ also occur independently. So, we can write
\begin{align}
P\left(\left\{k^{\left(H\right) \alpha}_{i,i', \tau, t}\right\}_t \vert \{x_{i,\tau,t}^\alpha\}_t \right) = \prod_{i,i',\tau, \alpha}P\left(  k^{\left(H\right) \alpha}_{i,i', \tau, t} \vert x_{i,\tau,t}^\alpha \right) .\label{sep2}
\end{align}
Similarly, reactions of different types $r$ and at different sites $i$ occur independently of one another. We also presume that reaction rates do not depend on the ages of the particles, only the total local concentration. Thus, we can write
\begin{align}
P\left(\left\{\ell^{\left(R\right)}_{i,r,t}\right\}_t \vert \{x_{i,\tau,t}^\alpha\}_t \right) = \prod_{i,r} P\left(\ell^{\left(R\right)}_{i,r,t} \vert \{x_{i,t}^\alpha\}_{i,t } \right) .\label{sep3}
\end{align}
Finally, the numbers of particles of various ages $\tau$ which are created/annihilated in reactions of type $r$ at location $i$ and time $t$ obey the following relation
\begin{align}
P\left(\left\{k^{\left(R\right) \alpha}_{i,r, \tau, t}\right\}_t \vert \left\{\ell^{\left(R\right)}_{i,r,t}\right\}_t , \{x_{i,\tau,t}^\alpha\}_t \right) = \prod_{i,r,\alpha} P\left( \{ k^{\left(R\right) \alpha}_{i,r, \tau, t} \}_{\alpha,i,r,t} | \ell^{\left(R\right)}_{i,r,t}, \{x_{i,t}^\alpha\}_{i,t } \right) .\label{sep4}
\end{align}

We note again that the condition $\sum_{\tau} k^{\left(R\right) \alpha}_{i,r, \tau, t} = \ell^{\left(R\right)}_{i,r,t} \lvert \nu_{r}^{\alpha} \rvert \theta\left(-\nu_r^\alpha\right) $ must be met, so the values of $\{ k^{\left(R\right) \alpha}_{i,r, \tau, t} \}_{\alpha,i,r,t}$ are interdependent. Using Eqs.~(\ref{sep1}-\ref{sep4}), we can write Eq.~(\ref{generating1}) in the form 
\begin{align}
Z\Big[\{ \Xi^{\alpha}_{i,\tau,t} \} \Big] &=  \int  \prod_{i,\alpha,\tau,t} \Bigg\{ \frac{dx^{\alpha}_{i,\tau, t} d\hat{x}^{\alpha}_{i,\tau, t}}{2\pi} 
\exp\left[i \Xi^{\alpha}_{i,\tau, t} x^{\alpha}_{i,\tau, t}\right] \exp \left[ i\hat{x}^{\alpha}_{i,\tau,t} \left(x^{\alpha}_{i,\tau+\Delta,t+\Delta} - x^{\alpha}_{i,\tau,t} \right) \right]  \exp \left[ i\hat{x}^{\alpha}_{i,0,t} x^{\alpha}_{i,0,t+\Delta}\right]\Bigg\}\nonumber \\ 
&\times\prod_{i,r,t} R_{i,r,t} \prod_{\alpha, i, i', \tau, t} H^{\alpha}_{i,i',\tau,t} ,\label{genfunct}
\end{align}
where we have gathered the reaction terms
\begin{align}
R_{i,r,t} =& \sum_{k^{\left( R\right)}_{i,r,t}} \sum_{ \{k^{\left( R\right) \alpha}_{i,r,\tau,t} \}_{i,r,t} } \Bigg[ P\left(\ell^{\left(R\right)}_{i,r,t} \vert \{x_{i,t}^\alpha\}_{i,t } \right) \exp\left( -i\sum_{\alpha} \frac{ \hat{x}^{\alpha}_{i,0,t} \ell^{\left( R\right)}_{i,r,t} \nu^{\alpha}_{r} \theta\left( \nu^{\alpha}_{r}\right)}{\Delta N} \right)  \nonumber \\
&\times \prod_{\alpha} \left( P\left( \{ k^{\left(R\right) \alpha}_{i,r, \tau, t} \}_{\alpha,i,r,t} | \ell^{\left(R\right)}_{i,r,t}, \{x_{i,t}^\alpha\}_{i,t } \right) \exp\left( i\sum_{\tau} \frac{\hat{x}^{\alpha}_{i, \tau, t} k^{\left(R\right)\alpha}_{i,r,\tau,t} \theta\left( -\nu^{\alpha}_{r} \right) }{\Delta N} \right) \right) \Bigg] \nonumber \\
=& \sum_{k^{\left( R\right)}_{i,r,t}}  \Bigg[ P\left(\ell^{\left(R\right)}_{i,r,t} \vert \{x_{i,t}^\alpha\}_{i,t } \right) \exp\left( -i\sum_{\alpha} \frac{ \hat{x}^{\alpha}_{i,0,t} \ell^{\left( R\right)}_{i,r,t} \nu^{\alpha}_{r} \theta\left( \nu^{\alpha}_{r}\right)}{\Delta N} \right)  \nonumber \\
&\times \prod_{\alpha} \sum_{ \{k^{\left( R\right) \alpha}_{i,r,\tau,t} \}_{\alpha,i,r,t} } \left( P\left( \{ k^{\left(R\right) \alpha}_{i,r, \tau, t} \}_{\alpha,i,r,t} | \ell^{\left(R\right)}_{i,r,t}, \{x_{i,t}^\alpha\}_{i,t } \right) \exp\left( i\sum_{\tau} \frac{\hat{x}^{\alpha}_{i, \tau, t} k^{\left(R\right)\alpha}_{i,r,\tau,t} \theta\left( -\nu^{\alpha}_{r} \right) }{\Delta N} \right) \right) \Bigg]. \label{reaction}
\end{align}
The hopping terms are given by 
\begin{align}
H_{i,i',\tau, t}^{\alpha} = \sum_{k_{i,i',\tau, t}^{ \left(H\right) \alpha}}P\left(  k^{\left(H\right) \alpha}_{i,i', \tau, t} \vert x_{i,\tau,t}^\alpha \right)  \exp\left[i \frac{ k^{\left(H\right) \alpha}_{i,i', \tau, t}}{\Delta N} \left( \hat{x}_{i,\tau,t}^{\alpha} -  \hat{x}_{i',0,t}^{\alpha}  \right) \right] . \label{hopping}
\end{align}
Now we evaluate the sums over $\left\{\ell^{\left(R\right)}_{i,r,t}\right\}$ ,$\left\{k^{\left(R\right) \alpha}_{i,r, \tau, t}\right\}$ and $\left\{k^{\left(H\right) \alpha}_{i,i', \tau, t}\right\}$ in Eqs.~(\ref{reaction}) and (\ref{hopping}). In order to do this, we first must specify the distributions $P\left(  k^{\left(H\right) \alpha}_{i,i', \tau, t} \vert x_{i,\tau,t}^\alpha \right)$, $P\left(\ell^{\left(R\right)}_{i,r,t} \vert \{x_{i,t}^\alpha\}_{i,t } \right)$ and $P\left( \{ k^{\left(R\right) \alpha}_{i,r, \tau, t} \}_{\alpha,i,r} | \ell^{\left(R\right)}_{i,r,t} , \{x_{i,t}^\alpha\}_{i,t }\right)$. After these sums are evaluated, we perform the system-size expansion in powers of $N^{-1}$. The calculation that follows is based on the assumption made for example in the context of Gillespie's $\tau$-leaping algorithm \cite{tauleap}. Reaction rates are assumed to be constant during each time interval $\Delta$. Hence the number of reactions of type $r$ firing in such an internal is a Poissonian random number with parameter $\lambda_{i,r,t}\times \Delta$. The statistics of the number of hopping events are Poissonian as well, with the appropriate rates. 

\smallskip

First, we deal with the hopping contribution. The distribution for $k^{\left(H\right) \alpha}_{i,i', \tau, t}$ is Poissonian with mean $m^{\alpha}_{i,i', \tau, t} = \Delta^2\phi_{i,i'} h^\alpha_{\tau} n_{i,\tau,t}^{\alpha}  $ where $h^\alpha_{\tau}$ is the rate of hopping for particles which have survived a time $\tau$. That is, we have
\begin{gather}
P\left(  k^{\left(H\right) \alpha}_{i,i', \tau, t} \right) = \frac{\left( m^{\alpha}_{i,i', \tau, t} \right)^{k^{\left(H\right) \alpha}_{i,i', \tau, t} }}{k^{\left(H\right) \alpha}_{i,i', \tau, t}  !} \exp\left(- m^{\alpha}_{i,i', \tau, t}\right) . \label{hopprob}
\end{gather}
Substituting Eq.~(\ref{hopprob}) into Eq.~(\ref{hopping}) and evaluating the sum over $k_{i,i',\tau, t}^{ \left(H\right) \alpha}$ , one then obtains
\begin{align}
H_{i,i',\tau, t}^{\alpha} &= \sum_{k_{i,i',\tau, t}^{ \left(H\right) \alpha}} \frac{\left( m^{\alpha}_{i,i', \tau, t} e^{i \frac{ 1}{\Delta N} \left( \hat{x}_{i,\tau,t}^{\alpha} -  \hat{x}_{i',0,t}^{\alpha}  \right) } \right)^{k^{\left(H\right) \alpha}_{i,i', \tau, t} }}{k^{\left(H\right) \alpha}_{i,i', \tau, t}  !} \exp\left(- m^{\alpha}_{i,i', \tau, t}\right)  \nonumber \\
=& \exp \left[m^{\alpha}_{i,i', \tau, t}\left(  e^{i \frac{ 1}{\Delta N} \left( \hat{x}_{i,\tau,t}^{\alpha} -  \hat{x}_{i',0,t}^{\alpha}  \right) } -1\right) \right]\nonumber \\
= & \exp \left[ m^{\alpha}_{i,i', \tau, t} \left( i \frac{\left( \hat{x}_{i,\tau,t}^{\alpha} -  \hat{x}_{i',0,t}^{\alpha}  \right) }{\Delta N} -
\frac{\left( \hat{x}_{i,\tau,t}^{\alpha} -  \hat{x}_{i',0,t}^{\alpha}  \right)^{2} }{2\Delta^2 N^{2}} \right)+{\cal O}(N^{-3}) \right] . \label{fullhop}
\end{align}
In the last step we have carried out the expansion in $N^{-1}$, and have retained terms up to and including sub-leading order. 

The procedure for the reaction events is more complicated, but follows similar lines. The distribution for $k^{\left(R\right) \alpha}_{i,r, \tau, t}$ is dependent on the number of firings of the different reaction types, $\ell^{\left(R\right)}_{i,r,t}$. The distribution for $\ell^{\left(R\right)}_{i,r,t}$ is Poissonian. The mean number of firings of a reaction of type $r$ in time $\Delta$ is $\Delta\lambda_{i,r,t}$. The rate $\lambda_{i,r,t}$ is dependent on the overall concentrations $\{ x_{i,t}^{\alpha} \}$. The distribution of the number of firings of type $r$ at position $i$ and time $t$ is therefore
\begin{gather}
P\left(\ell^{\left(R\right)}_{i,r,t} \vert \{x_{i,t}^\alpha\}_{i,t } \right) = \frac{\left(\Delta\lambda_{i,r,t} \right)^{\ell^{\left(R\right)}_{i,r,t}}}{\ell^{\left(R\right)}_{i,r,t} !} e^{-\Delta\lambda_{i,r,t}}. \label{reactprob2}
\end{gather}
The conditional joint distribution for $\{ k^{\left(R\right) \alpha}_{i,r, \tau, t} \}_{\alpha,i,r,t} $ is multinomial if particles of type $\alpha$ are annihilated in the reaction of type $r$. If particles of type $\alpha$ are not annihilated in reactions of type $r$, $\{ k^{\left(R\right) \alpha}_{i,r, \tau, t} \}_{\alpha,i,r,t} $ must all be zero. The distribution is multinomial because the ages $\tau$ of the particles to be annihilated are selected randomly, with each value of $\tau$ coming up with probability proportional to $\frac{x_{i,\tau,t}^{\alpha}}{x_{i,t}^{\alpha}}$. The values of $\{ k^{\left(R\right) \alpha}_{i,r, \tau, t} \}_{\alpha,i,r,t} $ must also be such that $\sum_{\tau} k^{\left(R\right) \alpha}_{i,r, \tau, t} = \ell^{\left(R\right)}_{i,r,t} \lvert \nu_{r}^{\alpha} \rvert \theta\left(-\nu_r^\alpha\right) $. So the conditional joint distribution for $\{ k^{\left(R\right) \alpha}_{i,r, \tau, t} \}_{\alpha,i,r,t} $ is
\BE
P\left( \{ k^{\left(R\right) \alpha}_{i,r, \tau, t} \}_{\alpha,i,r,t} | \ell^{\left(R\right)}_{i,r,t} , \{ x^\alpha_{i,\tau,t}\}_{i,t}\right) &=& 
\theta\left( -\nu^\alpha_r\right)\left(\ell^{\left(R\right)}_{i,r,t} \lvert \nu^{\alpha}_{r} \rvert \right)! \left( \prod_{\tau}  \frac{1}{k^{\left(R\right) \alpha}_{i,r, \tau, t} !}\left( \frac{x_{i,\tau,t}^{\alpha}}{x_{i,t}^{\alpha}} \right)^{k^{\left(R\right) \alpha}_{i,r, \tau, t}}  \right)\nonumber \\
&&~~~~~~\times \delta^{(K)}\left(\sum_{\tau}k^{\left(R\right) \alpha}_{i,r, \tau, t} - \ell^{\left(R\right)}_{i,r,t}\lvert \nu^{\alpha}_{r} \rvert  \right) \nonumber \\
&&+ \theta\left( \nu^\alpha_r\right) \prod_\tau \delta^{(K)}\left( k^{\left(R\right) \alpha}_{i,r, \tau, t}\right) , \label{reactprob1}
\EE
where we use $\delta^{(K)}\left( \cdot\right)$ to denote a Kronecker delta function which is equal to one when the argument is zero and is equal to zero otherwise. We note that the expression $\left(\ell^{\left(R\right)}_{i,r,t} \lvert \nu^{\alpha}_{r} \rvert \right)! \left( \prod_{\tau}  \frac{1}{k^{\left(R\right) \alpha}_{i,r, \tau, t} !}\left( \frac{x_{i,\tau,t}^{\alpha}}{x_{i,t}^{\alpha}} \right)^{k^{\left(R\right) \alpha}_{i,r, \tau, t}}  \right)$ in Eq.~(\ref{reactprob1}) reduces to the familiar binomial distribution if the number of values that $\tau$ can take is limited to two. As a sanity check we use the multinomial expansion to verify normalisation,
\begin{align}
\sum_{\{ k^{\left(R\right) \alpha}_{i,r, \tau, t} \}_{\alpha,i,r,t}} P\left( \{ k^{\left(R\right) \alpha}_{i,r, \tau, t} \}_{\alpha,i,r,t} | \ell^{\left(R\right)}_{i,r,t} , \{ x^\alpha_{i,\tau,t}\}_{i,t}\right) &= \theta\left(-\nu^\alpha_r \right)\left( \sum_{\tau} \frac{\Delta x_{i,\tau,t}^{\alpha}}{x_{i,t}^{\alpha}}  \right)^{|\nu^{\alpha}_{r}| \ell^{\left( R\right)}_{i,r,t} } + \theta\left(\nu^\alpha_r \right) \nonumber \\
&=\theta\left(-\nu^\alpha_r \right)+ \theta\left(\nu^\alpha_r \right) \nonumber \\
&= 1 .
\end{align}
Using equation (\ref{reactprob1}), one can evaluate the sum over $\{k^{\left( R\right) \alpha}_{i,r,\tau,t} \}_{i,r,t}$ in Eq.~(\ref{reaction}) to show that
\BE
&&\sum_{ \{k^{\left( R\right) \alpha}_{i,r,\tau,t} \}_{\alpha,i,r,t} } \left( P\left( \{ k^{\left(R\right) \alpha}_{i,r, \tau, t} \}_{\alpha,i,r,t} | \ell^{\left(R\right)}_{i,r,t} , \{ x^\alpha_{i,\tau,t}\}_{i,t} \right) \exp\left( i\sum_{\tau} \frac{\hat{x}^{\alpha}_{i, \tau, t} k^{\left(R\right)\alpha}_{i,r,\tau,t} \theta\left( -\nu^{\alpha}_{r} \right) }{\Delta N} \right) \right) \nonumber \\
&=& \theta\left( -\nu^\alpha_r\right)\sum_{ \{k^{\left( R\right) \alpha}_{i,r,\tau,t} \}_{\alpha,i,r,t} } \left(\ell^{\left(R\right)}_{i,r,t} \lvert \nu^{\alpha}_{r} \rvert \right)! \left( \prod_{\tau}  \frac{1}{k^{\left(R\right) \alpha}_{i,r, \tau, t} !}\left( \frac{x_{i,\tau,t}^{\alpha}}{x_{i,t}^{\alpha}} \exp\left( i \frac{\hat{x}^{\alpha}_{i, \tau, t}  \theta\left( -\nu^{\alpha}_{r} \right) }{\Delta N} \right) \right)^{k^{\left(R\right) \alpha}_{i,r, \tau, t}}   \right)\nonumber \\
&&~~~~~~~~~~~~~~~~~~~~~~~~~~~~~~~~~~~~~~~~~~~~\times \delta^{(K)}\left(\sum_{\tau}k^{\left(R\right) \alpha}_{i,r, \tau, t} - \ell^{\left(R\right)}_{i,r,t}\lvert \nu^{\alpha}_{r} \rvert  \right) \nonumber \\
&&+ \theta\left( \nu^\alpha_r\right) \sum_{ \{k^{\left( R\right) \alpha}_{i,r,\tau,t} \}_{\alpha,i,r,t} }\prod_\tau \delta^{(K)}\left( k^{\left(R\right) \alpha}_{i,r, \tau, t}\right) \exp\left( i\frac{\hat{x}^{\alpha}_{i, \tau, t} k^{\left(R\right)\alpha}_{i,r,\tau,t} \theta\left( -\nu^{\alpha}_{r} \right) }{\Delta N} \right) \nonumber \\
&=& \left( \sum_{\tau} \frac{\Delta x_{i,\tau,t}^{\alpha}}{x_{i,t}^{\alpha}} e^{i \frac{\hat{x}^{\alpha}_{i,\tau,t} \theta\left(-\nu^{\alpha}_{r}\right)}{\Delta N} } \right)^{|\nu^{\alpha}_{r}| \ell^{\left( R\right)}_{i,r,t} } ,
\EE
where again we have used the multinomial expansion to evaluate the sums over $\{k^{\left( R\right) \alpha}_{i,r,\tau,t}\}$. Combining this result with Eq.~(\ref{reaction}), one obtains
\begin{align}
R_{i,r,t} &= \sum_{\ell^{\left( R\right)}_{i,r,t}} \Bigg[P\left(\ell^{\left(R\right)}_{i,r,t} \vert \{x_{i,t}^\alpha\}_{i,t } \right) \exp\left( -i\sum_{\alpha} \frac{ \hat{x}^{\alpha}_{i,0,t} \ell^{\left( R\right)}_{i,r,t} |\nu^{\alpha}_{r}| \theta\left( \nu^{\alpha}_{r}\right)}{\Delta N} \right) \nonumber \\
&\times \prod_{\alpha} \left( \sum_{\tau} \frac{\Delta x_{i,\tau,t}^{\alpha}}{x_{i,t}^{\alpha}} e^{i \frac{\hat{x}^{\alpha}_{i,\tau,t} \theta\left(-\nu^{\alpha}_{r}\right)}{\Delta N} } \right)^{|\nu^{\alpha}_{r}| \ell^{\left( R\right)}_{i,r,t} } \Bigg] .
\end{align}
If one then uses the expression in Eq.~(\ref{reactprob2}), one can perform a similar deduction to that used to find Eq.~(\ref{fullhop}) to finally obtain
\begin{gather}
R_{i,r,t} = \exp\left( -\Delta\lambda_{i,r,t} + \Delta\lambda_{i,r,t} \prod_{\alpha} \left( \sum_{\tau} \frac{\Delta x_{i,\tau,t}^{\alpha}}{x_{i,t}^{\alpha}} e^{i \frac{\left( \hat{x}^{\alpha}_{i,\tau,t} \theta\left(-\nu^{\alpha}_{r}\right) - \hat{x}^{\alpha}_{i,0,t} \theta\left(\nu^{\alpha}_{r}\right) \right) }{\Delta N} } \right)^{|\nu^{\alpha}_{r}| } \right) .
\end{gather}
This can then be expanded in powers of $\frac{1}{N}$. To the lowest two orders one obtains
\BE
\ln\left(R_{i,r,t} \right) &\approx&\frac{i}{N} \sum_{\alpha \tau} \Delta\lambda_{i,r,t} |\nu^{\alpha}_{r}| \frac{x^{\alpha}_{i,\tau,t}}{x^{\alpha}_{i,t}}  \left( \hat{x}^{\alpha}_{i,\tau,t} \theta\left(-\nu^{\alpha}_{r}\right) - \hat{x}^{\alpha}_{i,0,t} \theta\left(\nu^{\alpha}_{r}\right) \right) \nonumber \\
&&- \frac{1}{2N^{2}} \sum_{\alpha \tau} \lambda_{i,r,t}|\nu^{\alpha}_{r}| \frac{x^{\alpha}_{i,\tau,t}}{x^{\alpha}_{i,t}} \left( \left( \hat{x}^{\alpha}_{i,\tau,t}\right) ^{2} \theta\left(-\nu^{\alpha}_{r}\right)  + \left( \hat{x}^{\alpha}_{i,0,t}\right) ^{2} \theta\left(\nu^{\alpha}_{r}\right) \right) \nonumber \\
&&- \frac{1}{2N^{2}} \sum_{\alpha \tau \tau'} \Delta\lambda_{i,r,t} |\nu^{\alpha}_{r}| \left( |\nu^{\alpha}_{r}| -1 \right) \frac{x^{\alpha}_{i,\tau,t}x^{\alpha}_{i,\tau',t}}{\left( x^{\alpha}_{i,t}\right)^{2}} \left( \hat{x}^{\alpha}_{i,\tau,t} \hat{x}^{\alpha}_{i,\tau',t} \theta\left(-\nu^{\alpha}_{r}\right)  + \left( \hat{x}^{\alpha}_{i,0,t}\right) ^{2} \theta\left(\nu^{\alpha}_{r}\right) \right) \nonumber \\
&&- \frac{1}{N^{2}} \sum_{{\rm pairs} \,\left(\alpha, \alpha' \right)} \sum_{ \tau \tau'} \Delta\lambda_{i,r,t} |\nu^{\alpha}_{r}||\nu^{\alpha'}_{r}| \frac{x^{\alpha}_{i,\tau,t}x^{\alpha'}_{i,\tau',t}}{ x^{\alpha}_{i,t} x^{\alpha '}_{i,t} } \left( \hat{x}^{\alpha}_{i,\tau,t} \theta\left(-\nu^{\alpha}_{r}\right) - \hat{x}^{\alpha}_{i,0,t} \theta\left(\nu^{\alpha}_{r}\right) \right)\nonumber \\
&&~~~~~~~~~~~~~~~~~~~~~~~~~~~~~~~~~~~~~~~~~~~~~~\times \left( \hat{x}^{\alpha'}_{i,\tau,t} \theta\left(-\nu^{\alpha'}_{r}\right) - \hat{x}^{\alpha'}_{i,0,t} \theta\left(\nu^{\alpha'}_{r}\right) \right) . \label{fullreact}
\EE
In the above, `${\rm pairs}(\alpha,\alpha')$' means each combination of $\alpha$ and $\alpha'$ with $\alpha \neq \alpha'$ such that $(\alpha,\alpha')$ and $(\alpha',\alpha)$ are the same `pair'. Neglected terms are of order $N^{-3}$. 
Having found approximate expressions for $R_{i,r,t}$ and $H^{\alpha}_{i,i',\tau,t}$ in Eq.~(\ref{fullhop}) and Eq.~(\ref{fullreact}), we can write down the final expression for the generating functional 
\begin{align}
Z&\Big[\{ \Xi^{\alpha}_{i,\tau,t} \} \Big] \approx  \int  \prod_{i,\alpha,\tau,t} \Bigg\{ \frac{dx^{\alpha}_{i,\tau, t} d\hat{x}^{\alpha}_{i,\tau, t}}{2\pi} 
\exp\left[i \Xi^{\alpha}_{i,\tau, t} x^{\alpha}_{i,\tau, t}\right] \exp \left[ i\hat{x}^{\alpha}_{i,\tau,t} \left(x^{\alpha}_{i,\tau+\Delta,t+\Delta} - x^{\alpha}_{i,\tau,t} \right) \right]  \exp \left[ i\hat{x}^{\alpha}_{i,0,t} x^{\alpha}_{i,0,t+\Delta}\right]\Bigg\}\nonumber \\ 
\times& \exp\left\{\frac{i}{N} \sum_{\alpha,i,\tau,t} \hat x^\alpha_{i,\tau,t}\left[ \sum_{i'} \Delta \phi_{i,i'} h^\alpha _\tau n^\alpha_{i,\tau,t} + \sum_{r} \Delta \lambda_{i,r,t} \lvert \nu_r^\alpha\rvert \frac{n^\alpha_{i,\tau,t}}{n^\alpha_{i,t}} \theta\left( -\nu_r^\alpha\right)\right]\right\} \nonumber \\
\times& \exp\left\{ -\frac{i}{N} \sum_{\alpha,i,t} \hat x^\alpha_{i,0,t}\left[\sum_{i',\tau} \Delta \phi_{i,i'} h^\alpha _\tau n^\alpha_{i',\tau,t} + \sum_{r,\tau} \Delta \lambda_{i,r,t} \lvert \nu_r^\alpha\rvert \frac{n^\alpha_{i,\tau,t}}{n^\alpha_{i,t}} \theta\left( \nu_r^\alpha\right) \right]\right\} \nonumber \\
\times& \exp\Bigg\{-\frac{1}{2 N^2}\sum_{\alpha,i,\tau,t}\sum_{\alpha',i',\tau',t'} \hat x^\alpha_{i,\tau,t} \hat x^{\alpha'}_{i',\tau',t'} \Bigg[\delta_{\alpha,\alpha'}\delta_{\tau,\tau'} \delta_{t,t'} h_\tau^\alpha n^\alpha_{i,\tau,t} + \delta_{i,i'}\delta_{\alpha \alpha'}\delta_{\tau \tau'}\delta_{t,t'} \sum_r \lambda_{i,r,t}|\nu^{\alpha}_{r}| \frac{n^{\alpha}_{i,\tau,t}}{n^{\alpha}_{i,t}} \theta\left( -\nu_r^\alpha\right) \nonumber \\
&+ \delta_{\alpha, \alpha'}\delta_{i,i'}\delta_{t,t'} \sum_r \Delta\lambda_{i,r,t} |\nu^{\alpha}_{r}| \left( |\nu^{\alpha}_{r}| -1 \right) \frac{n^{\alpha}_{i,\tau,t}n^{\alpha}_{i,\tau',t}}{\left( n^{\alpha}_{i,t}\right)^{2}}\theta\left( -\nu_r^\alpha\right) \nonumber \\
&+ \left(1-\delta_{\alpha,\alpha'} \right)\delta_{i,i'}\delta_{t,t'} \sum_r \Delta\lambda_{i,r,t} |\nu^{\alpha}_{r}||\nu^{\alpha'}_{r}| \frac{n^{\alpha}_{i,\tau,t}n^{\alpha'}_{i,\tau',t}}{ n^{\alpha}_{i,t} n^{\alpha '}_{i,t} }\theta\left( -\nu_r^\alpha\right) \theta\left( -\nu_r^{\alpha'}\right)\Bigg]\Bigg\} \nonumber \\
\times& \exp\Bigg\{-\frac{1}{N^2}\sum_{\alpha,i,\tau,t}\sum_{\alpha',i',t'} \hat x^\alpha_{i,\tau,t} \hat x^{\alpha'}_{i',0,t'} \Bigg[\left(1-\delta_{\alpha,\alpha'} \right)\delta_{i,i'}\delta_{t,t'} \sum_r \lambda_{i,r,t} |\nu^{\alpha}_{r}||\nu^{\alpha'}_{r}| \frac{n^{\alpha}_{i,\tau,t}}{ n^{\alpha}_{i,t}  } \theta\left( -\nu_r^\alpha\right)\theta\left( \nu_r^{\alpha'}\right) \nonumber \\
&-\delta_{\alpha,\alpha'}\delta_{t,t'}\phi_{i,i'}h_\tau^\alpha n^\alpha_{i,t}\Bigg]\Bigg\} \nonumber \\
\times& \exp\Bigg\{-\frac{1}{2 N^2}\sum_{\alpha,i,t}\sum_{\alpha',i',t'} \hat x^\alpha_{i,0,t} \hat x^{\alpha'}_{i',0,t'} \Bigg[ \delta_{\alpha,\alpha'} \delta_{i,i'} \frac{\delta_{t,t'}}{\Delta} \sum_{r} \lambda_{i,r,t} \lvert \nu_r^\alpha\rvert^2 \theta\left(\nu_r^\alpha \right) \nonumber \\
&+ \left(1-\delta_{\alpha,\alpha'} \right)\delta_{i,i'}\frac{\delta_{t,t'}}{\Delta} \sum_r \lambda_{i,r,t} |\nu^{\alpha}_{r}||\nu^{\alpha'}_{r}|  \theta\left( \nu_r^\alpha\right)\theta\left( \nu_r^{\alpha'}\right) + \delta_{\alpha,\alpha'}\delta_{i,i'}\frac{\delta_{t,t'}}{\Delta}\sum_{i'',\tau} \Delta \phi_{i,i''}h^\alpha_\tau n^\alpha_{i'',\tau,t}\Bigg] \Bigg\}. \label{finalgf}
\end{align}
We have used the fact that $\frac{n^{\alpha}_{i,\tau,t}}{ n^{\alpha}_{i,t}  } = \frac{x^{\alpha}_{i,\tau,t}}{ x^{\alpha}_{i,t}  }$. Eq.~(\ref{finalgf}) contains terms in the exponentials which are at most quadratic in the conjugate variables $\{ \hat x\}$. By comparing Eq.~(\ref{finalgf}) with Eq.~(\ref{generatingfunctionalsde}) we can read off expressions for the effective SDEs for the individual-based system in the Gaussian approximation. 

The terms which are linear in the conjugate variables (and with pre-factors of order $N^{-1}$) represent the dynamics in the deterministic limit. The quadratic terms in Eq.~(\ref{finalgf}) carry pre-factors $N^{-2}$; the coefficients in these terms represent the correlations between the noise variables. These correlators are thus given by 
\begin{align}
\Delta \langle \chi^{\alpha}_{i,\tau,t} \chi^{\alpha'}_{i',\tau',t'} \rangle =& \delta_{i i'} \delta_{t t'} \delta_{\alpha,\alpha'}\left[\delta_{\tau \tau'}\left( h^\alpha_{\tau} + p^\alpha_{i,t} \right)  n^{\alpha}_{i ,\tau ,t} + \Delta \sum_r \lvert \nu_r^\alpha \rvert \left( \lvert \nu_r^\alpha \rvert -1 \right) \lambda_{i,r,t} \frac{ n^{\alpha}_{i ,\tau, t} n^{\alpha}_{i', \tau' ,t'}}{\left( n^{\alpha}_{i ,t}\right)^2}\theta\left( - \nu_r^\alpha \right)\right] \nonumber \\
&+\left(1-\delta_{\alpha,\alpha'}\right)\delta_{i i'} \delta_{t t'} \Delta \sum_r \lvert \nu_r^\alpha \rvert \lvert \nu_r^{\alpha'}\rvert  \lambda_{i,r,t} \frac{ n^{\alpha}_{i ,\tau, t}}{ n^{\alpha}_{i ,t}}\frac{ n^{\alpha'}_{i' ,\tau', t'}}{ n^{\alpha'}_{i' ,t'}}\theta\left( - \nu_r^\alpha \right)\theta\left( - \nu_r^{\alpha'} \right),\nonumber \\
\langle \chi^{\alpha}_{i,\tau,t} \chi^{\alpha'}_{i',0,t'} \rangle =& - \delta_{t t'}\delta_{\alpha,\alpha'}\phi_{i,i'}h^{\alpha}_{\tau}  n^{\alpha}_{i \tau t} - \left(1-\delta_{\alpha,\alpha'}\right)\delta_{i i'} \delta_{t t'} \sum_r \lvert \nu_r^\alpha \rvert \lvert \nu_r^{\alpha'}\rvert  \lambda_{i,r,t} \frac{ n^{\alpha}_{i ,\tau, t}}{ n^{\alpha}_{i ,t}}\theta\left( - \nu_r^\alpha \right)\theta\left( \nu_r^{\alpha'} \right) , \nonumber \\
\frac{\langle \chi^{\alpha}_{i,0,t} \chi^{\alpha'}_{i',0,t'} \rangle}{\Delta} =& \delta_{\alpha,\alpha'}\left[\delta_{i i'}\frac{\delta_{t t'}}{\Delta} \sum_r \lambda_{i,r,t}  \left(\nu_r^\alpha\right)^2  \theta\left( \nu_r^\alpha \right) + \delta_{i i'} \frac{\delta_{t t'}}{\Delta} \sum_{i'',\tau} \Delta \phi_{i,i''}h^\alpha_\tau n_{i'',\tau,t}\right] \nonumber \\
&+\left(1-\delta_{\alpha,\alpha'}\right)\delta_{i i'} \frac{\delta_{t t'}}{\Delta} \sum_r \lvert \nu_r^\alpha \rvert \lvert \nu_r^{\alpha'}\rvert  \lambda_{i,r,t} \theta\left( \nu_r^\alpha \right)\theta\left( \nu_r^{\alpha'} \right) .
\end{align}

\subsection{Results of the generating functional approach}
Once continuous time is restored by sending $\Delta \to 0$, the effective SDE found from the comparison of Eq.~(\ref{finalgf}) with Eq.~(\ref{generatingfunctionalsde}) is given by
\begin{align}
\frac{\partial n^{\alpha}_{i,\tau , t} }{\partial t} + \frac{\partial n^{\alpha}_{i,\tau, t} }{\partial \tau} &= -  h^\alpha_{\tau} n_{i,\tau,t}^{\alpha} - p_{i,t}^{\alpha} n_{i,\tau,t}^\alpha + \xi^{\alpha}_{{i,\tau,t}} , \nonumber \\
n^{\alpha}_{i, 0 , t} &= \sum_{i'} \phi_{i,i'} \int_0^t  h^\alpha_{\tau} n_{i',\tau,t}^{\alpha} d\tau + \gamma_{i,t}^\alpha  + \xi^{\alpha}_{{i,0,t}} .\label{stochasticeqssuppl}
\end{align}
These are the expressions in Eq.~(11) in the main paper. When one takes the continuous-time limit one uses the fact that $\frac{\delta_{t t'}}{\Delta} \to \delta\left( t - t'\right)$ as $\Delta\to 0$, and that in going from Eqs.~(\ref{discretisedapp1}) and (\ref{discretisedapp2}) to Eq.~(\ref{stochasticeqssuppl}) the Gaussian variables are rescaled in the following manner $\frac{\chi_{i,\tau,t}^{\alpha}}{\sqrt{\Delta}} \to \xi_{i,\tau,t}^{\alpha}$. One finds
\begin{align}
\langle \xi^{\alpha}_{i,\tau,t} \xi^{\alpha'}_{i',\tau',t'} \rangle =& \delta_{i i'} \delta\left(t-t'\right) \delta_{\alpha,\alpha'}\left[\delta\left(\tau-\tau'\right)\left( h^\alpha_{\tau} + p^\alpha_{i,t} \right)  n^{\alpha}_{i ,\tau ,t} +  \sum_r \lvert \nu_r^\alpha \rvert \left( \lvert \nu_r^\alpha \rvert -1 \right) \lambda_{i,r,t} \frac{ n^{\alpha}_{i ,\tau, t} n^{\alpha}_{i', \tau' ,t'}}{\left( n^{\alpha}_{i ,t}\right)^2}\theta\left( - \nu_r^\alpha \right)\right] \nonumber \\
&+\left(1-\delta_{\alpha,\alpha'}\right)\delta_{i i'} \delta\left(t-t'\right) \sum_r \lvert \nu_r^\alpha \rvert \lvert \nu_r^{\alpha'}\rvert  \lambda_{i,r,t} \frac{ n^{\alpha}_{i ,\tau, t}}{ n^{\alpha}_{i ,t}}\frac{ n^{\alpha'}_{i' ,\tau', t'}}{ n^{\alpha'}_{i' ,t'}}\theta\left( - \nu_r^\alpha \right)\theta\left( - \nu_r^{\alpha'} \right),\nonumber \\
\langle \xi^{\alpha}_{i,\tau,t} \xi^{\alpha'}_{i',0,t'} \rangle =& - \delta\left(t-t'\right)\delta_{\alpha,\alpha'}\phi_{i,i'}h^{\alpha}_{\tau}  n^{\alpha}_{i \tau t} - \left(1-\delta_{\alpha,\alpha'}\right)\delta_{i i'} \delta\left(t-t'\right) \sum_r \lvert \nu_r^\alpha \rvert \lvert \nu_r^{\alpha'}\rvert  \lambda_{i,r,t} \frac{ n^{\alpha}_{i ,\tau, t}}{ n^{\alpha}_{i ,t}}\theta\left( - \nu_r^\alpha \right)\theta\left( \nu_r^{\alpha'} \right) , \nonumber \\
\langle \xi^{\alpha}_{i,0,t} \xi^{\alpha'}_{i',0,t'}\rangle  =& \delta_{\alpha,\alpha'}\left[\delta_{i i'}\delta\left(t-t'\right) \sum_r \lambda_{i,r,t}  \left(\nu_r^\alpha\right)^2  \theta\left( \nu_r^\alpha \right) + \delta_{i i'} \delta\left(t-t'\right) \sum_{i''} \int_0^t \phi_{i,i''}h^\alpha_\tau n_{i'',\tau,t} d\tau \right] \nonumber \\
&+\left(1-\delta_{\alpha,\alpha'}\right)\delta_{i i'} \delta\left(t-t'\right) \sum_r \lvert \nu_r^\alpha \rvert \lvert \nu_r^{\alpha'}\rvert  \lambda_{i,r,t} \theta\left( \nu_r^\alpha \right)\theta\left( \nu_r^{\alpha'} \right) . \label{correlatorsgeneral}
\end{align}
In the Markovian case the noise variables of concern are $\eta^\alpha_{i,t} = \int_0^t \xi^\alpha_{i,\tau,t} d\tau + \xi^\alpha_{i,0,t}$ (as stated in Eq.~(16) of the main paper) and the hazard rate $h^\alpha_\tau = h^\alpha$ is a constant with respect to $\tau$. As a check, we calculate the correlators of these noise variables and find
\begin{align}
\langle \eta^\alpha_{i,t} \eta^{\alpha'}_{i',t'} \rangle &= \delta\left( t-t'\right)\delta_{i,i'}\sum_r \nu_r^\alpha \nu^{\alpha'}_r \lambda_{i,r,t} \nonumber \\
&+ \delta_{\alpha,\alpha'}\delta\left( t-t'\right)h^\alpha \left(\delta_{i,i'}n^\alpha_{i,t} + \delta_{i,i'}\sum_{i''}\phi_{i,i''}n^\alpha_{i'',t} - \phi_{i,i'}n^\alpha_{i',t}- \phi_{i',i}n^\alpha_{i,t}\right) .
\end{align} 
This is the same result one would obtain by using a standard master equation approach and the Kramers-Moyal expansion or by the direct application of Kurtz' theorem \cite{brettgalla, kurtz, gardiner}.\\
To summarise, we have found the approximate generating functional for a non-Markovian reaction-diffusion system. This was done by identifying the probability distributions for the numbers of the various events that might occur, carrying out the summations over $\{k\}$ and, finally, by using a system-size expansion to obtain the same form as in Eq.~(\ref{generatingfunctionalsde}). This enabled us to read off the effective SDEs for the system, given in Eq.~(11) of the main paper, and the correlation coefficients for the stochastic variables by equating terms of the same order in $\frac{1}{N}$.
\section{Derivation of the anomalous reaction-diffusion equation with noise}\label{reactdiffnoise}
\setcounter{equation}{40}
Beginning with Eq.~(\ref{stochasticeqssuppl}), we wish to derive the general reaction-diffusion equation
 \be
\frac{\partial n_{i,t}^\alpha }{\partial t}  = \sum_{i'} \left\{\left( \phi_{i,i'} - \delta_{i,i'}\right)\left(t_0^\alpha\right)^{-\gamma^\alpha} e^{ -\int_0^t p_{i',T'}^{\alpha}dT' } {}_0D^{1-\gamma^\alpha}_t\left[ n_{i',t}^\alpha e^{ \int_0^t p_{i',T'}^{\alpha}dT'} \right]\right\} +f^\alpha_{i,t} +\eta^\alpha_{i,t} \label{reactsubdiffsuppl}
\ee
This is Eq.~(13) in the main paper. This is done using a similar method to those used in \cite{vladross}, \cite{fedotovfalconer1} or \cite{yadav}. Here, however, we also handle the noise variables. Using the method of characteristics, one can show from Eq.~(\ref{stochasticeqssuppl}) that
\begin{align}
n^\alpha_{i,\tau,t} &= n^\alpha_{i,0,t-\tau}\Psi^\alpha\left(\tau\right)\exp\left[ - \int_{t-\tau}^t p^\alpha_{i,s} ds \right] \nonumber \\ 
&+ \Psi^\alpha\left(\tau\right)\exp\left[- \int_{t-\tau}^t p^\alpha_{i,s} ds \right]\int_0^\tau \frac{\xi^\alpha_{i,T,T+t-\tau}}{\Psi^\alpha\left(T\right)\exp\left[- \int_{t-\tau}^{t-\tau+T} p^\alpha_{i,s} ds \right]}dT , \label{characteristics}
\end{align}
where we have used $\Psi^\alpha\left(\tau\right) = e^{-\int_0^\tau h^\alpha_s ds}$. Also, one can integrate the first of Eqs.~(\ref{stochasticeqssuppl}) with respect to $\tau$ in order to obtain
\begin{gather}
\frac{\partial n^\alpha_{i,t}}{\partial t} = \sum_{i'} \left( \phi_{i,i'} - \delta_{i,i'} \right) \int_0^t h^\alpha_\tau n^\alpha_{i',\tau,t} d\tau + f^\alpha_{i,t} +\int_0^t \xi^\alpha_{i,\tau,t} d\tau + \xi^\alpha_{i,0,t} . \label{integrated}
\end{gather}
All that remains is to eliminate the integral over $\tau$ in Eq. (\ref{integrated}) in favour of an expression involving only $n^\alpha_{i,t}$. We do this using the convolution theorem for Laplace transforms and Eq.~(\ref{characteristics}). Integrating Eq. (\ref{characteristics}) with respect to $\tau$ and then taking the Laplace transform we obtain
\begin{align}
\mathcal{L}_t\left[n^\alpha_{i,t} e^{\int_0^t p^\alpha_{i,s}ds}\right] &=\mathcal{L}_t\left[n^\alpha_{i,0,t} e^{\int_0^t p^\alpha_{i,s}ds}\right] \mathcal{L}_t\left[\Psi^\alpha\left(t\right) \right]\nonumber \\
&+\mathcal{L}_t\left[  \int_0^t\Psi^\alpha\left(\tau\right)\exp\left[ \int^{t-\tau}_0 p^\alpha_{i,s} ds \right]\int_0^\tau \frac{\xi^\alpha_{i,T,T+t-\tau}}{\Psi^\alpha\left(T\right)\exp\left[- \int_{t-\tau}^{t-\tau+T} p^\alpha_{i,s} ds \right]}dT d\tau\right] \label{laplaceconvolution1}
\end{align}
One the other hand we can also multiply both sides of Eq.~(\ref{characteristics}) by $h_\tau^\alpha$, then integrate over $\tau$, and then take the Laplace transform. We then arrive at 
\begin{align}
\mathcal{L}_t\left[e^{\int_0^t p^\alpha_{i,s}ds}\int_0^t h^\alpha_\tau n^\alpha_{i,\tau,t} d\tau \right]&= \mathcal{L}_t\left[n^\alpha_{i,0,t} e^{\int_0^t p^\alpha_{i,s}ds}\right] \mathcal{L}_t\left[\psi^\alpha\left(t\right) \right] \nonumber \\ &+\mathcal{L}_t\left[  \int_0^t\psi^\alpha\left(\tau\right)\exp\left[ \int^{t-\tau}_0 p^\alpha_{i,s} ds \right]\int_0^\tau \frac{\xi^\alpha_{i,T,T+t-\tau}}{\Psi^\alpha\left(T\right)\exp\left[- \int_{t-\tau}^{t-\tau+T} p^\alpha_{i,s} ds \right]}dT d\tau\right] \label{laplaceconvolution2}.
\end{align}
 One can use the combination of Eqs.~(\ref{laplaceconvolution1}) and (\ref{laplaceconvolution2}) to eliminate $\mathcal{L}_t\left[n^\alpha_{i,0,t} e^{\int_0^t p^\alpha_{i,s}ds}\right]$, and to find an expression for $\int_0^t h^\alpha_\tau n^\alpha_{i,\tau,t} d\tau$ in terms of $n^\alpha_{i,t}$.  Defining the memory kernel $K^\alpha_{t} = \mathcal{L}^{-1}_u\left(t\right) \left\{ \frac{\psi^\alpha\left(u\right)}{\Psi^\alpha\left(u\right)}\right\}$ and inverting the Laplace transforms, this finally leads one to 
\begin{gather}
\frac{\partial n_{i,t}^\alpha }{\partial t}  = \sum_{i'} \left\{ \left( \phi_{i',i} - \delta_{i',i}\right) e^{ -\int_0^t p_{i',T'}^{\alpha}dT'} \int_0^t K^\alpha_{t-T} n_{i',T}^\alpha e^{ \int_0^T p_{i',T'}^{\alpha}dT'} dT \right\} +f^\alpha_{i,t} +\eta^\alpha_{i,t} ~, \label{generalreactdiff}
\end{gather}
with the noise term
\begin{align}
\eta^\alpha_{i,t} &= \sum_{i'} \Bigg\{ \left( \phi_{i,i'} - \delta_{i,i'}\right) \Big( e^{ -\int_0^t p_{i',T'}^{\alpha}dT' } \int_0^t \psi^\alpha\left(\tau\right) e^{ \int_0^{t-\tau} p_{i',T'}^{\alpha}dT' }\int^\tau_0 \frac{\xi^\alpha_{i',T,T+t-\tau}}{\Psi^\alpha\left(T\right) e^{-\int_0^{T} p_{i',T'+t-\tau}^{\alpha}dT' }} dTd\tau \nonumber \\
&-   e^{ -\int_0^t p_{i',T'}^{\alpha}dT' } \int_0^t K^\alpha_{t-\tau} \int_0^\tau \Psi^\alpha\left(T\right) e^{ \int_0^{\tau-T} p_{i',T'}^{\alpha}dT' }\int^T_0 \frac{\xi^\alpha_{i',X,X+\tau-T}}{\Psi^\alpha\left(X\right) e^{-\int_0^{X} p_{i',T'+\tau-T}^{\alpha}dT' }} dXdTd\tau \Big) \Bigg\} \nonumber \\
&+\int_0^t \xi^\alpha_{i,\tau,t} d\tau + \xi^\alpha_{i,0,t} . \label{noiseterms}
\end{align}
If we now choose the waiting-time distribution to be the Mittag-Leffler function,
\begin{align}
\psi^{\alpha}\left(t\right) &= -\frac{d}{dt} \left\{ E_{\gamma^\alpha}\left[ - \left( \frac{t}{t_0^{\alpha}}\right)^{\gamma^\alpha}\right] \right\},\label{waitingtimessuppl}
\end{align}
then we obtain Eq.~(\ref{reactsubdiffsuppl}) since in this case $\int_0^t K^\alpha_{t-T} f\left(T\right) dT = \left( t_0^\alpha\right)^{-\gamma^\alpha}{}_0 D_t^{1-\gamma^\alpha} \left\{ f\left(t\right)\right\}$.
\section{Linear-noise approximation}\label{linearnoise}
\setcounter{equation}{48}
We define the deviation from the homogeneous fixed point as $\delta^\alpha_{i,\tau,t} = n^\alpha_{i,\tau,t} - \bar n^\alpha$. We presume that $\delta^\alpha_{i,\tau,t} \sim \mathcal{O}\left(\sqrt{N}\right)$. One can then expand Eq.~(\ref{generalreactdiff}) as a series in $\frac{1}{\sqrt{N}}$. We discard terms $\mathcal{O}\left(N^0\right)$. The leading terms are of order $N$, and correspond to the deterministic trajectory. The sub-leading terms are $\mathcal{O}\left(\sqrt{N}\right)$, and encapsulate the behaviour of the fluctuations to a linear approximation. Carrying out this expansion, we have
\begin{align}
\frac{\partial \tilde\delta^\alpha_{q,t}}{\partial t} = \left( \tilde \phi_q - 1 \right) e^{-\bar p^\alpha t} \Bigg[ &  \sum_\beta A_{\alpha \beta} \bar n^\alpha \int_0^t K^\alpha_{t-T} e^{\bar p^\alpha T} \int_0^T \tilde\delta^\beta_{q,t'} dt' dT -\sum_\beta A_{\alpha \beta} \bar n^\alpha \int_0^t \tilde\delta^\beta_{q,t'} dt'  \int_0^t K^\alpha_{t-T} e^{\bar p^\alpha T} dT \nonumber \\
&+\int_0^t K^\alpha_{t-T} \tilde\delta^\alpha_{q, T} e^{\bar p^\alpha T} dT \Bigg] + \sum_\beta f^{\alpha \beta} \tilde \delta^\beta_{q,t} + \tilde \eta^\alpha_{q t} ,\label{linearisedfrac}
\end{align}
where the concentration-dependent terms in the expression for the noise are now evaluated at the homogeneous fixed point such that
\BE
\tilde\eta^\alpha_{q,t} &= & \left( \tilde\phi_{q} - 1\right) \Big( \int_0^t \psi^\alpha\left(\tau\right) e^{ - \bar p^{\alpha} \tau }\int^\tau_0 \frac{\tilde\xi^\alpha_{q,T,T+t-\tau}}{\Psi^\alpha\left(T\right) e^{-\bar p^\alpha T }} dTd\tau \nonumber\nonumber \\
&& \hspace{4cm}- \int_0^t K^\alpha_{t-\tau} e^{-\bar p^\alpha \left(t-\tau\right)} \int_0^\tau \Psi^\alpha\left(T\right) e^{ \bar p^\alpha T }\int^T_0 \frac{\tilde\xi^\alpha_{q,X,X+\tau-T}}{\Psi^\alpha\left(X\right) e^{-\bar p^\alpha X }} dXdTd\tau \Big)  \nonumber \\
&&+\int_0^t \tilde \xi^\alpha_{q,\tau,t} d\tau + \tilde \xi^\alpha_{q,0,t} . \label{compnoise}
\EE
The correlators of $\{\xi^\alpha_{i,\tau,t} \}$ are also evaluated at the fixed point. In order to simplify the above expression for the noise, one notes that
\BE
&&\frac{\partial}{\partial \tau}  \int_0^\tau \Psi^\alpha\left(\tau - T\right) e^{-\bar p^\alpha \left(\tau -T\right) }\int^{\tau -T}_0 \frac{\tilde\xi^\alpha_{q,X,X+T}}{\Psi^\alpha\left(X\right) e^{-\bar p^\alpha X }} dXdT \nonumber \\
&= &- \int_0^\tau \psi^\alpha\left(\tau - T\right) e^{-\bar p^\alpha \left(\tau -T\right) }\int^{\tau -T}_0 \frac{\tilde\xi^\alpha_{q,X,X+T}}{\Psi^\alpha\left(X\right) e^{-\bar p^\alpha X }} dXdT \nonumber \\
&&- \bar p^\alpha \int_0^\tau \Psi^\alpha\left(\tau - T\right) e^{ -\bar p^\alpha \left(\tau -T\right) }\int^{\tau -T}_0 \frac{\tilde\xi^\alpha_{q,X,X+T}}{\Psi^\alpha\left(X\right) e^{-\bar p^\alpha X }} dXdT + \int_0^\tau \tilde\xi^\alpha_{q,T,\tau} dT .
\EE
Using the Laplace transform and the corresponding convolution theorem, this enables us to define the following simplified noise variables
\begin{align}
\tilde\epsilon^\alpha_{q,t} &= \int_0^t \Psi\left(t-\tau\right) e^{-\bar p^\alpha \left(t-\tau\right)} \tilde \eta^\alpha_{q,\tau} d\tau  \nonumber \\
&=  \int_0^t \Psi^\alpha\left(t-\tau\right) e^{ -\bar p^\alpha \left(t-\tau\right) } \Bigg[\left(1-\tilde \phi_q \right)\int^{t-\tau}_0 \frac{\tilde\xi^\alpha_{q,T,\tau+T}}{\Psi^\alpha\left(T\right) e^{-\bar p^\alpha T }} dT + \tilde\phi_q \int_0^\tau \tilde \xi^\alpha_{q,T,\tau} dT  + \tilde \xi^\alpha_{q,0,\tau} \Bigg] d\tau . \label{simpnoise}
\end{align}
The expression in Eq.~(\ref{simpnoise}) is preferable that in Eq.~(\ref{compnoise}) because it involves integrals over at most two variables, as opposed to three. This simplification makes the calculation of the correlators of the noise variables much more straightforward. Taking the Laplace transform of Eq.~(\ref{linearisedfrac}) and multiplying through by $\hat \Psi^\alpha\left(u+ \bar p^\alpha\right)$, one obtains
\begin{gather}
\sum_{\beta}\Bigg\{\left[ 1- \tilde \phi_q \hat\psi^\alpha\left(u+ \bar p^\alpha\right) - \bar p^\alpha \hat \Psi^\alpha\left(u+\bar p^\alpha\right)\right] \delta_{\alpha,\beta}- \hat\Psi^\alpha\left(u+\bar p^\alpha\right) \left[f^{\alpha \beta} + \left(1-\tilde\phi_q\right) \bar n^\alpha A^{\alpha \beta}  \right] \nonumber \\
+ \left(1-\tilde\phi_q\right)\bar n^\alpha\frac{1}{u}\left[1 - \frac{\hat \Psi^\alpha\left( u+\bar p^\alpha\right)}{\hat \Psi^\alpha\left( \bar p^\alpha\right)} \right] A^{\alpha \beta}\Bigg\}\hat{\tilde{\delta}}_{q,u}^{\beta} = \hat{\tilde\epsilon}_{q,u}^\alpha . \label{laplacespace}
\end{gather}
This can then be inverted to find an expression for the fluctuations in terms of the noise variables. Observing that Eq.~(\ref{laplacespace}) represents a matrix equation of the form $\underline{\underline{\hat{\tilde{m}}}}^{-1}_{q,u} \underline{\hat{\tilde{\delta}}}_{q,u} = \underline{\hat{\tilde{\epsilon}}}_{q,u}$, we obtain the solution
\begin{align}
\underline{\hat{\tilde{\delta}}}_{q,u} = \underline{\underline{\hat{\tilde{m}}}}_{q,u} \underline{\hat{\tilde{\epsilon}}}_{q,u} .\label{laplacem}
\end{align}
The elements of the matrix $\underline{\underline{\tilde{m}}}_{q,t}$ are the Green functions of Eq.~(\ref{linearisedfrac}) without the noise term.

\section{Correlators for the noise variables and fluctuations}\label{correlatorexpressions}
\setcounter{equation}{54}
The correlators of the noise variables in Eq.~(\ref{simpnoise}) can be derived using Eqs.~(\ref{correlatorsgeneral}) and are given by 
\begin{align}
\langle \tilde\epsilon^{\left(\alpha\right)}_{q,t}\tilde\epsilon^{\left(\alpha\right)\star}_{q',t'} \rangle &= \delta_{q,q'} \Bigg[ c^{\left(q\right)}_{\Psi\Psi}\left(\alpha\right) \int_{0}^{\min\left(t,t'\right)} \Psi_\star^{\left(\alpha\right)}\left( t-T\right) \Psi_\star^{\left(\alpha\right)}\left(t'-T\right)  dT  \nonumber \\
&+ c^{\left(q\right)}_{\chi}\left(\alpha\right)\chi^{\left(\alpha\right)}\left(\lvert t-t'\rvert \right) + c^{\left(q\right)}_{\chi\chi}\left(\alpha\right)\int_{0}^{\min\left(t,t'\right)} \chi^{\left(\alpha\right)}\left( t-T\right) \chi^{\left(\alpha\right)}\left(t'-T\right)  dT \Bigg],\label{long1}
\end{align}
and
\begin{align}
\langle \tilde\epsilon^{\left(\alpha\right)}_{q,t}\tilde\epsilon^{\left(\alpha'\right)\star}_{q',t'} \rangle &=  \delta_{q,q'} \Bigg[c'^{\left(q\right)}_{\Psi\Psi}\left(\alpha,\alpha'\right) \int_{0}^{\min\left(t,t'\right)} \Psi_\star^{\left(\alpha\right)}\left( t-T\right) \Psi_\star^{\left(\alpha'\right)}\left(t'-T\right)  dT \nonumber \\\
&+c'^{\left(q\right)}_{\chi\Psi}\left(\alpha,\alpha'\right)\int_0^{\min\left(t,t'\right)} \chi^{\left(\alpha'\right)}\left(t' -T\right) \Psi_\star^{\left(\alpha\right)}\left(t-T\right) dT \nonumber \\
&+c'^{\left(q\right)}_{\chi\Psi}\left(\alpha',\alpha\right)\int_0^{\min\left(t,t'\right)} \chi^{\left(\alpha\right)}\left(t -T\right) \Psi_\star^{\left(\alpha'\right)}\left(t'-T\right) dT \nonumber \\
&+ c'^{\left(q\right)}_{\chi\chi}\left(\alpha,\alpha'\right) \int_{0}^{\min\left(t,t'\right)} \chi^{\left(\alpha\right)}\left( t-T\right) \chi^{\left(\alpha'\right)}\left(t'-T\right)  dT \Bigg] .\label{long2}
\end{align}
We have defined the function $\chi^{\left(\alpha\right)}\left(t\right) = \int_0^{\infty} \Psi_\star^{\left(\alpha\right)}\left( t + T \right) dT$ and used the shorthand $\Psi_\star^{\left(\alpha\right)}\left(T\right) =\Psi^{\left(\alpha\right)}\left(T\right)e^{-\bar p^{\left(\alpha\right)} T}$. We note also that $\hat\chi^{\left(\alpha\right)}\left(u\right) = \frac{1}{u}\left(\hat\Psi\left(\bar p^{\left(\alpha\right)}\right)-\hat\Psi\left(u+\bar p^{\left(\alpha\right)}\right) \right)$. The remaining coefficients in Eqs. (\ref{long1}) and (\ref{long2}) are given by
\begin{align}
c^{\left(q\right)}_{\Psi\Psi}\left(\alpha\right) = & \sum_r \lambda_r \lvert\nu_r^{\left(\alpha\right)}\rvert^2 \theta\left(\nu_r^{\left(\alpha\right)}\right) + \tilde\phi_q^2 \sum_r \lambda_r \lvert\nu_r^{\left(\alpha\right)}\rvert^2 \theta\left(-\nu_r^{\left(\alpha\right)}\right)\nonumber \\
&+ \left( 1-\tilde\phi_q^2\right)\int_0^\infty h^{\left(\alpha\right)}_\tau \bar{n}^{\left(\alpha\right)}_\tau d\tau- \left(1 -  \tilde\phi_q\right)^2\bar{n}^{\left(\alpha\right)}_0 , \nonumber \\
c^{\left(q\right)}_{\chi}\left(\alpha\right) = & \left(1 -  \tilde\phi_q\right)^2\bar{n}^{\left(\alpha\right)}_0 + \tilde\phi_q\left( 1-\tilde\phi_q\right)\sum_r \lambda_r \lvert\nu_r^{\left(\alpha\right)}\rvert^2 \theta\left(-\nu_r^{\left(\alpha\right)}\right) , \nonumber \\
c^{\left(q\right)}_{\chi\chi}\left(\alpha\right) = & \left( 1-\tilde\phi_q\right)^2 \left(\frac{\bar{n}^{\left(\alpha\right)}_0}{\bar{n}^{\left(\alpha\right)}} \right)^2 \sum_r \lvert\nu_r^{\left(\alpha\right)}\rvert \left( \lvert\nu_r^{\left(\alpha\right)}\rvert - 1\right)\lambda_r \theta\left( -\nu_r^{\left(\alpha\right)}\right), \nonumber \\
c'^{\left(q\right)}_{\Psi\Psi}\left(\alpha,\alpha'\right) =& \sum_r \lambda_r \lvert\nu_r^{\left(\alpha\right)}\rvert \lvert\nu_r^{\left(\alpha'\right)}\rvert \theta\left(\nu_r^{\left(\alpha\right)}\right) \theta\left(\nu_r^{\left(\alpha'\right)}\right) + \tilde\phi_q^2 \sum_r \lambda_r \lvert\nu_r^{\left(\alpha\right)}\rvert\lvert\nu_r^{\left(\alpha'\right)}\rvert\theta\left(-\nu_r^{\left(\alpha\right)}\right) \theta\left(-\nu_r^{\left(\alpha'\right)}\right) \nonumber \\
&- \tilde\phi_q \sum_r \lambda_r \lvert\nu_r^{\left(\alpha\right)}\rvert\lvert\nu_r^{\left(\alpha'\right)}\rvert\theta\left(\nu_r^{\left(\alpha\right)}\right) \theta\left(-\nu_r^{\left(\alpha'\right)}\right)- \tilde\phi_q \sum_r \lambda_r \lvert\nu_r^{\left(\alpha\right)}\rvert\lvert\nu_r^{\left(\alpha'\right)}\rvert\theta\left(-\nu_r^{\left(\alpha\right)}\right) \theta\left(\nu_r^{\left(\alpha'\right)}\right),\nonumber \\
c'^{\left(q\right)}_{\chi\Psi}\left(\alpha,\alpha'\right)=&~\left(1-\tilde\phi_q\right)\tilde\phi_q \frac{\bar n_0^{\alpha'}}{\bar n^{\alpha'}} \sum_r  \lambda_r \lvert\nu_r^{\left(\alpha\right)}\rvert\lvert\nu_r^{\left(\alpha'\right)}\rvert\theta\left(-\nu_r^{\left(\alpha\right)}\right) \theta\left(-\nu_r^{\left(\alpha'\right)}\right)\nonumber \\
&- \left(1-\tilde\phi_q\right)\frac{\bar n_0^{\alpha'}}{\bar n^{\alpha'}} \sum_r  \lambda_r \lvert\nu_r^{\left(\alpha\right)}\rvert\lvert\nu_r^{\left(\alpha'\right)}\rvert\theta\left(\nu_r^{\left(\alpha\right)}\right) \theta\left(-\nu_r^{\left(\alpha'\right)}\right),\nonumber \\
c'^{\left(q\right)}_{\chi\chi}\left(\alpha,\alpha'\right) =& \left( 1-\tilde\phi_q\right)^2 \frac{\bar{n}^{\left(\alpha\right)}_0}{\bar{n}^{\left(\alpha\right)}} \frac{\bar{n}^{\left(\alpha'\right)}_0}{\bar{n}^{\left(\alpha'\right)}}   \sum_r  \lambda_r \lvert\nu_r^{\left(\alpha\right)}\rvert\lvert\nu_r^{\left(\alpha'\right)}\rvert\theta\left(-\nu_r^{\left(\alpha\right)}\right) \theta\left(-\nu_r^{\left(\alpha'\right)}\right) .
\end{align}

We have used the long-time limit to approximate the barred quantities as constant in time. Note that 
\begin{equation}
\bar n^\alpha_0 = \frac{\bar n^\alpha}{\hat\Psi\left(\bar p^{\left(\alpha\right)}\right)}, \; \;\int_0^\infty h^\alpha_\tau \bar n^\alpha_\tau d\tau  = \frac{\hat\psi\left(\bar p^{\left(\alpha\right)}\right)}{\hat\Psi\left(\bar p^{\left(\alpha\right)}\right)}\bar n^\alpha .
\end{equation}
The following identities for the double Laplace transform allow us to simplify our expressions considerably
\begin{align}
\int_0^{\min\left(t,t'\right)} f\left(t-T\right) g\left(t'-T\right) dT &= \int_0^t \int_0^{t'} f\left(t-T\right) g\left(t'-T'\right) \delta\left( T-T'\right) dT' dT \nonumber \\
&= \mathcal{L}^{-1}_{u,u'} \left\{ \frac{\hat f\left(u\right)\hat g\left( u'\right)}{u+u'}\right\}\left(t,t'\right) \nonumber \\
\mathcal{L}_{t,t'} \left\{ f\left( \lvert t - t'\rvert\right) \right\}\left(u,u'\right) &= \frac{\hat f\left(u\right) + \hat f\left(u'\right)}{u + u'} .
\end{align}
Using Eq.~(\ref{laplacem}) one can then use the expressions for the noise correlators to calculate the correlators and power spectra of the fluctuations $\delta^\alpha_{i,t}$. These are given by
\begin{align}
\langle \tilde\delta^{\left(\alpha\right)}_{q,t}\tilde\delta^{\left(\alpha'\right)\star}_{q,t'} \rangle &= \sum_{\beta, \beta'}\mathcal{L}^{-1}_{u,u'} \left\{  m^{\left(\alpha \beta\right)}_{q,u}m^{\left(\alpha' \beta'\right)}_{q,u'}\langle \hat{\tilde{\epsilon}}^{\left(\beta\right)}_{q,u}\hat{\tilde{\epsilon}}^{\left(\beta'\right)\star}_{q,u'} \rangle \right\} \left(t,t'\right) \nonumber \\
&= \sum_{\beta} c_{\Psi \Psi}^{\left(q\right)}\left(\beta\right) I^{\alpha, \alpha', \beta}_{\Psi \Psi}\left(q,t,t'\right) + c_{\chi}^{\left(q\right)}\left(\beta\right) I^{\alpha, \alpha', \beta}_{\chi}\left(q,t,t'\right) + c_{\chi\chi}^{\left(q\right)}\left(\beta\right) I^{\alpha, \alpha', \beta}_{\chi\chi}\left(q,t,t'\right) \nonumber \\
&+ \sum_{\beta \neq \beta'} \big[ c_{\Psi\Psi}^{\left(q\right)}\left(\beta, \beta'\right) I^{\alpha, \alpha', \beta, \beta'}_{\Psi\Psi}\left(q,t,t'\right) + c_{\chi\Psi}^{\left(q\right)}\left(\beta, \beta'\right) I^{\alpha, \alpha', \beta, \beta'}_{\chi\Psi}\left(q,t,t'\right) \nonumber \\
&+ c_{\chi\Psi}^{\left(q\right)}\left(\beta', \beta\right) I^{\alpha', \alpha, \beta, \beta'}_{\chi\Psi}\left(q,t',t\right) + c_{\chi\chi}^{\left(q\right)}\left(\beta, \beta'\right) I^{\alpha, \alpha', \beta, \beta'}_{\chi\chi}\left(q,t,t'\right) \big]
, \label{fluctfull}
\end{align}
\begin{align}
 I^{\alpha, \alpha', \beta}_{\Psi \Psi}\left(q,t,t'\right) =& \int_0^{\min\left(t,t'\right)} \mathcal{L}^{-1}_u\left\{ \hat\Psi^\beta\left(u+ \bar p^{\beta}\right)m^{\left(\alpha \beta\right)}_{q,u} \right\}\left(t-T\right)\mathcal{L}^{-1}_{u'}\left\{ \hat\Psi^{\beta}\left(u'+ \bar p^{\beta}\right)m^{\left(\alpha' \beta\right)}_{q,u'}\right\}\left(t'-T\right)dT \nonumber \\
 I^{\alpha, \alpha', \beta}_{\chi}\left(q,t,t'\right) =& \int_0^{\min\left(t,t'\right)} \mathcal{L}^{-1}_u\left\{ \hat\chi^\beta\left(u\right)m^{\left(\alpha \beta\right)}_{q,u} \right\}\left(t-T\right)\mathcal{L}^{-1}_{u'}\left\{ m^{\left(\alpha' \beta\right)}_{q,u'}\right\}\left(t'-T\right) dT \nonumber \\
&+ \int_0^{\min\left(t,t'\right)}\mathcal{L}^{-1}_u\left\{ m^{\left(\alpha \beta\right)}_{q,u} \right\}\left(t-T\right)\mathcal{L}^{-1}_{u'}\left\{ \hat\chi^{\beta}\left(u'\right)m^{\left(\alpha' \beta\right)}_{q,u'}\right\}\left(t'-T\right)dT \nonumber \\
 I^{\alpha, \alpha', \beta}_{\chi\chi}\left(q,t,t'\right) =& \int_0^{\min\left(t,t'\right)} \mathcal{L}^{-1}_u\left\{ \hat\chi^\beta\left(u\right)m^{\left(\alpha \beta\right)}_{q,u} \right\}\left(t-T\right)\mathcal{L}^{-1}_{u'}\left\{ \hat\chi^{\beta}\left(u'\right)m^{\left(\alpha' \beta\right)}_{q,u'}\right\}\left(t'-T\right) dT \nonumber \\
 I^{\alpha, \alpha', \beta, \beta'}_{\Psi\Psi}\left(q,t,t'\right) =& \int_0^{\min\left(t,t'\right)} \mathcal{L}^{-1}_u\left\{ \hat\Psi^\beta\left(u+ \bar p^{\beta}\right)m^{\left(\alpha \beta\right)}_{q,u} \right\}\left(t-T\right)\mathcal{L}^{-1}_{u'}\left\{ \hat\Psi^{\beta'}\left(u'+ \bar p^{\beta'}\right)m^{\left(\alpha' \beta'\right)}_{q,u'}\right\}\left(t'-T\right)dT \nonumber \\
 I^{\alpha, \alpha', \beta, \beta'}_{\chi\Psi}\left(q,t,t'\right) =& \int_0^{\min\left(t,t'\right)} \mathcal{L}^{-1}_u\left\{ \hat\chi^\beta\left(u\right)m^{\left(\alpha \beta\right)}_{q,u} \right\}\left(t-T\right)\mathcal{L}^{-1}_{u'}\left\{ \hat\Psi^{\beta'}\left(u'+ \bar p^{\beta'}\right)m^{\left(\alpha' \beta'\right)}_{q,u'}\right\}\left(t'-T\right)dT \nonumber \\
 I^{\alpha, \alpha', \beta, \beta'}_{\chi\chi}\left(q,t,t'\right) =& \int_0^{\min\left(t,t'\right)}\mathcal{L}^{-1}_u\left\{ \hat\chi^\beta\left(u\right)m^{\left(\alpha \beta\right)}_{q,u} \right\}\left(t-T\right)\mathcal{L}^{-1}_{u'}\left\{ \hat\chi^{\beta'}\left(u'\right)m^{\left(\alpha' \beta'\right)}_{q,u'}\right\}\left(t'-T\right) dT . \label{integrals}
\end{align}
In order to evaluate the integrals in Eq.~(\ref{integrals}), the inverse Laplace transforms of the functions in the integrands are taken numerically using the Zakian method \cite{halstedbrown} and the integrals are performed numerically.\\
We note that in the long-time limit the correlators for both the noise variables and the fluctuations can be shown to be time-translation invariant, i.e., they are a function of only $t-t'$. This is due to the fact that, since the functions in the integrands decay as their arguments are increased, the integrals can be approximated in the following way in the long-time limit
\begin{align}
\int_0^{\min\left(t,t'\right)} f\left(t-T\right)g\left(t'-T\right) dT \approx \int_0^\infty f\left[t-\min\left(t,t'\right)+T\right]g\left[t'-\min\left(t,t'\right)+T\right] dT .
\end{align} 
The evaluation of the equal-time correlator can be performed without the use of the numerical inverse Laplace transform. The equal-time correlator involves integrals of the form 
\be
\int_0^t f\left(t-T\right) g\left(t-T\right) dT = \int_0^t f\left(T\right) g\left(T\right) dT.
\ee
In the long-term, one can use the following theorem to evaluate the equal-time correlator
\BE
\lim_{t \to \infty} \int_0^t f\left(T\right)g\left(T\right) dT &=& \lim_{u \to 0} u \mathcal{L}_t \left\{ \int_0^t f\left(T\right)g\left(T\right) dT \right\}\left(u\right)\nonumber \\
&  =& \lim_{u \to 0} \mathcal{L}_t \left\{ f\left(t\right)g\left(t\right) \right\}\left(u\right) \nonumber \\
&=&  \frac{1}{2 \pi i}\int_{c -i \infty}^{c+i \infty} \hat f\left(-s\right) \hat g\left(s\right) ds, \label{parsevallaplace}
\EE
where $c$ is a real number to the right of all the poles of either function in the integrand and we have used the dual convolution theorem for Laplace transforms \cite{debnath}. This result is an analogue of Parseval's theorem, which applies to Fourier transforms. The result Eq.~(\ref{parsevallaplace}) allows one to calculate the equal-time power-spectra of the fluctuations without using a numerical Laplace transform. Instead, one merely has to evaluate a contour integral numerically. Let $A^{\beta \beta'}_q \left(u,u'\right) = \left(u + u'\right)\langle \hat{\tilde{\epsilon}}^{\left(\beta\right)}_{q,u}\hat{\tilde{\epsilon}}^{\left(\beta'\right)\star}_{q,u'} \rangle$. The equal-time power spectrum is then given by
\begin{align}
\lim_{t \to \infty}\langle \tilde\delta^{\left(\alpha\right)}_{q,t}\tilde\delta^{\left(\alpha'\right)\star}_{q,t} \rangle &= \sum_{\beta, \beta'} \frac{1}{2 \pi i}\int_{c -i \infty}^{c+i \infty} m^{\left(\alpha \beta\right)}_{q,u}m^{\left(\alpha' \beta'\right)}_{q,-u}  A^{\beta \beta'}_q \left(u,-u\right) du .
\end{align}
\pagebreak
\section{Accuracy of the system size expansion and low-$N$ behaviour}
\setcounter{figure}{0}
The system-size expansion performed in Section \ref{appendix:correlatorsrdsystem} assumes a large value of $N$. In this section, we briefly discuss how large the system-size is required to be for our theory to be accurate and whether the conclusions drawn in the main paper can be extended to lower system sizes.\\
Fig. \ref{fig:ndependence} demonstrates the convergence to the analytical result for the power spectrum as the system size is increased, in the case of the Brusselator model. One notices that the inaccuracy is greatest for lower $q$ and that, as a result, the stochastic patterns are more prominent in the system for lower $N$. That is, the peak of the power spectrum surpasses the value at $q=0$ to a greater extent for lower $N$ than for higher $N$.

\begin{figure}[h!]
	\centering
	\includegraphics[scale = 0.2]{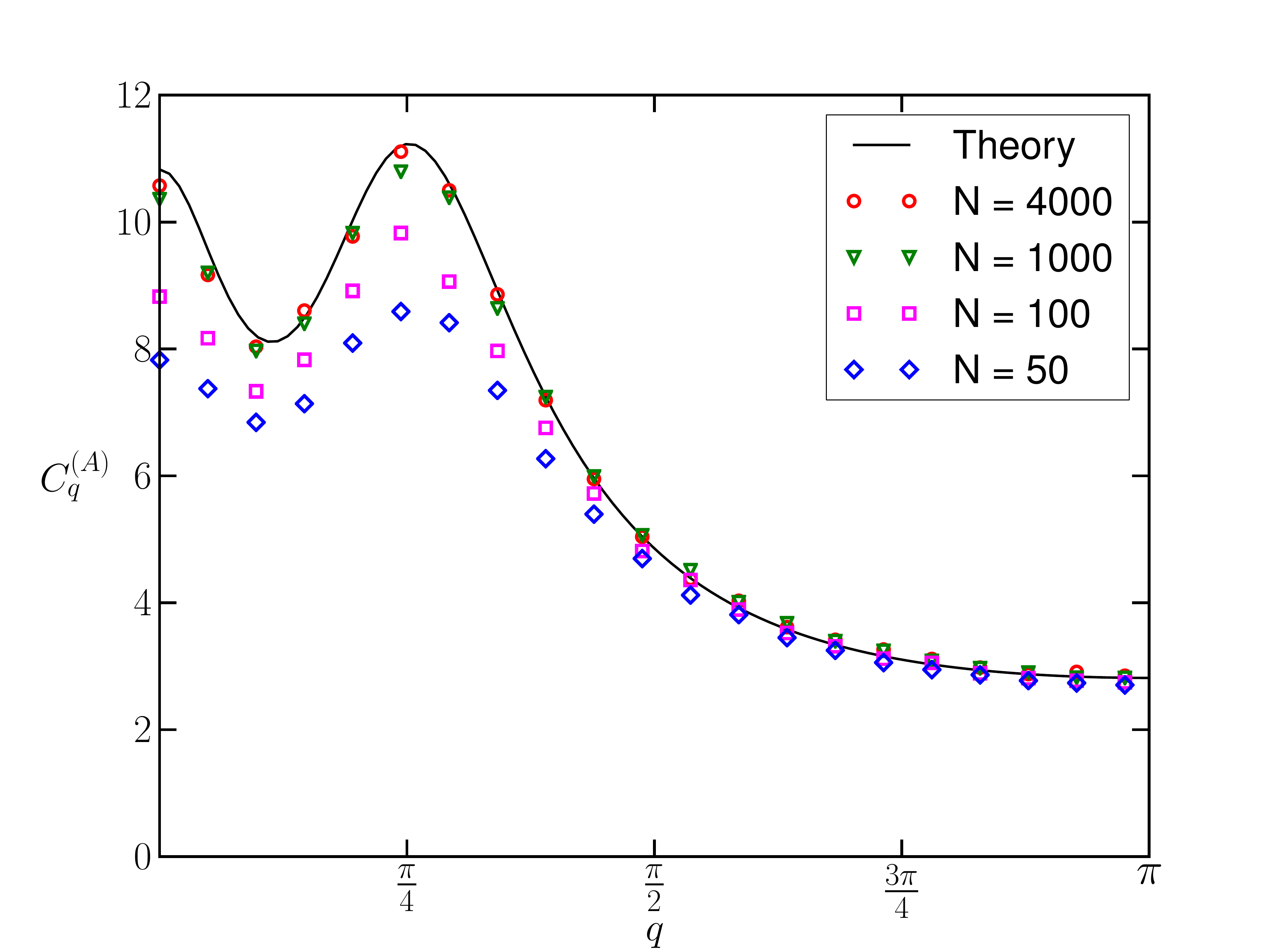}
	\caption{The dependence of the power spectrum of fluctuations on the system size $N$. The data for $N=4000$ is the same as that shown in Fig. 1a of the main paper. The agreement with the theory improves as $N$ is increased and the higher-order terms in the system-size expansion become more negligible. That being said, the relative size of the peak in the spectrum (compared to the value at $q=0$) increases as $N$ is reduced, indicating that the region of parameter space where noise-driven patterns are present would become larger as $N$ were made smaller. }
	\label{fig:ndependence}
\end{figure}

\begin{figure}[h!!]
	\centering
	\includegraphics[scale = 0.2]{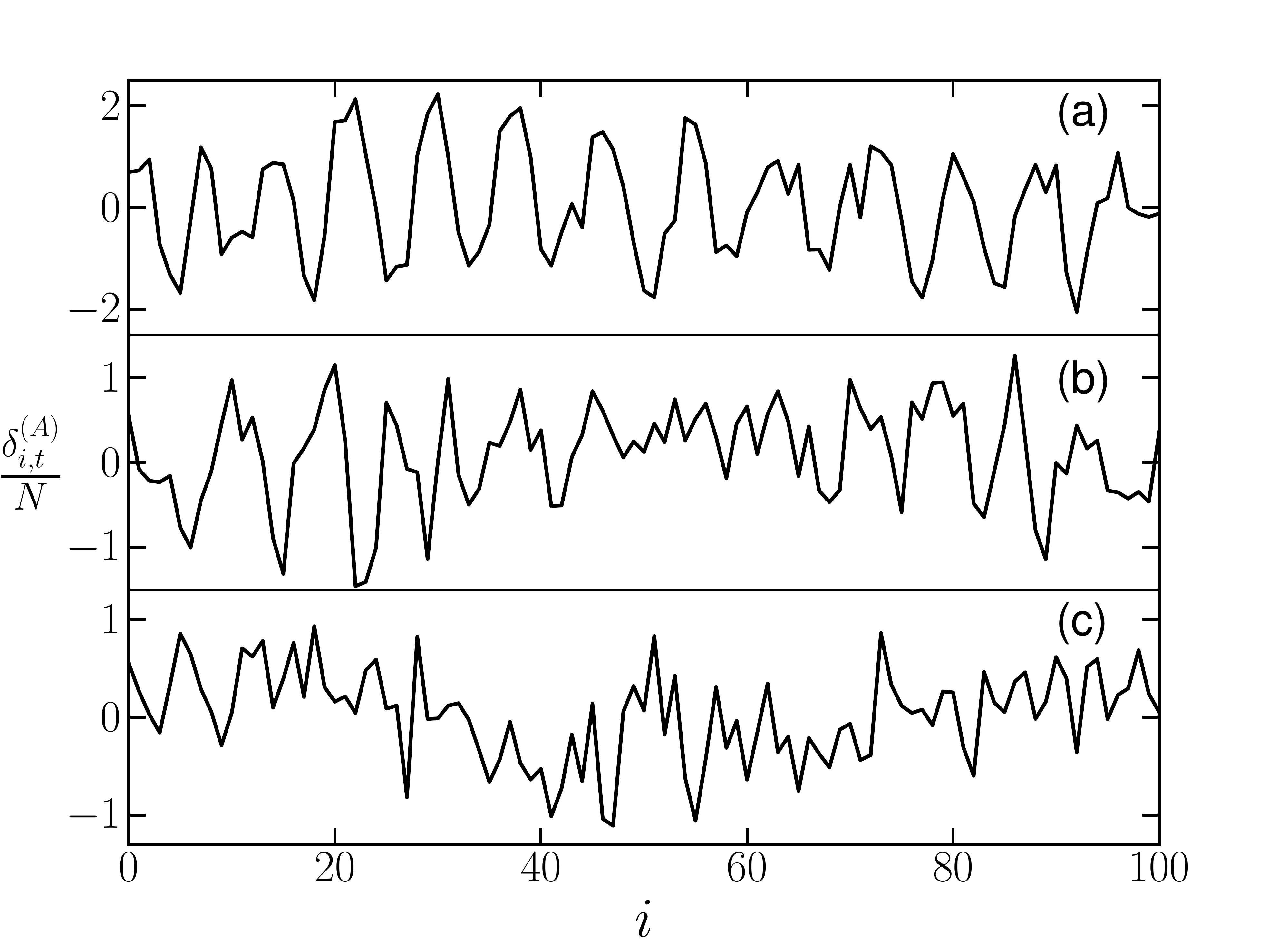}
	\caption{ A version of Fig. 3 of the main text with $N = 200$. This figure demonstrates that noise-driven patterns persist for smaller $N$ [panel (b)] and that their nature is very similar to the case of larger $N$. One notes that the patterns in the phase where the deterministic system is pattern-forming [panel (a)] appear somewhat more noisy here than for larger $N$.   }
	\label{fig:n200}
\end{figure}

For reference, we also include examples of how the deterministicly-driven and noise-driven patterns manifest in a system which has fewer particles. Fig. \ref{fig:n200} demonstrates that noise plays a more prominent role for smaller $N$, resulting in the deterministicly-driven patterns being far less orderly and uniform than for the large-$N$ case presented in the main text. The noise-driven patterns however are slightly less affected than the deterministicly-driven patterns. This could be attributed to the fact that the peak of the power spectrum is slightly more prominent for low $N$, as discussed above.

\begin{figure}[h!!]
\includegraphics[scale = 0.18]{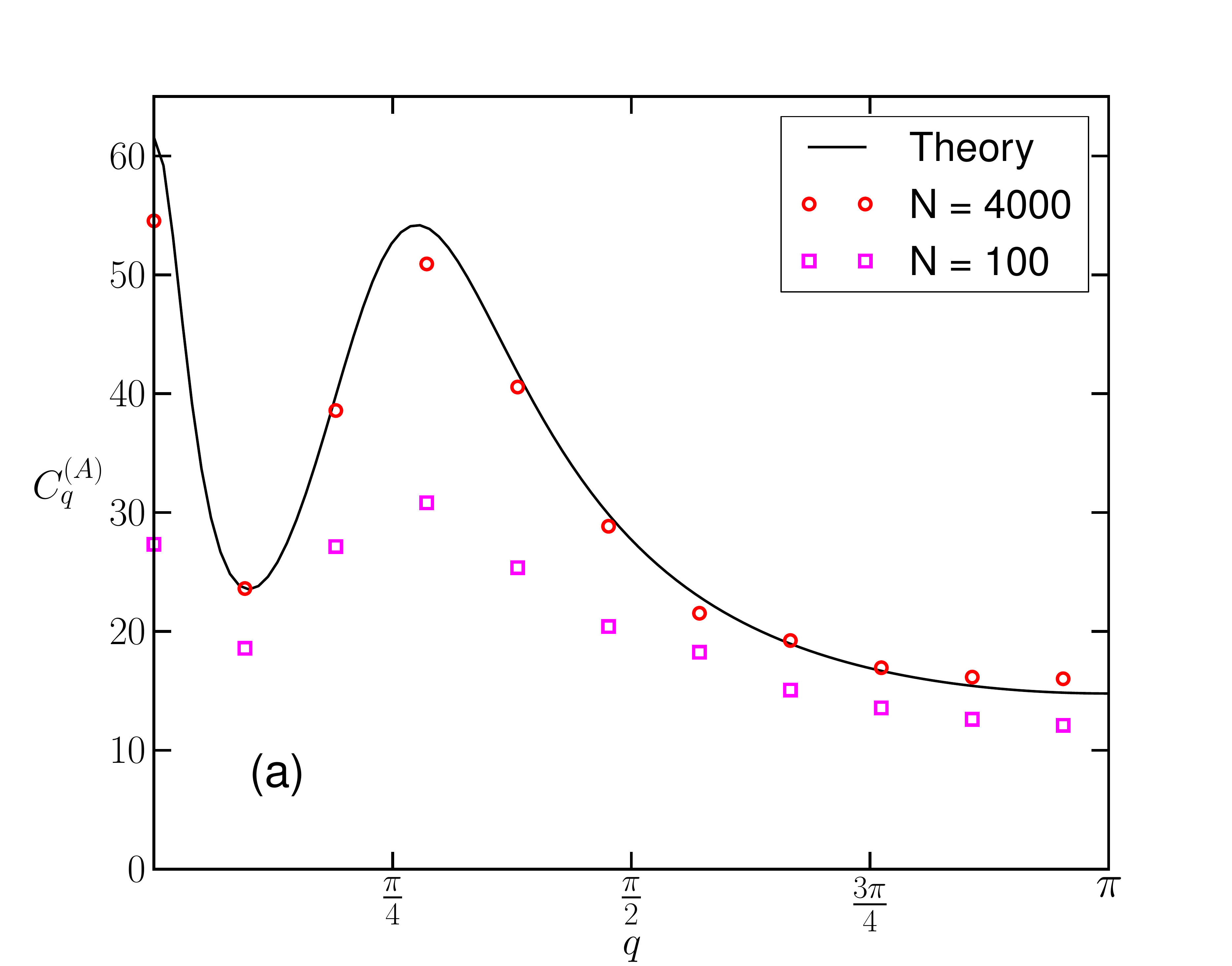}\hspace{2cm}
\includegraphics[scale = 0.18]{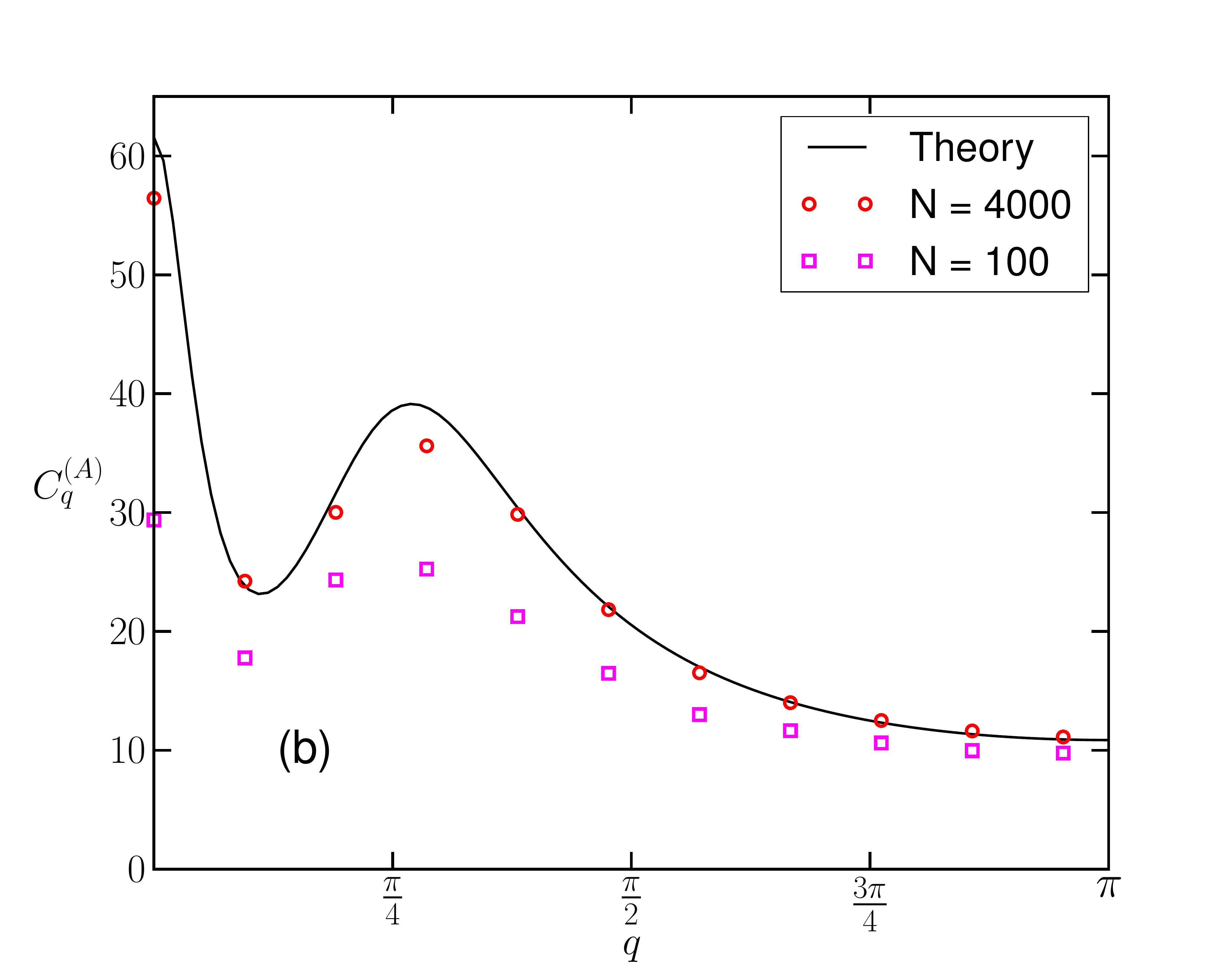}
 \caption{Example power spectra in the Lengyel-Epstein model with a subdiffusing activator, for the same ratio of effective diffusion coefficients $\theta = 13.7$ but differing $\gamma$. Here, the model parameters are $a = 5$, $b = 0.7$, $c = 1$, $d=1$. In (a), $\gamma = 0.5$ and in (b), $\gamma = 0.7$. One notes that noise-driven patterns are present only for one data set: $N=100$ and $\gamma = 0.5$ [the lower set of points in (a)]. The above figures demonstrate that the critical value of $\theta$ for the onset of noise-driven patterns still deceases as $\gamma$ deceases even when $N$ is small enough for our theory to be inaccurate. They also demonstrate that noise-driven patterns form for lower values of $\theta$ when $N$ is low enough for the higher-order terms in the system-size expansion to be relevant. Here, the results are averaged over 1000 trials. }
	\label{fig:lowN}
\end{figure}

In Fig. \ref{fig:lowN}, we show the power spectra for the Lengyel-Epstein model in four cases: $N=4000$, $\gamma = 0.5$; $N = 100$, $\gamma = 0.5$; $N=4000$, $\gamma = 0.7$; $N = 100$, $\gamma = 0.7$. All other system parameters, including the ratio of effective diffusion coefficients $\theta$, are held constant. We find that the peak in the power spectrum rises (relative to the value at $q = 0$) as $N$ is reduced (regardless of the value of $\gamma$) and as $\gamma$ is reduced (regardless of the value of $N$). This is a common feature for other sets of parameters. This verifies that the threshold value of $\theta$ required for stochastic pattern formation still reduces as the activator is made more subdiffusive for smaller values of $N$, as is the case in Fig. 4(a) of the main text. Furthermore, the critical value of $\theta$ actually reduces as $N$ is made smaller, meaning that the region of parameter space for which stochastic patterns occur actually appears to be larger for smaller $N$.

\begin{thebibliography}{65}%
\makeatletter
\providecommand \@ifxundefined [1]{%
 \@ifx{#1\undefined}
}%
\providecommand \@ifnum [1]{%
 \ifnum #1\expandafter \@firstoftwo
 \else \expandafter \@secondoftwo
 \fi
}%
\providecommand \@ifx [1]{%
 \ifx #1\expandafter \@firstoftwo
 \else \expandafter \@secondoftwo
 \fi
}%
\providecommand \natexlab [1]{#1}%
\providecommand \enquote  [1]{``#1''}%
\providecommand \bibnamefont  [1]{#1}%
\providecommand \bibfnamefont [1]{#1}%
\providecommand \citenamefont [1]{#1}%
\providecommand \href@noop [0]{\@secondoftwo}%
\providecommand \href [0]{\begingroup \@sanitize@url \@href}%
\providecommand \@href[1]{\@@startlink{#1}\@@href}%
\providecommand \@@href[1]{\endgroup#1\@@endlink}%
\providecommand \@sanitize@url [0]{\catcode `\\12\catcode `\$12\catcode
  `\&12\catcode `\#12\catcode `\^12\catcode `\_12\catcode `\%12\relax}%
\providecommand \@@startlink[1]{}%
\providecommand \@@endlink[0]{}%
\providecommand \url  [0]{\begingroup\@sanitize@url \@url }%
\providecommand \@url [1]{\endgroup\@href {#1}{\urlprefix }}%
\providecommand \urlprefix  [0]{URL }%
\providecommand \Eprint [0]{\href }%
\providecommand \doibase [0]{http://dx.doi.org/}%
\providecommand \selectlanguage [0]{\@gobble}%
\providecommand \bibinfo  [0]{\@secondoftwo}%
\providecommand \bibfield  [0]{\@secondoftwo}%
\providecommand \translation [1]{[#1]}%
\providecommand \BibitemOpen [0]{}%
\providecommand \bibitemStop [0]{}%
\providecommand \bibitemNoStop [0]{.\EOS\space}%
\providecommand \EOS [0]{\spacefactor3000\relax}%
\providecommand \BibitemShut  [1]{\csname bibitem#1\endcsname}%
\let\auto@bib@innerbib\@empty
\bibitem [{\citenamefont {Gierer}\ and\ \citenamefont
  {Meinhardt}(1972)}]{gierer}%
  \BibitemOpen
  \bibfield  {author} {\bibinfo {author} {\bibfnamefont {A.}~\bibnamefont
  {Gierer}}\ and\ \bibinfo {author} {\bibfnamefont {H.}~\bibnamefont
  {Meinhardt}},\ }\href@noop {} {\bibfield  {journal} {\bibinfo  {journal}
  {Kybernetik}\ }\textbf {\bibinfo {volume} {12}},\ \bibinfo {pages} {30}
  (\bibinfo {year} {1972})}\BibitemShut {NoStop}%
\bibitem [{\citenamefont {Murray}(1981{\natexlab{a}})}]{murray1}%
  \BibitemOpen
  \bibfield  {author} {\bibinfo {author} {\bibfnamefont {J.}~\bibnamefont
  {Murray}},\ }\href@noop {} {\bibfield  {journal} {\bibinfo  {journal} {Phil.
  Trans. R. Soc. Lond. B}\ }\textbf {\bibinfo {volume} {295}},\ \bibinfo
  {pages} {473} (\bibinfo {year} {1981}{\natexlab{a}})}\BibitemShut {NoStop}%
\bibitem [{\citenamefont {Murray}(1981{\natexlab{b}})}]{murray2}%
  \BibitemOpen
  \bibfield  {author} {\bibinfo {author} {\bibfnamefont {J.}~\bibnamefont
  {Murray}},\ }\href@noop {} {\bibfield  {journal} {\bibinfo  {journal} {J.
  Theor. Biol.}\ }\textbf {\bibinfo {volume} {88}},\ \bibinfo {pages} {161 }
  (\bibinfo {year} {1981}{\natexlab{b}})}\BibitemShut {NoStop}%
\bibitem [{\citenamefont {Petrov}\ \emph {et~al.}(1997)\citenamefont {Petrov},
  \citenamefont {Ouyang},\ and\ \citenamefont {Swinney}}]{bzreaction}%
  \BibitemOpen
  \bibfield  {author} {\bibinfo {author} {\bibfnamefont {V.}~\bibnamefont
  {Petrov}}, \bibinfo {author} {\bibfnamefont {Q.}~\bibnamefont {Ouyang}}, \
  and\ \bibinfo {author} {\bibfnamefont {H.~L.}\ \bibnamefont {Swinney}},\
  }\href@noop {} {\bibfield  {journal} {\bibinfo  {journal} {Nature}\ }\textbf
  {\bibinfo {volume} {388}},\ \bibinfo {pages} {655} (\bibinfo {year}
  {1997})}\BibitemShut {NoStop}%
\bibitem [{\citenamefont {Falcke}\ and\ \citenamefont
  {Levine}(1998)}]{slimemold}%
  \BibitemOpen
  \bibfield  {author} {\bibinfo {author} {\bibfnamefont {M.}~\bibnamefont
  {Falcke}}\ and\ \bibinfo {author} {\bibfnamefont {H.}~\bibnamefont
  {Levine}},\ }\href@noop {} {\bibfield  {journal} {\bibinfo  {journal} {Phys.
  Rev. Lett.}\ }\textbf {\bibinfo {volume} {80}},\ \bibinfo {pages} {3875}
  (\bibinfo {year} {1998})}\BibitemShut {NoStop}%
\bibitem [{\citenamefont {Van~de Koppel}\ \emph {et~al.}(2008)\citenamefont
  {Van~de Koppel}, \citenamefont {Gascoigne}, \citenamefont {Theraulaz},
  \citenamefont {Rietkerk}, \citenamefont {Mooij},\ and\ \citenamefont
  {Herman}}]{vandekoppel}%
  \BibitemOpen
  \bibfield  {author} {\bibinfo {author} {\bibfnamefont {J.}~\bibnamefont
  {Van~de Koppel}}, \bibinfo {author} {\bibfnamefont {J.~C.}\ \bibnamefont
  {Gascoigne}}, \bibinfo {author} {\bibfnamefont {G.}~\bibnamefont
  {Theraulaz}}, \bibinfo {author} {\bibfnamefont {M.}~\bibnamefont {Rietkerk}},
  \bibinfo {author} {\bibfnamefont {W.~M.}\ \bibnamefont {Mooij}}, \ and\
  \bibinfo {author} {\bibfnamefont {P.~M.}\ \bibnamefont {Herman}},\
  }\href@noop {} {\bibfield  {journal} {\bibinfo  {journal} {Science}\ }\textbf
  {\bibinfo {volume} {322}},\ \bibinfo {pages} {739} (\bibinfo {year}
  {2008})}\BibitemShut {NoStop}%
\bibitem [{\citenamefont {Rayleigh}(1916)}]{rayleigh}%
  \BibitemOpen
  \bibfield  {author} {\bibinfo {author} {\bibfnamefont {J.}~\bibnamefont
  {Rayleigh}},\ }\href@noop {} {\bibfield  {journal} {\bibinfo  {journal}
  {Philos. Mag.}\ }\textbf {\bibinfo {volume} {32}},\ \bibinfo {pages} {529}
  (\bibinfo {year} {1916})}\BibitemShut {NoStop}%
\bibitem [{\citenamefont {Turing}(1952)}]{turingchemical}%
  \BibitemOpen
  \bibfield  {author} {\bibinfo {author} {\bibfnamefont {A.~M.}\ \bibnamefont
  {Turing}},\ }\href@noop {} {\bibfield  {journal} {\bibinfo  {journal}
  {Philos. Trans. Royal Soc. B}\ }\textbf {\bibinfo {volume} {237}},\ \bibinfo
  {pages} {37} (\bibinfo {year} {1952})}\BibitemShut {NoStop}%
\bibitem [{\citenamefont {Murray}(2004)}]{murraybook}%
  \BibitemOpen
  \bibfield  {author} {\bibinfo {author} {\bibfnamefont {J.}~\bibnamefont
  {Murray}},\ }\href@noop {} {\emph {\bibinfo {title} {Mathematical
  Biology}}},\ \bibinfo {edition} {3rd}\ ed.\ (\bibinfo  {publisher}
  {Springer},\ \bibinfo {address} {New York},\ \bibinfo {year}
  {2004})\BibitemShut {NoStop}%
\bibitem [{\citenamefont {Butler}\ and\ \citenamefont
  {Goldenfeld}(2009)}]{butlergoldenfeld1}%
  \BibitemOpen
  \bibfield  {author} {\bibinfo {author} {\bibfnamefont {T.}~\bibnamefont
  {Butler}}\ and\ \bibinfo {author} {\bibfnamefont {N.}~\bibnamefont
  {Goldenfeld}},\ }\href@noop {} {\bibfield  {journal} {\bibinfo  {journal}
  {Phys. Rev. E}\ }\textbf {\bibinfo {volume} {80}},\ \bibinfo {pages} {030902}
  (\bibinfo {year} {2009})}\BibitemShut {NoStop}%
\bibitem [{\citenamefont {Butler}\ and\ \citenamefont
  {Goldenfeld}(2011)}]{butlergoldenfeld2}%
  \BibitemOpen
  \bibfield  {author} {\bibinfo {author} {\bibfnamefont {T.}~\bibnamefont
  {Butler}}\ and\ \bibinfo {author} {\bibfnamefont {N.}~\bibnamefont
  {Goldenfeld}},\ }\href@noop {} {\bibfield  {journal} {\bibinfo  {journal}
  {Phys. Rev. E}\ }\textbf {\bibinfo {volume} {84}},\ \bibinfo {pages} {011112}
  (\bibinfo {year} {2011})}\BibitemShut {NoStop}%
\bibitem [{\citenamefont {Biancalani}\ \emph {et~al.}(2017)\citenamefont
  {Biancalani}, \citenamefont {Jafarpour},\ and\ \citenamefont
  {Goldenfeld}}]{biancalani2}%
  \BibitemOpen
  \bibfield  {author} {\bibinfo {author} {\bibfnamefont {T.}~\bibnamefont
  {Biancalani}}, \bibinfo {author} {\bibfnamefont {F.}~\bibnamefont
  {Jafarpour}}, \ and\ \bibinfo {author} {\bibfnamefont {N.}~\bibnamefont
  {Goldenfeld}},\ }\href@noop {} {\bibfield  {journal} {\bibinfo  {journal}
  {Phys. Rev. Lett.}\ }\textbf {\bibinfo {volume} {118}},\ \bibinfo {pages}
  {018101} (\bibinfo {year} {2017})}\BibitemShut {NoStop}%
\bibitem [{\citenamefont {Biancalani}\ \emph {et~al.}(2010)\citenamefont
  {Biancalani}, \citenamefont {Fanelli},\ and\ \citenamefont
  {Di~Patti}}]{biancalani}%
  \BibitemOpen
  \bibfield  {author} {\bibinfo {author} {\bibfnamefont {T.}~\bibnamefont
  {Biancalani}}, \bibinfo {author} {\bibfnamefont {D.}~\bibnamefont {Fanelli}},
  \ and\ \bibinfo {author} {\bibfnamefont {F.}~\bibnamefont {Di~Patti}},\
  }\href@noop {} {\bibfield  {journal} {\bibinfo  {journal} {Phys. Rev. E}\
  }\textbf {\bibinfo {volume} {81}},\ \bibinfo {pages} {046215} (\bibinfo
  {year} {2010})}\BibitemShut {NoStop}%
\bibitem [{\citenamefont {Karig}\ \emph {et~al.}(2018)\citenamefont {Karig},
  \citenamefont {Martini}, \citenamefont {Lu}, \citenamefont {DeLateur},
  \citenamefont {Goldenfeld},\ and\ \citenamefont {Weiss}}]{stochpattexp}%
  \BibitemOpen
  \bibfield  {author} {\bibinfo {author} {\bibfnamefont {D.}~\bibnamefont
  {Karig}}, \bibinfo {author} {\bibfnamefont {K.~M.}\ \bibnamefont {Martini}},
  \bibinfo {author} {\bibfnamefont {T.}~\bibnamefont {Lu}}, \bibinfo {author}
  {\bibfnamefont {N.~A.}\ \bibnamefont {DeLateur}}, \bibinfo {author}
  {\bibfnamefont {N.}~\bibnamefont {Goldenfeld}}, \ and\ \bibinfo {author}
  {\bibfnamefont {R.}~\bibnamefont {Weiss}},\ }\href@noop {} {\bibfield
  {journal} {\bibinfo  {journal} {Proc. Natl. Acad. Sci.}\ }\textbf {\bibinfo
  {volume} {115}},\ \bibinfo {pages} {6572} (\bibinfo {year}
  {2018})}\BibitemShut {NoStop}%
\bibitem [{\citenamefont {Di~Patti}\ \emph {et~al.}(2018)\citenamefont
  {Di~Patti}, \citenamefont {Lavacchi}, \citenamefont {Arbel-Goren},
  \citenamefont {Schein-Lubomirsky}, \citenamefont {Fanelli},\ and\
  \citenamefont {Stavans}}]{dipatti}%
  \BibitemOpen
  \bibfield  {author} {\bibinfo {author} {\bibfnamefont {F.}~\bibnamefont
  {Di~Patti}}, \bibinfo {author} {\bibfnamefont {L.}~\bibnamefont {Lavacchi}},
  \bibinfo {author} {\bibfnamefont {R.}~\bibnamefont {Arbel-Goren}}, \bibinfo
  {author} {\bibfnamefont {L.}~\bibnamefont {Schein-Lubomirsky}}, \bibinfo
  {author} {\bibfnamefont {D.}~\bibnamefont {Fanelli}}, \ and\ \bibinfo
  {author} {\bibfnamefont {J.}~\bibnamefont {Stavans}},\ }\href@noop {}
  {\bibfield  {journal} {\bibinfo  {journal} {PLOS Biol.}\ }\textbf {\bibinfo
  {volume} {16}},\ \bibinfo {pages} {1} (\bibinfo {year} {2018})}\BibitemShut
  {NoStop}%
\bibitem [{\citenamefont {Metzler}\ and\ \citenamefont
  {Klafter}(2000)}]{randomwalkguide}%
  \BibitemOpen
  \bibfield  {author} {\bibinfo {author} {\bibfnamefont {R.}~\bibnamefont
  {Metzler}}\ and\ \bibinfo {author} {\bibfnamefont {J.}~\bibnamefont
  {Klafter}},\ }\href@noop {} {\bibfield  {journal} {\bibinfo  {journal} {Phys.
  Rep.}\ }\textbf {\bibinfo {volume} {339}},\ \bibinfo {pages} {1} (\bibinfo
  {year} {2000})}\BibitemShut {NoStop}%
\bibitem [{\citenamefont {Metzler}\ and\ \citenamefont
  {Klafter}(2004)}]{metzlerklafter2}%
  \BibitemOpen
  \bibfield  {author} {\bibinfo {author} {\bibfnamefont {R.}~\bibnamefont
  {Metzler}}\ and\ \bibinfo {author} {\bibfnamefont {J.}~\bibnamefont
  {Klafter}},\ }\href@noop {} {\bibfield  {journal} {\bibinfo  {journal} {J.
  Phys. A}\ }\textbf {\bibinfo {volume} {37}},\ \bibinfo {pages} {R161}
  (\bibinfo {year} {2004})}\BibitemShut {NoStop}%
\bibitem [{\citenamefont {Yadav}\ \emph {et~al.}(2008)\citenamefont {Yadav},
  \citenamefont {Milu},\ and\ \citenamefont {Horsthemke}}]{yadavmilu}%
  \BibitemOpen
  \bibfield  {author} {\bibinfo {author} {\bibfnamefont {A.}~\bibnamefont
  {Yadav}}, \bibinfo {author} {\bibfnamefont {S.~M.}\ \bibnamefont {Milu}}, \
  and\ \bibinfo {author} {\bibfnamefont {W.}~\bibnamefont {Horsthemke}},\
  }\href@noop {} {\bibfield  {journal} {\bibinfo  {journal} {Phys. Rev. E}\
  }\textbf {\bibinfo {volume} {78}},\ \bibinfo {pages} {026116} (\bibinfo
  {year} {2008})}\BibitemShut {NoStop}%
\bibitem [{\citenamefont {Baron}\ and\ \citenamefont
  {Galla}(2019)}]{barongalla}%
  \BibitemOpen
  \bibfield  {author} {\bibinfo {author} {\bibfnamefont {J.~W.}\ \bibnamefont
  {Baron}}\ and\ \bibinfo {author} {\bibfnamefont {T.}~\bibnamefont {Galla}},\
  }\href@noop {} {\bibfield  {journal} {\bibinfo  {journal} {Phys. Rev. E}\
  }\textbf {\bibinfo {volume} {99}},\ \bibinfo {pages} {012212} (\bibinfo
  {year} {2019})}\BibitemShut {NoStop}%
\bibitem [{\citenamefont {Ghosh}\ and\ \citenamefont {Webb}(1994)}]{ghoshwebb}%
  \BibitemOpen
  \bibfield  {author} {\bibinfo {author} {\bibfnamefont {R.}~\bibnamefont
  {Ghosh}}\ and\ \bibinfo {author} {\bibfnamefont {W.}~\bibnamefont {Webb}},\
  }\href@noop {} {\bibfield  {journal} {\bibinfo  {journal} {Biophys. J.}\
  }\textbf {\bibinfo {volume} {66}},\ \bibinfo {pages} {1301 } (\bibinfo {year}
  {1994})}\BibitemShut {NoStop}%
\bibitem [{\citenamefont {Schwille}\ \emph {et~al.}(1999)\citenamefont
  {Schwille}, \citenamefont {Korlach},\ and\ \citenamefont {Webb}}]{schwille}%
  \BibitemOpen
  \bibfield  {author} {\bibinfo {author} {\bibfnamefont {P.}~\bibnamefont
  {Schwille}}, \bibinfo {author} {\bibfnamefont {J.}~\bibnamefont {Korlach}}, \
  and\ \bibinfo {author} {\bibfnamefont {W.~W.}\ \bibnamefont {Webb}},\
  }\href@noop {} {\bibfield  {journal} {\bibinfo  {journal} {Cytometry}\
  }\textbf {\bibinfo {volume} {36}},\ \bibinfo {pages} {176} (\bibinfo {year}
  {1999})}\BibitemShut {NoStop}%
\bibitem [{\citenamefont {Saxton}(1994)}]{saxton1994}%
  \BibitemOpen
  \bibfield  {author} {\bibinfo {author} {\bibfnamefont {M.~J.}\ \bibnamefont
  {Saxton}},\ }\href@noop {} {\bibfield  {journal} {\bibinfo  {journal}
  {Biophys. J.}\ }\textbf {\bibinfo {volume} {67}},\ \bibinfo {pages} {2110}
  (\bibinfo {year} {1994})}\BibitemShut {NoStop}%
\bibitem [{\citenamefont {Saxton}(1996)}]{saxton1996}%
  \BibitemOpen
  \bibfield  {author} {\bibinfo {author} {\bibfnamefont {M.~J.}\ \bibnamefont
  {Saxton}},\ }\href@noop {} {\bibfield  {journal} {\bibinfo  {journal}
  {Biophys. J.}\ }\textbf {\bibinfo {volume} {70}},\ \bibinfo {pages} {1250}
  (\bibinfo {year} {1996})}\BibitemShut {NoStop}%
\bibitem [{\citenamefont {Hornung}\ \emph {et~al.}(2005)\citenamefont
  {Hornung}, \citenamefont {Berkowitz},\ and\ \citenamefont
  {Barkai}}]{hornung}%
  \BibitemOpen
  \bibfield  {author} {\bibinfo {author} {\bibfnamefont {G.}~\bibnamefont
  {Hornung}}, \bibinfo {author} {\bibfnamefont {B.}~\bibnamefont {Berkowitz}},
  \ and\ \bibinfo {author} {\bibfnamefont {N.}~\bibnamefont {Barkai}},\
  }\href@noop {} {\bibfield  {journal} {\bibinfo  {journal} {Phys. Rev. E}\
  }\textbf {\bibinfo {volume} {72}},\ \bibinfo {pages} {041916} (\bibinfo
  {year} {2005})}\BibitemShut {NoStop}%
\bibitem [{\citenamefont {Yuste}\ \emph {et~al.}(2010)\citenamefont {Yuste},
  \citenamefont {Abad},\ and\ \citenamefont {Lindenberg}}]{yustabadlindenberg}%
  \BibitemOpen
  \bibfield  {author} {\bibinfo {author} {\bibfnamefont {S.~B.}\ \bibnamefont
  {Yuste}}, \bibinfo {author} {\bibfnamefont {E.}~\bibnamefont {Abad}}, \ and\
  \bibinfo {author} {\bibfnamefont {K.}~\bibnamefont {Lindenberg}},\
  }\href@noop {} {\bibfield  {journal} {\bibinfo  {journal} {Phys. Rev. E}\
  }\textbf {\bibinfo {volume} {82}} (\bibinfo {year} {2010})}\BibitemShut
  {NoStop}%
\bibitem [{\citenamefont {Fedotov}\ and\ \citenamefont
  {Falconer}(2014)}]{fedotovfalconer2}%
  \BibitemOpen
  \bibfield  {author} {\bibinfo {author} {\bibfnamefont {S.}~\bibnamefont
  {Fedotov}}\ and\ \bibinfo {author} {\bibfnamefont {S.}~\bibnamefont
  {Falconer}},\ }\href@noop {} {\bibfield  {journal} {\bibinfo  {journal}
  {Phys. Rev. E}\ }\textbf {\bibinfo {volume} {89}},\ \bibinfo {pages} {012107}
  (\bibinfo {year} {2014})}\BibitemShut {NoStop}%
\bibitem [{\citenamefont {Fedotov}\ and\ \citenamefont
  {Falconer}(2013)}]{fedotovfalconer3}%
  \BibitemOpen
  \bibfield  {author} {\bibinfo {author} {\bibfnamefont {S.}~\bibnamefont
  {Fedotov}}\ and\ \bibinfo {author} {\bibfnamefont {S.}~\bibnamefont
  {Falconer}},\ }\href@noop {} {\bibfield  {journal} {\bibinfo  {journal}
  {Phys. Rev. E}\ }\textbf {\bibinfo {volume} {87}},\ \bibinfo {pages} {052139}
  (\bibinfo {year} {2013})}\BibitemShut {NoStop}%
\bibitem [{\citenamefont {Glansdorff}\ and\ \citenamefont
  {Prigogine}(1971)}]{brusselator}%
  \BibitemOpen
  \bibfield  {author} {\bibinfo {author} {\bibfnamefont {P.}~\bibnamefont
  {Glansdorff}}\ and\ \bibinfo {author} {\bibfnamefont {I.}~\bibnamefont
  {Prigogine}},\ }\href@noop {} {\emph {\bibinfo {title} {Thermodynamic theory
  of structure, stability and fluctuations}}}\ (\bibinfo  {publisher}
  {Wiley-Interscience},\ \bibinfo {address} {New York},\ \bibinfo {year}
  {1971})\BibitemShut {NoStop}%
\bibitem [{\citenamefont {Lengyel}\ \emph {et~al.}(1990)\citenamefont
  {Lengyel}, \citenamefont {Rabai},\ and\ \citenamefont
  {Epstein}}]{lengyelepstein1}%
  \BibitemOpen
  \bibfield  {author} {\bibinfo {author} {\bibfnamefont {I.}~\bibnamefont
  {Lengyel}}, \bibinfo {author} {\bibfnamefont {G.}~\bibnamefont {Rabai}}, \
  and\ \bibinfo {author} {\bibfnamefont {I.~R.}\ \bibnamefont {Epstein}},\
  }\href@noop {} {\bibfield  {journal} {\bibinfo  {journal} {J. Am. Chem.
  Soc.}\ }\textbf {\bibinfo {volume} {112}},\ \bibinfo {pages} {9104} (\bibinfo
  {year} {1990})}\BibitemShut {NoStop}%
\bibitem [{\citenamefont {Lengyel}\ and\ \citenamefont
  {Epstein}(1991)}]{lengyelepstein2}%
  \BibitemOpen
  \bibfield  {author} {\bibinfo {author} {\bibfnamefont {I.}~\bibnamefont
  {Lengyel}}\ and\ \bibinfo {author} {\bibfnamefont {I.~R.}\ \bibnamefont
  {Epstein}},\ }\href@noop {} {\bibfield  {journal} {\bibinfo  {journal}
  {Science}\ }\textbf {\bibinfo {volume} {251}},\ \bibinfo {pages} {650}
  (\bibinfo {year} {1991})}\BibitemShut {NoStop}%
\bibitem [{\citenamefont {McNaught}\ and\ \citenamefont
  {Wilkinson}(1997)}]{chemicalterminology}%
  \BibitemOpen
  \bibfield  {author} {\bibinfo {author} {\bibfnamefont {A.}~\bibnamefont
  {McNaught}}\ and\ \bibinfo {author} {\bibfnamefont {A.}~\bibnamefont
  {Wilkinson}},\ }\href@noop {} {\emph {\bibinfo {title} {Compendium of
  chemical terminology}}},\ Vol.\ \bibinfo {volume} {1669}\ (\bibinfo
  {publisher} {Blackwell Science},\ \bibinfo {address} {Oxford},\ \bibinfo
  {year} {1997})\BibitemShut {NoStop}%
\bibitem [{\citenamefont {Mainardi}\ and\ \citenamefont
  {Gorenflo}(2000)}]{gorenflo}%
  \BibitemOpen
  \bibfield  {author} {\bibinfo {author} {\bibfnamefont {F.}~\bibnamefont
  {Mainardi}}\ and\ \bibinfo {author} {\bibfnamefont {R.}~\bibnamefont
  {Gorenflo}},\ }\href@noop {} {\bibfield  {journal} {\bibinfo  {journal} {J.
  Comput. Appl. Math.}\ }\textbf {\bibinfo {volume} {118}},\ \bibinfo {pages}
  {283 } (\bibinfo {year} {2000})}\BibitemShut {NoStop}%
\bibitem [{\citenamefont {{Michael F. Shlesinger }}(2017)}]{montrollweiss}%
  \BibitemOpen
  \bibfield  {author} {\bibinfo {author} {\bibnamefont {{Michael F. Shlesinger
  }}},\ }\href@noop {} {\bibfield  {journal} {\bibinfo  {journal} {Eur. Phys.
  J. B}\ }\textbf {\bibinfo {volume} {90}},\ \bibinfo {pages} {93} (\bibinfo
  {year} {2017})}\BibitemShut {NoStop}%
\bibitem [{\citenamefont {Montroll}\ and\ \citenamefont
  {Weiss}(1965)}]{montrollweissoriginal}%
  \BibitemOpen
  \bibfield  {author} {\bibinfo {author} {\bibfnamefont {E.~W.}\ \bibnamefont
  {Montroll}}\ and\ \bibinfo {author} {\bibfnamefont {G.~H.}\ \bibnamefont
  {Weiss}},\ }\href@noop {} {\bibfield  {journal} {\bibinfo  {journal} {J.
  Math. Phys.}\ }\textbf {\bibinfo {volume} {6}},\ \bibinfo {pages} {167}
  (\bibinfo {year} {1965})}\BibitemShut {NoStop}%
\bibitem [{\citenamefont {Klages}\ \emph {et~al.}(2008)\citenamefont {Klages},
  \citenamefont {Radons},\ and\ \citenamefont {Sokolov}}]{anomaloustransport}%
  \BibitemOpen
  \bibfield  {author} {\bibinfo {author} {\bibfnamefont {R.}~\bibnamefont
  {Klages}}, \bibinfo {author} {\bibfnamefont {G.}~\bibnamefont {Radons}}, \
  and\ \bibinfo {author} {\bibfnamefont {I.~M.}\ \bibnamefont {Sokolov}},\
  }\href@noop {} {\emph {\bibinfo {title} {Anomalous transport: foundations and
  applications}}}\ (\bibinfo  {publisher} {Wiley-VCH},\ \bibinfo {address}
  {Weinheim},\ \bibinfo {year} {2008})\BibitemShut {NoStop}%
\bibitem [{\citenamefont {Mendez}\ \emph {et~al.}(2010)\citenamefont {Mendez},
  \citenamefont {Fedotov},\ and\ \citenamefont {Horsthemke}}]{mendezfedotov}%
  \BibitemOpen
  \bibfield  {author} {\bibinfo {author} {\bibfnamefont {V.}~\bibnamefont
  {Mendez}}, \bibinfo {author} {\bibfnamefont {S.}~\bibnamefont {Fedotov}}, \
  and\ \bibinfo {author} {\bibfnamefont {W.}~\bibnamefont {Horsthemke}},\
  }\href@noop {} {\emph {\bibinfo {title} {Reaction-transport systems:
  mesoscopic foundations, fronts, and spatial instabilities}}}\ (\bibinfo
  {publisher} {Springer},\ \bibinfo {address} {New York},\ \bibinfo {year}
  {2010})\BibitemShut {NoStop}%
\bibitem [{\citenamefont {Tyson}(1976)}]{tysonbelousov}%
  \BibitemOpen
  \bibfield  {author} {\bibinfo {author} {\bibfnamefont {J.~J.}\ \bibnamefont
  {Tyson}},\ }\href@noop {} {\emph {\bibinfo {title} {The Belousov-Zhabotinskii
  reaction}}}\ (\bibinfo  {publisher} {Springer-Verlag},\ \bibinfo {address}
  {Berlin - Heidelberg},\ \bibinfo {year} {1976})\BibitemShut {NoStop}%
\bibitem [{\citenamefont {Castets}\ \emph {et~al.}(1990)\citenamefont
  {Castets}, \citenamefont {Dulos}, \citenamefont {Boissonade},\ and\
  \citenamefont {De~Kepper}}]{castets}%
  \BibitemOpen
  \bibfield  {author} {\bibinfo {author} {\bibfnamefont {V.}~\bibnamefont
  {Castets}}, \bibinfo {author} {\bibfnamefont {E.}~\bibnamefont {Dulos}},
  \bibinfo {author} {\bibfnamefont {J.}~\bibnamefont {Boissonade}}, \ and\
  \bibinfo {author} {\bibfnamefont {P.}~\bibnamefont {De~Kepper}},\ }\href@noop
  {} {\bibfield  {journal} {\bibinfo  {journal} {Phys. Rev. Lett.}\ }\textbf
  {\bibinfo {volume} {64}},\ \bibinfo {pages} {2953} (\bibinfo {year}
  {1990})}\BibitemShut {NoStop}%
\bibitem [{\citenamefont {{\'E}rdi}\ and\ \citenamefont
  {T{\'o}th}(1989)}]{erdi}%
  \BibitemOpen
  \bibfield  {author} {\bibinfo {author} {\bibfnamefont {P.}~\bibnamefont
  {{\'E}rdi}}\ and\ \bibinfo {author} {\bibfnamefont {J.}~\bibnamefont
  {T{\'o}th}},\ }\href@noop {} {\emph {\bibinfo {title} {Mathematical models of
  chemical reactions: theory and applications of deterministic and stochastic
  models}}}\ (\bibinfo  {publisher} {Manchester University Press},\ \bibinfo
  {address} {Manchester},\ \bibinfo {year} {1989})\BibitemShut {NoStop}%
\bibitem [{\citenamefont {Van~Kampen}(1992)}]{vankampen}%
  \BibitemOpen
  \bibfield  {author} {\bibinfo {author} {\bibfnamefont {N.~G.}\ \bibnamefont
  {Van~Kampen}},\ }\href@noop {} {\emph {\bibinfo {title} {Stochastic processes
  in physics and chemistry}}},\ Vol.~\bibinfo {volume} {1}\ (\bibinfo
  {publisher} {Elsevier},\ \bibinfo {address} {London},\ \bibinfo {year}
  {1992})\BibitemShut {NoStop}%
\bibitem [{\citenamefont {Gardiner}(2009)}]{gardiner}%
  \BibitemOpen
  \bibfield  {author} {\bibinfo {author} {\bibfnamefont {C.~W.}\ \bibnamefont
  {Gardiner}},\ }\href@noop {} {\emph {\bibinfo {title} {Handbook of stochastic
  methods for physics, chemistry and the natural sciences}}},\ \bibinfo
  {edition} {4th}\ ed.\ (\bibinfo  {publisher} {Springer},\ \bibinfo {address}
  {New York},\ \bibinfo {year} {2009})\BibitemShut {NoStop}%
\bibitem [{\citenamefont {Cox}(2017)}]{cox}%
  \BibitemOpen
  \bibfield  {author} {\bibinfo {author} {\bibfnamefont {D.~R.}\ \bibnamefont
  {Cox}},\ }\href@noop {} {\emph {\bibinfo {title} {The theory of stochastic
  processes}}}\ (\bibinfo  {publisher} {Routledge},\ \bibinfo {address}
  {London},\ \bibinfo {year} {2017})\BibitemShut {NoStop}%
\bibitem [{\citenamefont {Gillespie}(2001)}]{tauleap}%
  \BibitemOpen
  \bibfield  {author} {\bibinfo {author} {\bibfnamefont {D.~T.}\ \bibnamefont
  {Gillespie}},\ }\href@noop {} {\bibfield  {journal} {\bibinfo  {journal} {J.
  Chem. Phys.}\ }\textbf {\bibinfo {volume} {115}},\ \bibinfo {pages} {1716}
  (\bibinfo {year} {2001})}\BibitemShut {NoStop}%
\bibitem [{\citenamefont {Risken}(1996)}]{risken}%
  \BibitemOpen
  \bibfield  {author} {\bibinfo {author} {\bibfnamefont {H.}~\bibnamefont
  {Risken}},\ }\href@noop {} {\emph {\bibinfo {title} {The Fokker-Planck
  Equation}}}\ (\bibinfo  {publisher} {Springer},\ \bibinfo {address} {New
  York},\ \bibinfo {year} {1996})\BibitemShut {NoStop}%
\bibitem [{\citenamefont {McKane}\ and\ \citenamefont
  {Newman}(2005)}]{mckanenewman}%
  \BibitemOpen
  \bibfield  {author} {\bibinfo {author} {\bibfnamefont {A.~J.}\ \bibnamefont
  {McKane}}\ and\ \bibinfo {author} {\bibfnamefont {T.~J.}\ \bibnamefont
  {Newman}},\ }\href@noop {} {\bibfield  {journal} {\bibinfo  {journal} {Phys.
  Rev. Lett.}\ }\textbf {\bibinfo {volume} {94}},\ \bibinfo {pages} {218102}
  (\bibinfo {year} {2005})}\BibitemShut {NoStop}%
\bibitem [{\citenamefont {Martin}\ \emph {et~al.}(1973)\citenamefont {Martin},
  \citenamefont {Siggia},\ and\ \citenamefont {Rose}}]{MSR}%
  \BibitemOpen
  \bibfield  {author} {\bibinfo {author} {\bibfnamefont {P.~C.}\ \bibnamefont
  {Martin}}, \bibinfo {author} {\bibfnamefont {E.~D.}\ \bibnamefont {Siggia}},
  \ and\ \bibinfo {author} {\bibfnamefont {H.~A.}\ \bibnamefont {Rose}},\
  }\href@noop {} {\bibfield  {journal} {\bibinfo  {journal} {Phys. Rev. A}\
  }\textbf {\bibinfo {volume} {8}} (\bibinfo {year} {1973})}\BibitemShut
  {NoStop}%
\bibitem [{\citenamefont {Janssen}(1976)}]{J}%
  \BibitemOpen
  \bibfield  {author} {\bibinfo {author} {\bibfnamefont {H.-K.}\ \bibnamefont
  {Janssen}},\ }\href@noop {} {\bibfield  {journal} {\bibinfo  {journal} {Z.
  Phys. B}\ }\textbf {\bibinfo {volume} {23}},\ \bibinfo {pages} {377}
  (\bibinfo {year} {1976})}\BibitemShut {NoStop}%
\bibitem [{\citenamefont {De~Dominicis}(1976)}]{D}%
  \BibitemOpen
  \bibfield  {author} {\bibinfo {author} {\bibfnamefont {C.}~\bibnamefont
  {De~Dominicis}},\ }in\ \href@noop {} {\emph {\bibinfo {booktitle} {J.
  Phys.(Paris), Colloq.}}},\ Vol.~\bibinfo {volume} {37}\ (\bibinfo {year}
  {1976})\ p.~\bibinfo {pages} {C1}\BibitemShut {NoStop}%
\bibitem [{\citenamefont {Brett}\ and\ \citenamefont
  {Galla}(2013)}]{brettgalla}%
  \BibitemOpen
  \bibfield  {author} {\bibinfo {author} {\bibfnamefont {T.}~\bibnamefont
  {Brett}}\ and\ \bibinfo {author} {\bibfnamefont {T.}~\bibnamefont {Galla}},\
  }\href@noop {} {\bibfield  {journal} {\bibinfo  {journal} {Phys. Rev. Lett.}\
  }\textbf {\bibinfo {volume} {110}},\ \bibinfo {pages} {250601} (\bibinfo
  {year} {2013})}\BibitemShut {NoStop}%
\bibitem [{\citenamefont {Brett}\ and\ \citenamefont
  {Galla}(2014)}]{brettgalla2}%
  \BibitemOpen
  \bibfield  {author} {\bibinfo {author} {\bibfnamefont {T.}~\bibnamefont
  {Brett}}\ and\ \bibinfo {author} {\bibfnamefont {T.}~\bibnamefont {Galla}},\
  }\href@noop {} {\bibfield  {journal} {\bibinfo  {journal} {J. Chem. Phys.}\
  }\textbf {\bibinfo {volume} {140}},\ \bibinfo {pages} {124112} (\bibinfo
  {year} {2014})}\BibitemShut {NoStop}%
\bibitem [{\citenamefont {Altland}\ and\ \citenamefont
  {Simons}(2010)}]{altlandsimons}%
  \BibitemOpen
  \bibfield  {author} {\bibinfo {author} {\bibfnamefont {A.}~\bibnamefont
  {Altland}}\ and\ \bibinfo {author} {\bibfnamefont {B.~D.}\ \bibnamefont
  {Simons}},\ }\href@noop {} {\emph {\bibinfo {title} {Condensed matter field
  theory}}}\ (\bibinfo  {publisher} {Cambridge University Press},\ \bibinfo
  {address} {Cambridge},\ \bibinfo {year} {2010})\BibitemShut {NoStop}%
\bibitem [{\citenamefont {Vlad}\ and\ \citenamefont {Ross}(2002)}]{vladross}%
  \BibitemOpen
  \bibfield  {author} {\bibinfo {author} {\bibfnamefont {M.~O.}\ \bibnamefont
  {Vlad}}\ and\ \bibinfo {author} {\bibfnamefont {J.}~\bibnamefont {Ross}},\
  }\href@noop {} {\bibfield  {journal} {\bibinfo  {journal} {Phys. Rev. E}\
  }\textbf {\bibinfo {volume} {66}},\ \bibinfo {pages} {061908} (\bibinfo
  {year} {2002})}\BibitemShut {NoStop}%
\bibitem [{\citenamefont {Fedotov}\ and\ \citenamefont
  {Falconer}(2012)}]{fedotovfalconer1}%
  \BibitemOpen
  \bibfield  {author} {\bibinfo {author} {\bibfnamefont {S.}~\bibnamefont
  {Fedotov}}\ and\ \bibinfo {author} {\bibfnamefont {S.}~\bibnamefont
  {Falconer}},\ }\href@noop {} {\bibfield  {journal} {\bibinfo  {journal}
  {Phys. Rev. E}\ }\textbf {\bibinfo {volume} {85}},\ \bibinfo {pages} {031132}
  (\bibinfo {year} {2012})}\BibitemShut {NoStop}%
\bibitem [{\citenamefont {Halsted}\ and\ \citenamefont
  {Brown}(1972)}]{halstedbrown}%
  \BibitemOpen
  \bibfield  {author} {\bibinfo {author} {\bibfnamefont {D.}~\bibnamefont
  {Halsted}}\ and\ \bibinfo {author} {\bibfnamefont {D.}~\bibnamefont
  {Brown}},\ }\href@noop {} {\bibfield  {journal} {\bibinfo  {journal} {Chem.
  Eng. J.}\ }\textbf {\bibinfo {volume} {3}},\ \bibinfo {pages} {312 }
  (\bibinfo {year} {1972})}\BibitemShut {NoStop}%
\bibitem [{\citenamefont {Gillespie}(1976)}]{gillespie}%
  \BibitemOpen
  \bibfield  {author} {\bibinfo {author} {\bibfnamefont {D.~T.}\ \bibnamefont
  {Gillespie}},\ }\href@noop {} {\bibfield  {journal} {\bibinfo  {journal} {J.
  Comput. Phys.}\ }\textbf {\bibinfo {volume} {22}},\ \bibinfo {pages} {403 }
  (\bibinfo {year} {1976})}\BibitemShut {NoStop}%
\bibitem [{\citenamefont {Anderson}(2007)}]{anderson}%
  \BibitemOpen
  \bibfield  {author} {\bibinfo {author} {\bibfnamefont {D.~F.}\ \bibnamefont
  {Anderson}},\ }\href@noop {} {\bibfield  {journal} {\bibinfo  {journal} {J.
  Chem. Phys.}\ }\textbf {\bibinfo {volume} {127}},\ \bibinfo {pages} {214107}
  (\bibinfo {year} {2007})}\BibitemShut {NoStop}%
\bibitem [{\citenamefont {Alonso}\ \emph {et~al.}(2007)\citenamefont {Alonso},
  \citenamefont {McKane},\ and\ \citenamefont {Pascual}}]{alonsomckane}%
  \BibitemOpen
  \bibfield  {author} {\bibinfo {author} {\bibfnamefont {D.}~\bibnamefont
  {Alonso}}, \bibinfo {author} {\bibfnamefont {A.~J.}\ \bibnamefont {McKane}},
  \ and\ \bibinfo {author} {\bibfnamefont {M.}~\bibnamefont {Pascual}},\
  }\href@noop {} {\bibfield  {journal} {\bibinfo  {journal} {J. Royal Soc.
  Interface}\ }\textbf {\bibinfo {volume} {4}},\ \bibinfo {pages} {575}
  (\bibinfo {year} {2007})}\BibitemShut {NoStop}%
\bibitem [{\citenamefont {Lotka}(1909)}]{lotka}%
  \BibitemOpen
  \bibfield  {author} {\bibinfo {author} {\bibfnamefont {A.~J.}\ \bibnamefont
  {Lotka}},\ }\href@noop {} {\bibfield  {journal} {\bibinfo  {journal} {J.
  Phys. Chem.}\ }\textbf {\bibinfo {volume} {14}},\ \bibinfo {pages} {271}
  (\bibinfo {year} {1909})}\BibitemShut {NoStop}%
\bibitem [{\citenamefont {Field}\ and\ \citenamefont
  {Noyes}(1974)}]{fieldnoyes}%
  \BibitemOpen
  \bibfield  {author} {\bibinfo {author} {\bibfnamefont {R.~J.}\ \bibnamefont
  {Field}}\ and\ \bibinfo {author} {\bibfnamefont {R.~M.}\ \bibnamefont
  {Noyes}},\ }\href@noop {} {\bibfield  {journal} {\bibinfo  {journal} {J.
  Chem. Phys.}\ }\textbf {\bibinfo {volume} {60}},\ \bibinfo {pages} {1877}
  (\bibinfo {year} {1974})}\BibitemShut {NoStop}%
\bibitem [{\citenamefont {Tyson}\ and\ \citenamefont {Fife}(1980)}]{tyson}%
  \BibitemOpen
  \bibfield  {author} {\bibinfo {author} {\bibfnamefont {J.~J.}\ \bibnamefont
  {Tyson}}\ and\ \bibinfo {author} {\bibfnamefont {P.~C.}\ \bibnamefont
  {Fife}},\ }\href@noop {} {\bibfield  {journal} {\bibinfo  {journal} {J. Chem.
  Phys.}\ }\textbf {\bibinfo {volume} {73}},\ \bibinfo {pages} {2224} (\bibinfo
  {year} {1980})}\BibitemShut {NoStop}%
\bibitem [{\citenamefont {Jahnke}\ \emph {et~al.}(1989)\citenamefont {Jahnke},
  \citenamefont {Skaggs},\ and\ \citenamefont {Winfree}}]{jahnkeskaggs}%
  \BibitemOpen
  \bibfield  {author} {\bibinfo {author} {\bibfnamefont {W.}~\bibnamefont
  {Jahnke}}, \bibinfo {author} {\bibfnamefont {W.~E.}\ \bibnamefont {Skaggs}},
  \ and\ \bibinfo {author} {\bibfnamefont {A.~T.}\ \bibnamefont {Winfree}},\
  }\href@noop {} {\bibfield  {journal} {\bibinfo  {journal} {J. Phys. Chem.}\
  }\textbf {\bibinfo {volume} {93}},\ \bibinfo {pages} {740} (\bibinfo {year}
  {1989})}\BibitemShut {NoStop}%
\bibitem [{\citenamefont {Schnakenberg}(1979)}]{schnakenberg}%
  \BibitemOpen
  \bibfield  {author} {\bibinfo {author} {\bibfnamefont {J.}~\bibnamefont
  {Schnakenberg}},\ }\href@noop {} {\bibfield  {journal} {\bibinfo  {journal}
  {J. Theor. Biol.}\ }\textbf {\bibinfo {volume} {81}},\ \bibinfo {pages} {389
  } (\bibinfo {year} {1979})}\BibitemShut {NoStop}%
\bibitem [{\citenamefont {Biancalani}\ \emph {et~al.}(2011)\citenamefont
  {Biancalani}, \citenamefont {Galla},\ and\ \citenamefont
  {McKane}}]{gallawaves}%
  \BibitemOpen
  \bibfield  {author} {\bibinfo {author} {\bibfnamefont {T.}~\bibnamefont
  {Biancalani}}, \bibinfo {author} {\bibfnamefont {T.}~\bibnamefont {Galla}}, \
  and\ \bibinfo {author} {\bibfnamefont {A.~J.}\ \bibnamefont {McKane}},\
  }\href@noop {} {\bibfield  {journal} {\bibinfo  {journal} {Phys. Rev. E}\
  }\textbf {\bibinfo {volume} {84}},\ \bibinfo {pages} {026201} (\bibinfo
  {year} {2011})}\BibitemShut {NoStop}%
\bibitem [{\citenamefont {Fulger}\ \emph {et~al.}(2008)\citenamefont {Fulger},
  \citenamefont {Scalas},\ and\ \citenamefont {Germano}}]{fulger}%
  \BibitemOpen
  \bibfield  {author} {\bibinfo {author} {\bibfnamefont {D.}~\bibnamefont
  {Fulger}}, \bibinfo {author} {\bibfnamefont {E.}~\bibnamefont {Scalas}}, \
  and\ \bibinfo {author} {\bibfnamefont {G.}~\bibnamefont {Germano}},\
  }\href@noop {} {\bibfield  {journal} {\bibinfo  {journal} {Phys. Rev. E}\
  }\textbf {\bibinfo {volume} {77}},\ \bibinfo {pages} {021122} (\bibinfo
  {year} {2008})}\BibitemShut {NoStop}%
\bibitem [{\citenamefont {Kozubowski}(2000)}]{kozubowski}%
  \BibitemOpen
  \bibfield  {author} {\bibinfo {author} {\bibfnamefont {T.~J.}\ \bibnamefont
  {Kozubowski}},\ }\href@noop {} {\bibfield  {journal} {\bibinfo  {journal} {J.
  Comput. Appl. Math.}\ }\textbf {\bibinfo {volume} {116}},\ \bibinfo {pages}
  {221} (\bibinfo {year} {2000})}\BibitemShut {NoStop}%
\end{thebibliography}
\end{document}